\documentclass[paper,notoc,nohyper]{JHEP3}
\usepackage{epsfig}
\usepackage{amsbsy} 
\def\S{S_{\epsilon}}
\def\C{{\cal C}}
\def\T{{\cal T}}

\def\M{{\cal M}}
\def\P{{\cal P}}
\def\bzero{\beta_0}

\def\Bab{\,{\rm Bub}(\sab)}
\def\Bac{\,{\rm Bub}(\sac)}
\def\Bbc{\,{\rm Bub}(\sbc)}

\def\Boxabac{\,{\rm Box}^6(\sab,\sac)}
\def\Boxacbc{\,{\rm Box}^6(\sac,\sbc)}
\def\Boxabbc{\,{\rm Box}^6(\sab,\sbc)}
\def\dd{d}
\def\dma{(\dd-1)}
\def\dmb{(\dd-2)}
\def\dmc{(\dd-3)}
\def\dmd{(\dd-4)}
\def\dmf{(\dd-6)}

\def\sab{s_{12}}
\def\sac{s_{13}}
\def\sbc{s_{23}}

\def\Taa{T^\dagger_{11}}
\def\Tab{T^\dagger_{12}}

\def\Tba{T^\dagger_{21}}
\def\Tbb{T^\dagger_{22}}

\def\T{T^\dagger}
\def\e{\epsilon}

\def\lnx{X}
\def\lny{Y}
\def\Ls{S}

\def\Libx{{\rm Li}_2(x)}
\def\Liby{{\rm Li}_2(y)}
\def\Licx{{\rm Li}_3(x)}
\def\Licy{{\rm Li}_3(y)}
\def\Lidx{{\rm Li}_4(x)}
\def\Lidy{{\rm Li}_4(y)}
\def\Lidz{{\rm Li}_4\Biggl(\frac{x-1}{x}\Biggr)}
\def\IPI{i\pi}

\def\tttouuu{\frac{t^3}{u^3}}
\def\ttouu{\frac{t^2}{u^2}}
\def\tou{\frac{t}{u}}
\def\one{}
\def\sot{\frac{s}{t}}
\def\tos{\frac{t}{s}}
\def\ssott{\frac{s^2}{t^2}}
\def\ttoss{\frac{t^2}{s^2}}

\def\CA{C_A}
\def\CF{C_F}

\def\TR{T_R}
\def\NF{N_F}

\def\M{|{\cal M}\rangle}

\def\C{{\cal C}}
\renewcommand\O[1]{{\cal O}\left(#1\right)}
\def\as{\ensuremath{\alpha_{s}}}
\def\a0{\alpha_0}

\def\Re{\mathop{\rm Re}}

\def\beq{\begin{equation}}
\def\eeq{\end{equation}}

\def\beqn{\begin{eqnarray}}
\def\eeqn{\end{eqnarray}}

\def\lq{\left[}
\def\rq{\right]}

\def\({\left(}
\def\){\right)}

\def\ket#1{|{#1}\rangle}
\def\bra#1{\langle{#1}|}
\def\braket#1#2{\langle #1 |#2 \rangle}
\def\cm{{\cal M}}

\def\MSbar{$\overline{{\rm MS}}$}

\def\bom#1{{\mbox{\boldmath $#1$}}}

\def\fs{\(-\frac{\mu^2}{\sab}\)^\ep }
\def\ft{\(-\frac{\mu^2}{\sbc}\)^\ep }
\def\fu{\(-\frac{\mu^2}{\sac}\)^\ep }
\def\fsd{\(-\frac{\mu^2}{\sab}\)^{2\ep} }

\def \ep{\epsilon}

\def\Ls{L_s}

\def\Lx{L_x}
\def\Ly{L_y}
\def\Lx{X}
\def\Ly{Y}
\def\Ls{S}

\def\Libx{{\rm Li}_2(x)}
\def\Liby{{\rm Li}_2(y)}
\def\Licx{{\rm Li}_3(x)}
\def\Licy{{\rm Li}_3(y)}

\def\Lidx{{\rm Li}_4(x)}
\def\Lidy{{\rm Li}_4(y)}
\def\Lidz{{\rm Li}_4(z)}

\def\astu{{\bom A}(\ep,\sab,\sbc,\sac)}
\def\asut{{\bom A}(\ep,\sab,\sac,\sbc)}
\def\bstu{{\bom B}(\ep,\sab,\sbc,\sac)}
\def\bsut{{\bom B}(\ep,\sab,\sac,\sbc)}
\def\buts{{\bom B}(\ep,\sac,\sbc,\sab)}
\def\btus{{\bom B}(\ep,\sbc,\sac,\sab)}
\def\cs{{\bom C}(\ep,\sab)}

\title{\boldmath 
Two-loop QCD helicity amplitudes for massless quark-massless gauge boson scattering
}
\author{
E.~W.~N.~Glover$^a$ and M.~E.~Tejeda-Yeomans$^{b}$\\
$^a$Department of Physics, 
University of Durham, 
Durham DH1 3LE, 
England\\ 
$^b$C.N.~Yang Institute for Theoretical Physics,
Stony Brook University,
Stony Brook, NY 11794-3840.
U.S.A.\\ 
E-mail:  \email{E.W.N.Glover@durham.ac.uk}, 
\email{tejeda@insti.physics.sunysb.edu}}
\abstract{
We present the two-loop helicity amplitudes for the scattering of massless
quarks and massless gauge bosons in QCD.
We use projector techniques to compute the coefficients of the general
tensor describing the two-quark two-boson amplitude in conventional
dimensional regularisation and use these coefficients to derive the
helicity amplitudes in the 't Hooft-Veltman scheme. The structure of the
infrared divergences agrees with that predicted by Catani while
expressions for the finite remainders for $q \bar q \to gg$, $q \bar q \to
g\gamma$,
$q \bar q \to \gamma\gamma$ and the processes related by crossing symmetry
 are given in
terms of logarithms and  polylogarithms that are real in the physical
region.  We have checked that the interference of tree and two-loop helicity
amplitudes, summed over helicities and colours, reproduces 
the previous results for the finite remainders for 
interference of tree and two-loop amplitudes
given in Refs.~[1,2].  We also find agreement with the
two-loop helicity amplitudes for quark-gluon scattering presented in
Ref.~[3].  
}

\preprint{{DCTP/03/44}, {IPPP/03/22}, {YITP-SB-03-14}, {hep-ph/yymmnnn}}

\begin{document}
%%%%%to make all the references appear 
%%%%%eventhough they are not in use
%%%%%\nocite{*}
\nocite{qqgg,qqpp,BFDqqgg}

\section{Introduction}

QCD perturbation theory has proven to be a reliable tool with which to generate theoretical predictions of
hard scattering processes that can be compared with experimental results from electron-positron
annihilation, electron-proton and hadron-hadron scattering.  In such studies, the theoretical analysis of
an observable at next-to-leading order (NLO) in $\as$ has become a {\it standard candle},  the best
theoretical prediction to compare against experiment. These NLO calculations are also used to provide an
estimate of the uncalculated theoretical uncertainties via the variation in the renormalisation
(and factorisation) scale. 

However, the large amounts of data soon to become available from for example, high-energy hadronic collider
runs place a demand for more accurate theoretical predictions.  To go beyond NLO to the next order in QCD
perturbation theory, next-to-next-to-leading order (NNLO),  has a direct impact on several aspects of the
theoretical predictions. For example~\cite{nigelRCLL}, the renormalisation (and factorization) scale 
uncertainty is significantly diminished,  there is a better matching of the parton level jet algorithm used
in theory and the hadron level jet algorithm used by experiment and the contributions
attributed to non-perturbative power corections may also be reduced.

There are several ingredients that need to be considered to calculate the NNLO corrections for a specific $
2 \to 2$ scattering process,
\begin{itemize}
\item tree-level $2 \to 4$ partonic amplitudes
\item one-loop $2\to 3$ partonic amplitudes,
\item two-loop $2\to 2$ partonic amplitudes.
\end{itemize}
The tree-level six-point amplitudes~\cite{6g1,6g2,6g3,6g4,4g2q1,4g2q2,2g4q,6q}
 and one-loop five point amplitudes~\cite{5g,3g2q,1g4q} have been known for some time while
there has been recent progress in computing two-loop amplitudes. Parton distribution functions from a
global fit to available data  with observables computed consistently at NNLO~\cite{nigelRCLL} coupled 
with a complete knowledge of the three-loop  splitting functions~\cite{moms1,moms2,moms3,Gra1,NV1,NV2,NVplb,VMVdis,carola}
to generate the evolution of the parton  distribution functions at this order are also required.

In this particular paper we are concerned with processes involving two massless 
quarks and two massless gauge bosons,
$$ q\bar q \to gg, \qquad q\bar q \to g \gamma, \qquad q\bar q \to \gamma\gamma$$
and the processes related  by time reversal and crossing symmetry.
In previous work~\cite{qqgg,qqpp}, we computed both the self-interference of the one-loop amplitudes and
the interference of  the tree and two-loop amplitudes for these
processes in conventional dimensional regularisation (CDR). In that work, 
the matrix-elements  were averaged over all external helicities and summed over
colours in the high energy limit where all quark masses can be neglected.  Here, we extend these
calculations to now compute the helicity amplitudes for these processes.

The one-loop helicity amplitudes for parton-parton scattering processes have been known for some time~\cite{KST}.
More recently, there have been a few two-loop helicity amplitudes calculated within CDR with $d=4-2\ep$ dimensions: 
\begin{itemize} 
\item for the $gg \to \gamma
\gamma$~\cite{BFDggpp}, $\gamma \gamma \to \gamma \gamma$~\cite{BFDGpppp,BGMBpppp} and  $gg \to gg$
~\cite{BDKgggg,BFDgggg}
scattering processes, the explicit contraction of the amplitude with the external  polarization vectors was
used. 
\item for the for the $\gamma \gamma \to \gamma \gamma$~\cite{BGMBpppp} and 
$e^+ e^- \to 3$~jets~\cite{helamp3j}  scattering processes, the individual components of the Lorentz-invariant amplitude
decomposition were  projected out. 
\end{itemize}

After the explicit helicity structure is obtained via any of these methods, one needs to tackle a
large number of two-loop integrals with different topologies and tensor structure.  It is now well
known (see for example~\cite{houchesQCD,thomasRCLL,thomasHAD} and references therein), that these
integrals can be reduced~\cite{laporta} to a handful of {\it master integrals} using
integration-by-parts~\cite{IBP1,IBP2}  and Lorentz-invariance~\cite{diffeqLI} identities. For the
process at hand, the relevant master integrals are 10  massless four-point functions with all legs
on-shell which have been calculated in the past few years using a wide variety of
methods~\cite{xtri1,xtri2,AGO1,AGO2,AGO3,planarA,nonplanarA,planarB,nonplanarB,planarIR}.

In this paper, we follow a similar appoach to extract the helicity structure of the amplitude as
the one outlined in Refs.~\cite{BGMBpppp,helamp3j}. We start with a general tensor that describes the two
quark-two boson amplitude and use gauge invariance to find relations among the tensor
coefficients. These coefficients can be extracted using a set of projectors that are valid at any
order in perturbation theory.  The projectors saturate
the tensor structure and produce scalar integrals. The
methodology of reducing these scalar integrals to master integrals is exactly the same as for the
calculation of the spin summed matrix elements~\cite{qqgg,qqpp}.  Note that the tensor coefficients
are evaluated in CDR.
 
Then, we provide the perturbative expansion for the one- and two-loop tensor coefficients and
remove ultraviolet  (UV) divergences at each order in $\as$ by renormalisation within the
\MSbar~scheme. The infrared (IR) divergent structure is shown to agree with the one obtained in
Refs.~\cite{catani}. Note that in extracting the coefficients of the general tensor structure,
conventional dimensional regularisation is preserved.   

Finally, the helicity amplitudes are obtained from the general tensor using standard helicity
techniques and within the t'Hooft-Veltman (HV) scheme where the external states are four-dimensional
and the internal states are kept in 4-2$\epsilon$-dimensions~\cite{hv1,hv2}. 
The finite helicity amplitudes
are the main new results presented in this paper and we give explicit analytic expressions valid
for each helicity configuration and process in terms of logarithms and polylogarithms that are
real in the physical domain.

During the writing of this paper, we became aware of an independent calculation of the
two-loop helicity amplitudes for quark-gluon scattering~\cite{BFDqqgg}.  
This paper~\cite{BFDqqgg} uses a rather different method to extract the helicity amplitudes in a variety
of schemes.    Where the two calculations overlap (in the HV scheme), we have compared the finite remainder
terms for the two-loop $q\bar q \to gg$ processes  and find complete agreement.

The paper is organised as follows. In Section~\ref{sec:notation} we introduce notation and discuss
the general tensor structure of the two quark-two boson amplitude.  Gauge invariance is used to
reduce the number of independent tensor structures.   We give projectors to isolate the various
coefficients and relate the individual coefficients to the interference of amplitudes. 
Section~\ref{sec:perturbative}  shows the perturbative expansion for the tensor coefficients
computed within CDR and presents an analysis of their UV and IR divergent structure. The
coefficients are renormalised in the \MSbar~scheme to remove all UV divergences, while the IR
poles are predicted using the results of Ref.~\cite{catani}.   Expressions for the one-loop
tensor coefficients to all orders in $\epsilon$ appear in Section~\ref{subsec:oneloop} in terms of
one-loop bubble integral and the one-loop box integral in $d=6-2\epsilon$.   
The relationship between the tensor coefficients and the helicity amplitudes 
in the HV scheme as described in Section~\ref{sec:helamp}.   The UV and IR
structure of the helicity amplitudes is exactly the same as for the tensor coefficients, and
finite $\O{\ep^0}$ remainders are obtained by subtracting the  predicted pole structure from the
explicit calculation both at one-loop and two-loop.   The main results
of this paper are the finite remainders of the two-loop helicity amplitudes.  
The expressions are rather lengthy and we confine ourselves to presenting the
amplitudes for the $q\bar q \to gg$ process in Appendix~\ref{app:twoloop}. 
For completeness, the finite one-loop remainders are
collected in Appendix~\ref{app:oneloop}.  Results for the processes related by crossing symmetry and
those processes involving photons can be obtained from the authors.
Finally, in
Section~\ref{sec:conclusions} we present a discussion of our results and conclude.

\section{Notation}
\label{sec:notation}

The generic process we consider in this paper is
\begin{equation} 0 \to q(p_1,\lambda_1) + \bar q(p_2,\lambda_2) +
b_3(p_3,\lambda_3) + b_4(p_4,\lambda_4),
\end{equation}
where the bosons $b_3$ and $b_4$ are either gluons or photons.
All particles are light-like with momenta satisfying,
\begin{equation}
0 \to p_1^\mu +p_2^\mu +p_3^\mu +p_4^\mu, \qquad \qquad p_i^2 = 0.
\end{equation}
Physical processes are obtained by crossing particles into the initial state.
The associated Mandelstam variables are given by
\begin{equation}
s_{12} = (p_1+p_2)^2, \qquad s_{23} = (p_2+p_3)^2, \qquad s_{13} = (p_1+p_3)^2, 
\qquad s_{12}+s_{23}+s_{13} = 0.
\end{equation}

We work in conventional dimensional regularisation  treating all external 
quark and boson states in $d$ dimensions  and renormalise the ultraviolet 
divergences in the \MSbar\ scheme. The bare coupling $\a0$ is related to 
the running coupling $\as \equiv \alpha_s(\mu^2)$  at renormalisation 
scale $\mu$, by
\beq
\label{eq:alpha}
\a0 \mu_0^{2\e} \,  \S = \as \,  \mu^{2\e} \, \lq 1 - \frac{\beta_0}{\ep}  
\, \left(\frac{\as}{2\pi}\right) + \( \frac{\beta_0^2}{\ep^2} - \frac{\beta_1}{2\ep} \)  \, 
\left(\frac{\as}{2\pi}\right)^2
+\O{\as^3} \rq,
\eeq
where
\beq
\S = (4 \pi)^\ep e^{-\ep \gamma},  \quad\quad \gamma=0.5772\ldots=
{\rm Euler\ constant}
\eeq
is the typical phase-space volume factor in $d=4-2\ep$ dimensions
and $\mu_0^2$ is the mass parameter introduced 
in dimensional regularisation~\cite{dreg1,dreg2,hv1,hv2} to maintain a 
dimensionless coupling 
in the bare QCD Lagrangian density.

The first two coefficients of the QCD beta function, 
$\beta_0$ and $\beta_1$, for $N_F$ (massless) quark flavours, are
\beq
\label{betas}
\beta_0 = \frac{11 \CA - 4 T_R \NF}{6} \;\;, \;\; \;\;\;\;
\beta_1 = \frac{17 \CA^2 - 10 \CA T_R \NF - 6 \CF T_R \NF}{6} \;\;,
\eeq
where $N$ is the number of colours and 
\beq
\CF = \frac{N^2-1}{2N}, \qquad \CA = N, \qquad T_R = \frac{1}{2}.
\eeq

\subsection{The general tensor}
\label{subsec:genten}
The most general tensor structure for the amplitude, $\M$, is
\begin{eqnarray}
\label{eq:gentensor}
\M &=&  
\bar u(p_1)\slash \!\!\! p_3 u(p_2) \left(
  A_{11} \epsilon_3.p_1~\epsilon_4.p_1 
+ A_{12} \epsilon_3.p_1~\epsilon_4.p_2
\right)
\nonumber \\
&+& \bar u(p_1)\slash \!\!\!p_3 u(p_2) \left(
  A_{21} \epsilon_3.p_2~\epsilon_4.p_1
+ A_{22}  \epsilon_3.p_2 ~\epsilon_4.p_2
\right)
\nonumber \\
&+& \bar u(p_1)\slash \!\!\!\epsilon_4 u(p_2) \left(
  B_{1} \epsilon_3.p_1
+ B_{2} \epsilon_3.p_2
\right)
\nonumber \\
&+&\bar u(p_1)\slash \!\!\!\epsilon_3 u(p_2) \left( 
 C_{1} \epsilon_4.p_1
+C_{2} \epsilon_4.p_2
\right)
\nonumber \\
&+& D_{1} \bar u(p_1)\slash \!\!\!\epsilon_3 \slash \!\!\!p_3 \slash \!\!\!\epsilon_4 u(p_2) 
\nonumber \\
&+& D_{2} \bar u(p_1)\slash \!\!\!\epsilon_4 \slash \!\!\!p_3 \slash \!\!\!\epsilon_3 u(p_2) \; ,
\end{eqnarray}
where the constraints $\epsilon_3 \cdot p_3 = \epsilon_4 \cdot p_4 = 0$ have been applied.
All coefficients depend on the identity of the gauge bosons and
are functions of $\sac$ and $\sbc$ (and implicitly $\sab =
-\sac-\sbc$) where $s_{ij}= (p_i+p_j)^2$.
This tensor structure is a priori $\dd$-dimensional, since the 
dimensionality of the external states has not yet been specified. 
The amplitude $\M$ obeys the Ward identity when one of the 
gauge boson polarization vectors is replaced with the 
corresponding boson momentum,
\begin{equation}
\M  (\epsilon_3 \to p_3) = \M  (\epsilon_4 \to p_4) = 0.
\end{equation}
These constraints yield relations amongst the 10 distinct tensor structures and 
applying
these identities gives the gauge-invariant form of the tensor,
\begin{eqnarray}
\label{eq:tensor}
\M ~~=&& A_{11}(\sac,\sbc) \,T_{11} 
                     + A_{12}(\sac,\sbc)\, T_{12} \nonumber \\
&+&
A_{21}(\sac,\sbc)  \,T_{21}
+ A_{22}(\sac,\sbc) \, T_{22}\nonumber \\
&+&B(\sac,\sbc)  \,T,
\end{eqnarray}
where $A_{IJ}$ and $B$ are five gauge-independent functions and 
the tensor structures $T_{IJ}$ and $T$ are given by
\begin{eqnarray}
T_{1J} &=&\bar u(p_1)\slash \!\!\! p_3 u(p_2)
\epsilon_3.p_1\epsilon_4.p_{J} 
-\frac{s_{13}}{2}
\bar u(p_1)\slash \!\!\! \epsilon_3 u(p_2)\epsilon_4.p_{J}
-\frac{s_{J4}}{4} 
\bar u(p_1)\slash \!\!\!\epsilon_3 \slash \!\!\!p_3 \slash \!\!\!\epsilon_4
u(p_2),\\
T_{2J} &=&\bar u(p_1)\slash \!\!\! p_3 u(p_2)
\epsilon_3.p_2\epsilon_4.p_{J} 
-\frac{s_{23}}{2}
\bar u(p_1)\slash \!\!\! \epsilon_3 u(p_2)\epsilon_4.p_{J}
-\frac{s_{J4}}{4} 
\bar u(p_1)\slash \!\!\!\epsilon_4 \slash \!\!\!p_3 \slash \!\!\!\epsilon_3
u(p_2),\\
T&=& s_{23}\left(
\bar u(p_1)\slash \!\!\!\epsilon_4 u(p_2)  
\epsilon_3.p_1
+\frac{1}{2}
\bar u(p_1)\slash \!\!\!\epsilon_3 \slash \!\!\!p_3 \slash \!\!\!\epsilon_4
u(p_2) \right)
\nonumber \\
&-&s_{13}\left(
\bar u(p_1)\slash \!\!\!\epsilon_4 u(p_2)  
\epsilon_3.p_2
+\frac{1}{2}
\bar u(p_1)\slash \!\!\!\epsilon_4 \slash \!\!\!p_3 \slash \!\!\!\epsilon_3
u(p_2) \right),
\end{eqnarray}
with $s_{14} = s_{23}$.
Each of the tensor structures satisfies the  Ward identities.
The tensor coefficients are 
further related by symmetry under the interchange of the momenta (and colour
labels) of the two gauge bosons,
\begin{eqnarray}
A_{21}(\sac,\sbc) &=& - A_{12}(\sbc,\sac),\nonumber \\
A_{22}(\sac,\sbc) &=& - A_{11}(\sbc,\sac),\nonumber \\
B(\sac,\sbc) &=& B(\sbc,\sac).
\end{eqnarray}

\subsection{Projectors for the tensor coefficients}
\label{subsec:projectors}

The coefficients $A_{IJ}$ and $B$  may be easily extracted from a Feynman diagram calculation,
using projectors such that
\begin{equation}
\sum_{\rm spins} {\cal P}(X) ~\M  = X(\sac,\sbc ).
\end{equation}
The explicit forms for the projectors in $\dd$ space-time dimensions are,
\begin{eqnarray}
{\cal P}(A_{11}) &=& 
\frac{\dd}{2\sab^2\sac\sbc\dmc}\Taa
+\frac{\dmd}{2\sab^2\sac\sbc\dmc}\Tbb
+\frac{1}{2\sab^2\sac\sbc\dmc}\T\nonumber \\
&&
+\frac{(2\sac-\dmd\sbc)}{2\sab^2\sac^2\sbc\dmc}\Tab
+\frac{(2\sbc-\dmd\sac)}{2\sab^2\sac\sbc^2\dmc}\Tba,
\nonumber \\
%{\cal P}(A_{22}) &=& 
%\frac{\dmd}{2\sab^2\sac\sbc\dmc}\Taa
%+\frac{\dd}{2\sab^2\sac\sbc\dmc}\Tbb
%-\frac{1}{2\sab^2\sac\sbc\dmc}\T\nonumber \\
%&&+\frac{(2\sac-\dmd\sbc)}{2\sab^2\sac^2\sbc\dmc}\Tab
%+\frac{(2\sbc-\dmd\sac)}{2\sab^2\sac\sbc^2\dmc}\Tba,
%\nonumber \\
{\cal P}(A_{12}) &=& 
\frac{(2\sac-\dmd\sbc)}{2\sab^2\sac^2\sbc\dmc}\Taa
+\frac{(2\sac-\dmd\sbc)}{2\sab^2\sac^2\sbc\dmc}\Tbb
-\frac{(2\sac-\dmd\sbc)}{2\sab^2\sac^2\sbc\dmc\dmd}\T\nonumber \\
&&
+\frac{(4\dmc\sac^2+4\dmd\sac\sbc+\dd\dmd\sbc^2)}{2\sab^2\sac^3\sbc\dmc\dmd}\Tab\nonumber \\
&&
+\frac{(2\dmd\sab^2+\dmb\dmf\sac\sbc)}{2\sab^2\sac^2\sbc^2\dmc\dmd}\Tba,
\nonumber \\
%{\cal P}(A_{21}) &=& 
%\frac{(2\sbc-\dmd\sac)}{2\sab^2\sac\sbc^2\dmc}\Taa
%+\frac{(2\sbc-\dmd\sac)}{2\sab^2\sac\sbc^2\dmc}\Tbb
%+\frac{(2\sbc-\dmd\sac)}{2\sab^2\sac\sbc^2\dmc\dmd}\T\nonumber \\
%&&
%+\frac{(2\dmd\sab^2+\dmb\dmf\sac\sbc)}{2\sab^2\sac^2\sbc^2\dmc\dmd}\Tab\nonumber \\
%&&
%+\frac{(4\dmc\sbc^2+4\dmd\sac\sbc+\dd\dmd\sac^2)}{2\sab^2\sac\sbc^3\dmc\dmd}\Tba,
%\nonumber \\
{\cal P}(B) &=& 
\frac{1}{2\sab^2\sac\sbc\dmc}\Taa
-\frac{1}{2\sab^2\sac\sbc\dmc}\Tbb
+\frac{1}{2\sab^2\sac\sbc\dmd}\T\nonumber \\
&&
-\frac{(2\sac-\dmd\sbc)}{2\sab^2\sac^2\sbc\dmc\dmd}\Tab
+\frac{(2\sbc-\dmd\sac)}{2\sab^2\sac\sbc^2\dmc\dmd}\Tba.
\nonumber \\
\end{eqnarray} 
The projectors for $A_{21}$ and $A_{22}$ are obtained by the permutations
\begin{eqnarray}
{\cal P}(A_{22}) &=& {\cal P}(A_{11}) \left \{ \sac \leftrightarrow \sbc, \Taa
\leftrightarrow \Tbb, \Tab \leftrightarrow \Tba, \T \to -\T\right\},
\nonumber \\
{\cal P}(A_{21}) &=& {\cal P}(A_{12}) \left \{ \sac \leftrightarrow \sbc, \Taa
\leftrightarrow \Tbb, \Tab \leftrightarrow \Tba, \T \to -\T\right\}.
\nonumber \\
\end{eqnarray}

\subsection{Reconstruction of the matrix elements}

Starting from the tensor coefficients, the interference (or square) of 
amplitudes can be easily extracted in a variety of schemes.   
In the HV scheme, the external particles are treated in 4-dimensions and we
find that,  
\begin{eqnarray}
\label{eq:square}
\sum_{\rm spins} \langle {\cal M} \M  &=& 
+ \frac{\sac^3\sbc}{2} \left(2B-A_{11}+A_{12}\right)
\left(2B-A_{11}+A_{12}\right)^\dagger\nonumber \\
&&+ \frac{\sac\sbc^3}{2} \left(2B-A_{21}+A_{22}\right)
\left(2B-A_{21}+A_{22}\right)^\dagger\nonumber \\
&&+\frac{\sac\sbc\sab^2}{2} \left(A_{11}A_{11}^\dagger+A_{22}A_{22}^\dagger\right).
\end{eqnarray}
Note that the tensor coefficients are in general matrices in colour space and
the colour sum has still to be performed in evaluating, for example,  
$A_{11}A_{11}^\dagger$. 
In CDR, additional terms of order $\epsilon$
and $\epsilon^2$ are straightforwardly generated.  The resulting formula is,
however,
rather lengthy and we do not give it here.

\section{The perturbative expansion of the tensor coefficients}
\label{sec:perturbative}

The three processes that we wish to consider in detail are,
\begin{eqnarray}
\label{eq:qqgg}
0 &\to& q \bar q g g,\\
\label{eq:qqgp}
0 &\to& q \bar q g \gamma ,\\
\label{eq:qqpp}
0 &\to& q \bar q \gamma \gamma.
\end{eqnarray}

The amplitude for the process $\P$ (where $\P = gg, ~g\gamma$ and $\gamma\gamma$)
has the perturbative expansion of the form,
\begin{eqnarray}
\label{eq:perturb}
|{\cal M}^\P\rangle &=&  c^\P \left[
|{\cal M}^{\P,(0)}  \rangle
+ \left(\frac{\alpha_s}{2\pi}\right)|{\cal M}^{\P,(1)}  \rangle
+ \left(\frac{\alpha_s}{2\pi}\right)^2 |{\cal M}^{\P,(2)} \rangle
+ {\cal O}(\alpha_s^3) \right],
\end{eqnarray}
where the overall coupling
factor is given by
\begin{equation}
\label{eq:alphaq}
c^{\P} = 4\pi \,\alpha_s^q \,(\alpha e_q^2)^{1-q},
\end{equation}
with $q=1,1/2,0$ for $\P=gg,g\gamma,\gamma\gamma$ respectively.
$ |{\cal M}^{\P,(n)}\rangle$ denotes the $n$-loop contribution to the amplitude.
Denoting the tensor coefficients as $X^\P$ (for $X = A_{IJ}$ and $B$), we find the 
perturbative expansions of the form
\begin{eqnarray}
\label{eq:perturbX}
X_{IJ}^\P &=&  c^\P \left[
X_{IJ}^{\P,(0)}  
+ \left(\frac{\alpha_s}{2\pi}\right) X_{IJ}^{\P,(1)}  
+ \left(\frac{\alpha_s}{2\pi}\right)^2 X_{IJ}^{\P,(2)} 
+ {\cal O}(\alpha_s^3) \right],
\end{eqnarray}
where the dependence on the kinematic scales is implicit.
The $n$-loop perturbative coefficients are vectors in colour space and can be further
decomposed as,
\begin{eqnarray}
\label{eq:colour}
X^{gg,(n)} &=& 
\sum_{i=1}^3 \;\C^{gg,[i]}\;X^{gg,(n),[i]},\nonumber\\
X^{g\gamma,(n)} &=& 
\C^{g\gamma,[1]}\;X^{g\gamma,(n),[1]},\nonumber\\
\C^{\gamma\gamma,[1]} &=&\delta_{i_1i_2}\;X^{\gamma\gamma,(n),[1]},
\end{eqnarray}
where
\begin{equation}
\C^{gg,[1]} = \left({\bom T}^{a_3}{\bom T}^{a_4}\right)_{i_2i_1}, \qquad
\C^{gg,[2]} = \left({\bom T}^{a_4}{\bom T}^{a_3}\right)_{i_2i_1}, \qquad
\C^{gg,[3]} = {\bom \delta^{a_3a_4}\delta_{i_1i_2}},
\end{equation}
and
\begin{equation}
\C^{g\gamma,[1]}=\left({\bom T}^{a_3}\right)_{i_2i_1},\qquad 
\C^{\gamma\gamma,[1]} = \delta_{i_1i_2}.
\end{equation}
Index $i$ denotes a colour label in the fundamental representation while index $a$ is in the adjoint.
Note that the normalisation of the colour matrices is given by $Tr(\bom{T}^{a_1}\bom{T}^{a_2}) =
\frac{1}{2}{\bom \delta}^{a_1a_2}$.

Symmetry under interchange of the two gluons
gives the additional relations,
\begin{eqnarray}
\label{eq:Xsymgg}
A_{IJ}^{gg,(n),[2]} (\sac,\sbc) &=&  -A_{JI}^{gg,(n),[1]} (\sbc,\sac),\nonumber \\
A_{IJ}^{gg,(n),[3]} (\sac,\sbc) &=&  -A_{JI}^{gg,(n),[3]} (\sbc,\sac),\nonumber \\
B^{gg,(n),[2]} (\sac,\sbc) &=&  B^{gg,(n),[1]} (\sbc,\sac),\nonumber \\
B^{gg,(n),[3]} (\sac,\sbc) &=&  B^{gg,(n),[3]} (\sbc,\sac).
\end{eqnarray}
For the processes involving one or more photons, 
\begin{eqnarray}
\label{eq:Xsympp}
A_{IJ}^{g\gamma,(n),[1]} (\sac,\sbc) &=&  -A_{JI}^{g\gamma,(n),[1]} (\sbc,\sac),\nonumber \\
A_{IJ}^{\gamma\gamma,(n),[1]} (\sac,\sbc) &=&  -A_{JI}^{\gamma\gamma,(n),[1]} (\sbc,\sac),\nonumber \\
B^{g\gamma,(n),[1]} (\sac,\sbc) &=&  B^{g\gamma,(n),[1]} (\sbc,\sac),\nonumber \\
B^{\gamma\gamma,(n),[1]} (\sac,\sbc) &=&  B^{\gamma\gamma,(n),[1]} (\sbc,\sac).
\end{eqnarray}

As in previous work, we use QGRAF~\cite{QGRAF} to generate the Feynman diagrams and then use
MAPLE~\cite{MAPLE} and FORM~\cite{FORM3} to manipulate the Dirac structures and traceology 
and then reduce the scalar loop integrals to master integrals.

\subsection{Tree level results}
\label{sec:treelevel}

At tree level, none of the $T_{IJ}$ structures are present so that,
\begin{equation}
A_{IJ}^{\P,(0),[i]}(\sac,\sbc) = 0,
\end{equation}
while only the first two colour structures contribute to
$B^{gg,(0)}$,
\begin{eqnarray}
%B^{gg,(0),[1]}(\sac,\sbc) 
% &=& -\frac{2}{\sab\sac},\nonumber \\
%B^{gg,(0),[2]}(\sac,\sbc) 
% &=& -\frac{2}{\sab\sbc},\nonumber \\
B^{gg,(0),[1]}(\sac,\sbc) 
 &=& -\frac{2}{\sab\sbc},\nonumber \\
B^{gg,(0),[2]}(\sac,\sbc) 
 &=& -\frac{2}{\sab\sac},\nonumber \\
B^{gg,(0),[3]}(\sac,\sbc) 
 &=& 0.
\end{eqnarray}
For the other processes,
\begin{eqnarray}
B^{g\gamma,(0),[1]}(\sac,\sbc) =B^{\gamma\gamma,(0),[1]}(\sac,\sbc) 
&=& B^{gg,(0),[1]}(\sac,\sbc)  + B^{gg,(0),[2]}(\sac,\sbc) \nonumber \\
&=&
\frac{2}{\sac\sbc}.
\end{eqnarray}

\subsection{One-loop results}
\label{subsec:oneloop}

At one-loop, all of the tensor and colour structures contribute.
For the two-gluon process, we find the following (unrenormalised) structure,
\begin{eqnarray}
A_{11}^{gg,(1),[1]}(\sac,\sbc) &=& \CF \,f_{1}(\sac,\sbc) 
+ \CA \,g_{1}(\sac,\sbc) + (2\NF-\CA(\dd-2))\,h(\sab),\nonumber \\
A_{11}^{gg,(1),[2]}(\sac,\sbc) &=& -A_{11}^{gg,(1),[1]}(\sbc,\sac),\nonumber \\
A_{11}^{gg,(1),[3]}(\sac,\sbc) &=& 0,\nonumber \\
A_{12}^{gg,(1),[1]}(\sac,\sbc) &=& \CF \,f_{2}(\sac,\sbc) 
+ \CA \,g_{2}(\sac,\sbc) + (2\NF-\CA(\dd-2))\,h(\sab),\nonumber \\
A_{12}^{gg,(1),[2]}(\sac,\sbc) &=& \CF \,f_{3}(\sac,\sbc) 
+ \CA \,g_{3}(\sac,\sbc) -(2\NF-\CA(\dd-2))\,h(\sab),\nonumber \\
A_{12}^{gg,(1),[3]}(\sac,\sbc) &=& h_1(\sac,\sbc),\nonumber \\
B^{gg,(1),[1]}(\sac,\sbc) &=& \CF \,f_{4}(\sac,\sbc) 
+ \CA \,g_{4}(\sac,\sbc),\nonumber \\
B^{gg,(1),[2]}(\sac,\sbc) &=& B^{gg,(1),[1]}(\sbc,\sac) ,\nonumber \\
B^{gg,(1),[3]}(\sac,\sbc) &=& h_2(\sac,\sbc), 
\end{eqnarray}
while for the gluon-photon and photon-photon processes we have,
\begin{eqnarray}
A_{11}^{g\gamma,(1),[1]}(\sac,\sbc) &=& \CF \left(f_{1}(\sac,\sbc) 
-f_{1}(\sbc,\sac)\right) \nonumber \\
&&+\frac{\CA}{2}\left(g_{1}(\sac,\sbc) -g_{1}(\sbc,\sac)
-4h_3(\sac,\sbc) \right),
\nonumber \\
A_{12}^{g\gamma,(1),[1]}(\sac,\sbc) &=& \CF \left(f_{2}(\sac,\sbc) 
+\,f_{3}(\sac,\sbc)\right)\nonumber \\
&& + \frac{\CA}{2}\left(g_2(\sac,\sbc)+g_3(\sac,\sbc)-2h_1(\sac,\sbc)\right),\nonumber \\
B^{g\gamma,(1),[1]}(\sac,\sbc) &=& \CF \left(f_{4}(\sac,\sbc) 
+ \,f_{4}(\sbc,\sac)\right)\nonumber \\
&& + \frac{\CA}{2}\left(g_4(\sac,\sbc)+g_4(\sbc,\sac)-2h_2(\sac,\sbc)-2h_3(\sac,\sbc)\right),\nonumber \\ 
\end{eqnarray}
and,
\begin{eqnarray}
A_{11}^{\gamma\gamma,(1),[1]}(\sac,\sbc) &=& \CF \left(f_{1}(\sac,\sbc) 
-f_{1}(\sbc,\sac)\right), \nonumber \\
A_{12}^{\gamma\gamma,(1),[1]}(\sac,\sbc) &=& \CF \left(f_{2}(\sac,\sbc) 
+\,f_{3}(\sac,\sbc)\right) ,\nonumber \\
B^{\gamma\gamma,(1),[1]}(\sac,\sbc) &=& \CF \left(f_{4}(\sac,\sbc) 
+ \,f_{4}(\sbc,\sac)\right).
\end{eqnarray}
The functions $f_i$, $g_i$, $h_i$ and $h$ can be written in terms of the one-loop
box integral in $\dd=6-2\epsilon$ dimensions, Box$^6(s_{ij},s_{ik})$, and the
one-loop bubble, Bub$(s_{ij})$.   
For all orders in $\epsilon = (4-\dd)/2$, we find that,
\begin{equation}
h(\sab)= \frac{2\dmd}{\sab^2\dma\dmb}\Bab,
\end{equation}
while,
\begin{eqnarray}
f_1(\sac,\sbc) &=& \frac{\dmb\dmd}{\sac}\Boxabbc
-\frac{2\dmd}{\sac\sbc}\Bbc-\frac{2\dmd}{\sab\sac}\Bab,\nonumber\\
f_2(\sac,\sbc) &=& \frac{\sbc^2\dd\dmb-\sac\sbc\dmb(\dd-14)-2\sac^2(\dd-10)}{\sab\sac^2}\Boxabbc \nonumber \\
&&+\frac{2(\dd-12)\sac-2\sbc\dd}{\sab\sac^2}\Bbc+\frac{2\sbc\dd-4(\dd-8)\sac}{\sab\sac^2}\Bab,\nonumber \\
f_3(\sac,\sbc) &=& \frac{\dmd(\dd-10)\sbc^2+(3\dd^2-36\dd+92)\sac\sbc-2\dmb\sac^2}{\sab\sac\sbc}\Boxabac
\nonumber \\
&&+\frac{4\sac-4(\dd-7)\sbc}{\sab\sac\sbc}\Bac+\frac{2(\dd-10)\sbc-4\sac}{\sab\sac\sbc}\Bab,\nonumber \\
f_4(\sac,\sbc) &=&+\frac{4\dmc\sac^2-(\dd^2-15\dd+38)\sbc\sac+\dmb\sbc^2}{\sab\sac\sbc}\Boxabbc
\nonumber \\
&&+\frac{2(\dd-7)\sac-2\sbc}{\sab\sac\sbc}\Bbc + \frac{2\dmd\sbc+8\dmc\sac}{\dmd\sab\sac\sbc}\Bab,\nonumber \\
g_1(\sac,\sbc) &=& -f_1(\sac,\sbc),\nonumber \\
g_2(\sac,\sbc) &=& -\frac{\sbc\dmb(4\sac+\dd\sbc)}{\sab\sac^2}\Boxabbc + \frac{2\sbc\dd+4\sac}{\sab\sac^2}\Bbc 
\nonumber \\
&&+\frac{2(\dd-6)\sac-2\dd\sbc}{\sab\sac^2}\Bab,\nonumber \\
g_3(\sac,\sbc) &=& \frac{\dmb(2\sac-(\dd-6)\sbc)}{\sab\sbc}\Boxabac + \frac{4}{\sab\sbc}\Bab \nonumber \\
&&+\frac{2\dmd\sbc-4\sac}{\sab\sac\sbc}\Bac,\nonumber \\
g_4(\sac,\sbc) &=& \frac{(\dd^2-16\dd+36)\sac-2\dmb\sbc}{2\sab\sac}\Boxabbc\nonumber \\
&&-\frac{(\dd^2-16\dd+40)\sac-2\dmd\sbc}{\dmd\sab\sac\sbc}\Bbc-\frac{2}{\sab\sac}\Bab, \nonumber \\
h_1(\sac,\sbc) &=& -\frac{(\dd-10)(\dmb\sbc+2\sac)}{2\sab\sac}\Boxabbc 
\nonumber \\
&&+\frac{\dmd(\dd-10)(2\sac+\sbc)}{2\sab\sac}\Boxabac\nonumber \\
&&-\frac{(\dd-10)(2\sac-\dmd\sbc)}{2\sab\sac}\Boxacbc,\nonumber \\
h_2(\sac,\sbc) &=&\phantom{+}\frac{(8\dmc\sac-\dmd(\dd-10)\sbc)}{4\sab\sbc}
\Biggr(\Boxabbc+\Boxacbc\Biggl)\nonumber \\
&&+\frac{(8\dmc\sbc-\dmd(\dd-10)\sac)}{4\sab\sac}\Biggr(\Boxabac+\Boxacbc\Biggl)\nonumber \\
&&+\frac{4 + 3\dmd^2}{4\sab}\Boxacbc\nonumber \\
&& -\frac{4\dmc}{\dmd}\left(\frac{1}{\sab\sac}\Bbc+\frac{1}{\sab\sbc}\Bac+\frac{1}{\sbc\sac}\Bab\right),\nonumber \\
h_3(\sac,\sbc) &=& \frac{1}{4}f_1(\sab,\sbc).
\end{eqnarray}

\subsection{Ultraviolet renormalisation}
\label{subsec:renorm}

The renormalisation of the matrix element is carried out by replacing 
the bare coupling $\alpha_0$ with the renormalised coupling 
$\alpha_s\equiv \alpha_s(\mu^2)$,
evaluated at the renormalisation scale $\mu^2$
according to Eq.~(\ref{eq:alpha}).

We denote the $i$-loop contribution to the unrenormalised tensor
coefficients for process $\P$ by 
$A_{IJ}^{\P,(i),{\rm un}}$ and $B^{\P,(i),{\rm un}}$, using the same normalization as 
for the decomposition of the tensor coefficients (\ref{eq:perturb}); the 
dependence on $(\sac,\sbc)$ is always understood implicitly. 
The renormalised coefficients are then obtained as
\begin{eqnarray}
\label{eq:Aren}
A_{IJ}^{\P,(0)}  &=&0,
 \nonumber \\
A_{IJ}^{\P,(1)}  &=& 
S_\e^{-1} A_{IJ}^{\P,(1),{\rm un}} 
 ,  \nonumber \\
A_{IJ}^{\P,(2)} &=& 
S_\e^{-2} A_{IJ}^{\P,(2),{\rm un}}  
-\frac{(q+1)\beta_0}{\e} S_\e^{-1}
A_{IJ}^{\P,(1),{\rm un}}  ,
\end{eqnarray}
and
\begin{eqnarray}
\label{eq:Bren}
B^{\P,(0)}  &=& B^{\P,(0),{\rm un}} ,
 \nonumber \\
B^{\P,(1)}  &=& 
S_\e^{-1} B^{\P,(1),{\rm un}} 
-\frac{q\beta_0}{\e} B^{\P,(0),{\rm un}}  ,  \nonumber \\
B^{\P,(2)} &=& 
S_\e^{-2} B^{\P,(2),{\rm un}}  
-\frac{(q+1)\beta_0}{\e} S_\e^{-1}
B^{\P,(1),{\rm un}}  
-\frac{q}{2}\left(\frac{\beta_1}{\e}-\frac{(q+1)\beta_0^2}{\e^2}\right)
B^{\P,(0),{\rm un}},\nonumber \\
\end{eqnarray}
where $q$ is the number of powers of the strong coupling for process $\P$
as described in Eq.~(\ref{eq:alphaq}).

\subsection{Infrared behaviour of the tensor coefficients}
\label{subsec:infrared}

After performing ultraviolet renormalisation,
the amplitudes still
contain singularities which are of infrared origin and will be  analytically
cancelled by those occurring in radiative processes of the
same order.
Catani~\cite{catani} has shown how to organize the 
infrared pole structure of the one- and two-loop contributions renormalised in the 
\MSbar\ scheme in terms of the tree and renormalised one-loop amplitudes.
The same procedure applies to the tensor coefficients.

In particular, the infrared behaviour of the one-loop coefficients is given by
\begin{eqnarray}
\label{eq:IR1L}
A_{IJ}^{\P,(1)} &=& A_{IJ}^{\P,(1),{\rm finite}},\nonumber \\
B^{\P,(1)} &=& {\bom I}^{\P,(1)}(\epsilon) B^{\P,(0)} + B^{\P,(1),{\rm finite}},
\end{eqnarray}
while the two-loop singularity structure is
\begin{eqnarray}
\label{eq:IR2L}
A_{IJ}^{\P,(2)} &=& {\bom I}^{\P,(1)}(\epsilon) A_{IJ}^{\P,(1)}
+A_{IJ}^{\P,(2),{\rm finite}},\nonumber \\
B^{\P,(2)} &=& {\bom I}^{\P,(2)}(\epsilon)  B^{\P,(0)}+ {\bom I}^{\P,(1)}(\epsilon) B^{(1)}+ B^{(2),{\rm finite}},
\end{eqnarray}
where ${\bom I}^{\P,(2)}(\epsilon)$ is given by,
\begin{eqnarray}
{\bom I}^{\P,(2)}(\epsilon)&=&
-\frac{1}{2}  {\bom I}^{\P,(1)}(\epsilon) {\bom I}^{\P,(1)}(\epsilon)
-\frac{\beta_0}{\epsilon} {\bom I}^{\P,(1)}(\epsilon) 
+e^{-\epsilon \gamma } \frac{ \Gamma(1-2\epsilon)}{\Gamma(1-\epsilon)} 
\left(\frac{\beta_0}{\epsilon} + K\right)
{\bom I}^{\P,(1)}(2\epsilon) \nonumber \\
&& \qquad + {\bom H}^{\P,(2)}(\epsilon) 
\end{eqnarray}
and the constant $K$ is
\begin{equation}
K = \left( \frac{67}{18} - \frac{\pi^2}{6} \right) \CA - 
\frac{10}{9} T_R \NF.
\end{equation}

In QCD the ${\bom I}^{\P,(1)}(\epsilon)$ operator is given by~\cite{catani}
\begin{eqnarray}
{\bom I^{\P,(1)}}(\ep) &=& \frac{1}{2}\frac{e^{\ep \gamma }}{\Gamma(1-\ep)}
     \sum_i \nu^{sing}_i(\ep) \sum_{j \neq i} {\bf T}_i
   \cdot {\bf T}_j \left(-\frac{\mu^2}{2p_i \cdot p_j}
\right)^{\epsilon}
\end{eqnarray}
where the sum runs over the pairs of external coloured particles involved in process $\P$ and
the singular function is
\begin{equation}
\nu^{sing}_i(\ep) = \frac{1}{\ep^2} + \gamma_i \frac{1}{\ep},\qquad
\gamma_{q,\bar q}=\frac{3}{2}, \qquad \gamma_g=\frac{\beta_0}{\CA}
\end{equation}
The colour charge algebra is defined as
$$
{\bf T}_i \cdot {\bf T}_j = \left\{ \begin{array}{ll}
                                   {\bf T}_j \cdot {\bf T}_i & \mbox{if $i
                                     \neq j$,} \\
                                   {\bf T}_i^2 = C_i & \mbox{otherwise,}
                                   \end{array} \right.
$$
with the Casimir $C_i =C_F$ ($C_i =C_A$) if parton $i$ is a quark (gluon) and where
the colour charge ${\bf T}_i$ is  $t^a_{cb}$ ($-t^a_{cb}$) for a quark (anti-quark) and $if_{cab}$
for a gluon.

Then, given the colour basis for the two-quark two-gluon process presented in Eq.~(\ref{eq:colour}), we have the 
following infrared-singularity operator 
\begin{eqnarray}
\label{eq:opIgg}
\bom{I}^{gg,(1)}(\ep) = \frac{e^{\ep\gamma}}{\Gamma(1-\ep)}
\left(
\begin{array}{ccc}
\asut & 0 & \btus \\
0 & \astu & \buts \\
\bsut & \bstu & \cs
\end{array}
\right) \nonumber \\
\end{eqnarray}
where
\begin{eqnarray}
\label{eq:Adefqg}
\astu &=&-\Biggl({1\over \ep^2}+{3\over 2  \ep}\Biggr){}\,
{}\Biggl[{N}\,{\fu}+\CF \,{\fs}\Biggr]{}\nonumber\\
&&\phantom{{ }-{}\Biggl\{}
+{N}\,{}\Biggl ({\bzero\over 2N  \ep}-{3\over 4\ep}\Biggr
){}\,{}\Biggl[{\fs}+{\fu}\Biggr], \\
\label{eq:Bdefqg}
\bstu &=&{ }{}\Biggl ({1\over \ep^2}+{3\over 4  \ep}+\,{\bzero\over 2 N
\ep}\Biggr 
)\,{}\Biggl[ {\ft} -{\fs}\Biggr],\\
\label{eq:Cdefqg}
\cs &=&{ }-{}\Biggl [\Biggl ({1\over \ep^2}+{3\over 2 \ep}\Biggr )\CF
+{}\Biggl ({1\over \ep^2}+{\bzero\over N \ep}\Biggr){}
\,{N}\Biggr ]{\fs}. 
\end{eqnarray}

Similarly, for the other processes we have
\begin{eqnarray}
\label{eq:opIgph}
\bom{I}^{g \gamma ,(1)}(\ep) &=&
\frac{e^{\ep\gamma}}{2 \Gamma(1-\ep)}
\left\{ 
\frac{1}{N} \Biggl( {1\over \ep^2}+{3\over 2  \ep}\Biggr )\,{}{\fs}
\right.\nonumber \\
&&\left.-N \Biggl( {1\over \ep^2}+{3\over 4  \ep}+\,{\bzero\over 2 N \ep}\Biggr )
\,{}\Biggl[ {\ft} + {\fu}\Biggr] 
\right\}
\end{eqnarray}
and
\begin{eqnarray}
\label{eq:opIphph}
\bom{I}^{\gamma \gamma ,(1)}(\ep) &=&
-\CF \frac{e^{\ep\gamma}}{\Gamma(1-\ep)}
\Biggl( {1\over \ep^2}+{3\over 2  \ep}\Biggr )\,{}{\fs}.
\end{eqnarray}

In Eq.~(\ref{eq:IR2L}) the function ${\bom H}^{\P,(2)}$ contains poles of $\O{1/\ep}$ and is
process and renormalisation scheme dependent. The specific structure of this term  was not given
in Ref.~\cite{catani}. However, there it was pointed out that in the case of a QCD process that
involves a $q \bar q$ pair, the  function ${\bom H}^{\P,(2)}$ is related to the quark
electromagnetic form factor. Now,  we know that the form factor is not merely related, but is the
backbone of the singular  structure of multi-loop QCD amplitudes~\cite{stermanTY}. 

It was only after the direct Feynman diagram evaluation of two-loop matrix elements 
~\cite{BDGbha,qqunlike,qqlike,qqgg,gggg,qqpp}, that the colour uncorrelated part
of the ${\bom H}^{\P,(2)}$ function for all 
$2 \to 2$ partonic  processes was fully identified. More precisely, it was found that each external coloured
leg in the  partonic process contributes independently to the matrix element given by
\begin{eqnarray}
\bra{\cm_0} {\bom H}^{\P,(2)} \ket{\cm_0} = \frac{e^{\ep\gamma}}{4 \ep \Gamma(1-\ep)}
H^{\P,(2)}\braket{\cm_0}{\cm_0}
\end{eqnarray}
with
\begin{eqnarray}
\label{eq:nqng}
H^{\P,(2)} = n_q H_{q}^{(2)} + n_g H_{g}^{(2)}
\end{eqnarray}
where $n_g$ is the number of external gluons, $n_q$ is the number of external quarks and 
anti-quarks for the process $\P$ and 
\begin{eqnarray}
H_{q}^{(2)} =H_{\bar q}^{(2)} &=&\left(\frac{\pi^2}{2}-6 ~\zeta_3 
-\frac{3}{8}\right) \CF^2
+\left(\frac{13}{2}\zeta_3 +\frac{245}{216}-\frac{23}{48} \pi^2 \right) \CA \CF
\nonumber \\
&& + \left(-\frac{25}{54}+\frac{\pi^2}{12} \right) \TR \NF \CF, \\
H_{g}^{(2)} &=& 
\frac{20}{27} \TR^2 \NF^2
+ \TR \CF \NF
-\left(\frac{ \pi^2}{36}+\frac{58}{27} \right)\TR \NF\CA
\nonumber \\ &&
+\left(\frac{\zeta_3}{2}+\frac{5}{12}
+\frac{11}{144}\pi^2 \right) \CA^2 .
\end{eqnarray}

Already in Ref.~\cite{catani}, Catani pointed out the possibility of  ${\bom H}^{\P,(2)}$ having
additional non-trivial colour correlations which could be explored  by calculating the hard scattering
amplitude, rather than the interference of two-loop and tree amplitudes.\footnote{The sum over
colours causes the colour correlations to vanish}  Such  colour correlations were found for the
four gluon interaction, which is the only partonic process (involving four coloured particles) for
which  the two-loop helicity amplitudes~\cite{BFDgggg} exist.   According to
Ref.~\cite{BFDgggg}, the colour
correlations form a non-trivial matrix in colour space and is the same for all helicities.

For the processes we study, we find that the colour correlations are indeed helicity independent
and are given by
\begin{eqnarray}\label{eq:h2coloura}
{\bom H}^{\P,(2)}(\ep) = \frac{e^{\ep\gamma}}{4 \ep \Gamma(1-\ep)}{\fsd}
\Biggl( H^{\P,(2)}~{\bf 1} + {\bf \hat H}^{\P,(2)} \Biggr)
\end{eqnarray}
with $H^{\P,(2)}$ as given in Eq.~(\ref{eq:nqng}) and
\begin{eqnarray}\label{eq:h2colourb}
{\bf \hat H}^{gg,(2)} &=&
-4~i  ~f_{abc} ~{\bom T}^{a[1]} {\bom T}^{b[2]} {\bom T}^{c[3]}
~{\rm ln}\left(\frac{-\sab}{-\sbc} \right) 
{\rm ln}\left(\frac{-\sbc}{-\sac} \right) 
{\rm ln}\left(\frac{-\sac}{-\sab} \right), \\
{\bf \hat H}^{g\gamma,(2)} &=& 0,\\
{\bf \hat H}^{\gamma \gamma,(2)} &=& 0.
\end{eqnarray}
These results are in complete agreement with the results found in Ref.~\cite{BFDgggg} and show that
the colour correlations are also independent of the external particles.    
This also agrees 
with the analysis of Ref.~\cite{stermanTY} where it was shown that the contributions to 
${\bf \hat H}^{\P,(2)}$ are of soft nature and directly related to soft anomalous 
dimension matrices obtained by studying the evolution of colour exchange in QCD hard 
scattering (see ~\cite{KOS} and references therein). It is worth mentioning that these type of 
colour correlations between three different partons also appeared in the analysis of higher-order 
contributions to the soft-gluon current in Ref.~\cite{sone3}.

Note that the matrix $f_{abc} ~{\bom T}^{a[1]} {\bom T}^{b[2]} {\bom T}^{c[3]}$ collects all 
colour weights that result from the analysis of triple gluon vertices connecting any three 
different external partons, projected into the colour basis given in Eq.~(\ref{eq:colour}), 
so that for the two quark-two gluon process,
\begin{eqnarray}
i~f_{abc} ~{\bom T}^{a[1]} {\bom T}^{b[2]} {\bom T}^{c[3]} = 
-\frac{1}{8}
\left(
\begin{array}{ccc}
2 & 2 & 4N \\
-2 & -2 & -4N \\
-N & N & 0
\end{array}
\right).
\end{eqnarray}
Given the tree level results presented in Section~\ref{sec:treelevel} and the divergent 
structure presented in Eq.(\ref{eq:IR2L}), we can see that in fact the
third column of this matrix is not needed at $\O{1/\ep}$, since the 
${\bom \delta^{a_3a_4}\delta_{i_1i_2}}$ colour structure is not present at tree level.

\section{Helicity amplitudes}
\label{sec:helamp}

The helicity amplitudes $\M_{\lambda_1\lambda_2\lambda_3\lambda_4}$ 
can be obtained from the general $d$-dimensional tensor of
Eq.~(\ref{eq:tensor}) 
using standard four-dimensional helicity techniques corresponding to working in
the 't Hooft-Veltman scheme.  Conventionally,
the two helicity states of a
four-dimensional light-like spinor $\psi(p)$ are denoted by,
\begin{equation}
\psi_\pm(p)= \frac{1}{2}(1\pm \gamma_5)\psi(p),
\end{equation}
with the further notation,
\begin{equation}
|p\pm\rangle = \psi_\pm(p), \qquad\qquad
\langle p\pm | = \overline{\psi_\pm(p)}.
\end{equation}
Particles may thus be crossed to the initial state by reversing the sign of the
helicity.
The basic quantity is the spinor product,
\begin{equation}
\langle pq\rangle = \langle p- | q+\rangle, \qquad\qquad
[ pq]= [ p+ | q-],
\end{equation}
such that 
\begin{equation}
\langle pq\rangle  [qp] = 2 p.q.
\end{equation}
In this notation, the polarization vector of the out-going light-like gauge
boson of momentum $p$ is defined with respect to a reference momentum $k$ as,
\begin{equation}
\epsilon^\pm(p,k) = \pm \frac{\langle p \pm | \gamma_\mu | k \pm
\rangle}{\sqrt{2} \langle k\mp| p\pm\rangle}.
\end{equation}

Fixing the helicity of the (final-state) quark to be positive, $\lambda_1 = +$ 
 we find that the helicity amplitudes, upto a helicity dependent phase, are given by
\begin{eqnarray}
\label{eq:helamp}
\M_{+-++} &\propto& \frac{\sqrt{\sab^2\sac\sbc}}{2}A_{11}(\sac,\sbc),\nonumber \\
\M_{+-+-} &\propto& 
~\frac{\sqrt{\sac^3\sbc}}{2}
\left( 
2 B(\sac,\sbc)+A_{12}(\sac,\sbc)-A_{11}(\sac,\sbc)\right),\nonumber \\
\M_{+--+} &\propto& 
~\frac{\sqrt{\sac\sbc^3}}{2}
\left( 
2 B(\sac,\sbc)+A_{22}(\sac,\sbc)-A_{21}(\sac,\sbc)\right),\nonumber \\
\M_{+---} &\propto& -\frac{\sqrt{\sab^2\sac\sbc}}{2}A_{22}(\sac,\sbc).
\end{eqnarray}
By inspection of Eqs.~(\ref{eq:Xsymgg}) and (\ref{eq:Xsympp}) we see that, as expected, there are two 
independent helicity amplitudes and that the other two 
helicity amplitudes can be obtained by Bose symmetry and,
\begin{eqnarray}
\M_{+--+} 
&=& 
\M_{+-+-} (\sac \leftrightarrow \sbc),\nonumber \\
\M_{+---}
&=& \M_{+-++} (\sac \leftrightarrow \sbc).
\end{eqnarray}
Helicity amplitudes where the quark helicity is negative are obtained through parity,
\begin{equation}
\M_{-+-\lambda_3-\lambda_4} 
= 
\left(\M_{+-\lambda_3\lambda_4}\right)^*.
\end{equation}
By inspection, we see that ``squaring'' and summing over the helicities generates
the interference of amplitudes given in Eq.~(\ref{eq:square}).
Thus, the helicity amplitudes can be obtained directly from a Feynman diagram
calculation by projecting out the individual tensor coefficients and taking the appropriate linear combination.

Just as for the individual tensor coefficients, the helicity amplitudes for each
process $\P$ can be perturbatively decomposed as,
\begin{eqnarray}
\lefteqn{|{\cal M}^\P\rangle_{\lambda_1\lambda_2\lambda_3\lambda_4}}\nonumber \\
 &=&  c^\P \left[
|{\cal M}^{\P,(0)}\rangle_{\lambda_1\lambda_2\lambda_3\lambda_4}  
+ \left(\frac{\alpha_s}{2\pi}\right) |{\cal M}^{\P,(1)}\rangle_{\lambda_1\lambda_2\lambda_3\lambda_4} 
+ \left(\frac{\alpha_s}{2\pi}\right)^2 |{\cal M}^{\P,(2)}\rangle_{\lambda_1\lambda_2\lambda_3\lambda_4}
+ {\cal O}(\alpha_s^3) \right],\nonumber \\
\end{eqnarray}
where the overall coupling factor is given by Eq.~(\ref{eq:alphaq}).
As in Eqs.~(\ref{eq:Aren}) and (\ref{eq:Bren}),
the renormalised helicity amplitudes are obtained from the unrenormalised
amplitudes by 
\begin{eqnarray}
\label{eq:helren}
|{\cal M}^{\P,(0)}\rangle_{\lambda_1\lambda_2\lambda_3\lambda_4}
  &=& |{\cal M}^{\P,(0),{\rm un}}\rangle_{\lambda_1\lambda_2\lambda_3\lambda_4},
 \nonumber \\
|{\cal M}^{\P,(1)}\rangle_{\lambda_1\lambda_2\lambda_3\lambda_4} &=& 
S_\e^{-1} |{\cal M}^{\P,(1),{\rm un}}\rangle_{\lambda_1\lambda_2\lambda_3\lambda_4} 
-\frac{q\beta_0}{\e} |{\cal M}^{\P,(0),{\rm un}}\rangle_{\lambda_1\lambda_2\lambda_3\lambda_4}  ,  \nonumber \\
|{\cal M}^{\P,(2)}\rangle_{\lambda_1\lambda_2\lambda_3\lambda_4} &=& 
S_\e^{-2} |{\cal M}^{\P,(2),{\rm un}}\rangle_{\lambda_1\lambda_2\lambda_3\lambda_4}  
-\frac{(q+1)\beta_0}{\e} S_\e^{-1}
|{\cal M}^{\P,(1),{\rm un}}\rangle_{\lambda_1\lambda_2\lambda_3\lambda_4}  \nonumber \\
&&
-\frac{q}{2}\left(\frac{\beta_1}{\e}-\frac{(q+1)\beta_0^2}{\e^2}\right)
|{\cal M}^{\P,(0),{\rm un}}\rangle_{\lambda_1\lambda_2\lambda_3\lambda_4},\nonumber \\
\end{eqnarray}
while the infrared singularity structure is given by,
\begin{eqnarray}
\label{eq:remainder1}
|{\cal M}^{\P,(1)}\rangle_{\lambda_1\lambda_2\lambda_3\lambda_4} &=& 
{\bom I}^{\P,(1)}(\epsilon) |{\cal M}^{\P,(0)}\rangle_{\lambda_1\lambda_2\lambda_3\lambda_4} 
+|{\cal M}^{\P,(1),{\rm finite}}\rangle_{\lambda_1\lambda_2\lambda_3\lambda_4},\\
\label{eq:remainder2}
|{\cal M}^{\P,(2)}\rangle_{\lambda_1\lambda_2\lambda_3\lambda_4} &=& {\bom I}^{\P,(2)}(\epsilon)  
|{\cal M}^{\P,(0)}\rangle_{\lambda_1\lambda_2\lambda_3\lambda_4}
+ {\bom I}^{\P,(1)}(\epsilon) 
|{\cal M}^{\P,(1)}\rangle_{\lambda_1\lambda_2\lambda_3\lambda_4}\nonumber \\
&&+ |{\cal M}^{\P,(2),{\rm finite}}\rangle_{\lambda_1\lambda_2\lambda_3\lambda_4}, 
\end{eqnarray}
where the operators ${\bom I}^{\P,(i)}$ are the process dependent matrices in colour space given in Section~\ref{subsec:infrared}.

Similarly, the $n$-loop helicity amplitudes are vectors in colour space and can be further
decomposed as in Eq.~(\ref{eq:colour})
\begin{eqnarray}
\label{eq:helcolour}
|{\cal M}^{gg,(n)}\rangle_{\lambda_1\lambda_2\lambda_3\lambda_4} &=&  
\sum_{i=1}^3 \;\C^{gg,[i]}\;|{\cal M}^{gg,(n),[i]}\rangle_{\lambda_1\lambda_2\lambda_3\lambda_4},\nonumber\\
|{\cal M}^{g\gamma,(n)}\rangle &=&  \C^{g\gamma,[1]}\; |{\cal M}^{g\gamma,(n),[1]}\rangle_{\lambda_1\lambda_2\lambda_3\lambda_4},\nonumber \\
|{\cal M}^{\gamma\gamma,(n)}\rangle &=&  \C^{\gamma\gamma,[1]}\; |{\cal M}^{\gamma\gamma,(n),[1]}\rangle_{\lambda_1\lambda_2\lambda_3\lambda_4}.
\end{eqnarray}

\subsection{Helicity amplitudes for physical processes}
\label{sec:results}
The physically relevant scattering amplitudes 
we wish to describe are the 
quark-antiquark annihilation processes,
\begin{eqnarray}
\label{eq:schannel1}
q(p_2,+) + \bar q(p_1,-) &\to& b_3(p_3,+)+b_4(p_4,+),\\
\label{eq:schannel2}
q(p_2,+) + \bar q(p_1,-) &\to& b_3(p_3,+)+b_4(p_4,-),\\
\label{eq:schannel3}
q(p_2,+) + \bar q(p_1,-) &\to& b_3(p_3,-)+b_4(p_4,+),\\
\label{eq:schannel4}
q(p_2,+) + \bar q(p_1,-) &\to& b_3(p_3,-)+b_4(p_4,-), 
\end{eqnarray}
and the quark-boson scattering processes,
\begin{eqnarray}
\label{eq:uchannel1}
q(p_2,+) + b_3(p_1,+) &\to& q(p_3,+)+b_4(p_4,+),\\
\label{eq:uchannel2}
q(p_2,+) + b_3(p_1,+) &\to& q(p_3,+)+b_4(p_4,-),\\
\label{eq:uchannel3}
q(p_2,+) + b_3(p_1,-) &\to& q(p_3,+)+b_4(p_4,+),\\
\label{eq:uchannel4}
q(p_2,+) + b_3(p_1,-) &\to& q(p_3,+)+b_4(p_4,-), 
\end{eqnarray}
\begin{eqnarray}
\label{eq:tchannel1}
q(p_2,+) + b_4(p_1,+) &\to& b_3(p_3,+)+q(p_4,+),\\
\label{eq:tchannel2}
q(p_2,+) + b_4(p_1,+) &\to& b_3(p_3,-)+q(p_4,+),\\
\label{eq:tchannel3}
q(p_2,+) + b_4(p_1,-) &\to& b_3(p_3,+)+q(p_4,+),\\
\label{eq:tchannel4}
q(p_2,+) + b_4(p_1,-) &\to& b_3(p_3,-)+q(p_4,+). 
\end{eqnarray}
where the bosons $b_3$ or $b_4$ are gluons or photons.
Processes where the initial state quark has negative helicity are obtained by a
parity transformation, while processes with two bosons in the initial state
are given by time reversal.  There are additional symmetries when the gauge
bosons are identical and further symmetries between the various 
colour components for the two gluon process.

For convenience, we denote the processes of Eqs.~(\ref{eq:schannel1})--(\ref{eq:schannel4}) as the
$s$-channel, the processes of Eqs.~(\ref{eq:uchannel1})--(\ref{eq:uchannel4}) as the
$u$-channel and the processes of Eqs.~(\ref{eq:tchannel1})--(\ref{eq:tchannel4}) as
the $t$-channel.  $s$, $t$ and $u$ are the usual Mandelstam variables,
$s = s_{12} > 0$, 
$t = s_{23} < 0$, $u = s_{13} < 0$ and $s+t+u = 0$.

To present the helicity amplitudes for the various processes, it is convenient to organise the amplitude where particles 1 and 2 are in the initial state
in terms of a spinor factor ${\cal S}$ and the colour factor $\C$  so that,
\begin{equation}
\label{eq:Mdecomp}
|{\cal M}_c^{\P,(n)}\rangle_{\lambda_2\lambda_1\lambda_3\lambda_4} =
\sum_i ~{\cal S}^{\P,[i]}_{c\lambda_2\lambda_1\lambda_3\lambda_4}  \times 
\C_c^{\P,[i]}\times
|\widehat {\cal M}_c^{\P,(n),[i]}\rangle_{\lambda_2\lambda_1\lambda_3\lambda_4},
\end{equation}
where $c$ (= $s$, $t$, $u$) denotes the channel and 
the sum runs over the number of colour structures: three for the two-gluon processes 
and one otherwise.

Explicitly, we have 
\begin{eqnarray}
\label{eq:tfactor}
\C_s^{gg,[1]} &=& \left({\bom T}^{a_3}{\bom T}^{a_4}\right)_{i_1i_2},\nonumber \\
\C_s^{gg,[2]} &=& \left({\bom T}^{a_4}{\bom T}^{a_3}\right)_{i_1i_2},\nonumber \\
\C_s^{gg,[3]} &=& {\bom \delta}^{a_3a_4}\delta_{i_1i_2}, 
\end{eqnarray}
while for the processes with photons we have,
\begin{eqnarray}
\label{eq:tfactor2}
\C_s^{g\gamma,[1]} &=& \left({\bom T}^{a_3}\right)_{i_1i_2},\nonumber \\
\C_s^{\gamma\gamma,[1]} &=& \delta_{i_1i_2},
\end{eqnarray}
The $u$- and $t$-channel colour factors are obtained by crossing symmetry,
\begin{equation}
\C_u^{\P,[i]} = \C_s^{\P,[i]} ~~(1 \leftrightarrow 3),\qquad\qquad
\C_t^{\P,[i]} = \C_s^{\P,[i]} ~~(1 \leftrightarrow 4).
\end{equation}

There is considerable freedom in the definition of the helicity dependent spinor factors.
For the $s$-channel, for process~(\ref{eq:schannel2}) we choose,
\begin{eqnarray}
\label{eq:s+-+-}
{\cal S}_{s+-+-}^{gg,[1]} &=& -i\frac{\langle 24\rangle^3\langle14\rangle}
{\langle12\rangle\langle23\rangle\langle34\rangle\langle41\rangle},\nonumber \\
{\cal S}_{s+-+-}^{gg,[2]} &=& -i\frac{\langle 24\rangle^3\langle14\rangle}
{\langle12\rangle\langle24\rangle\langle43\rangle\langle31\rangle},\nonumber \\
{\cal S}_{s+-+-}^{gg,[3]}={\cal S}_{+-+-}^{g\gamma} ={\cal S}_{+-+-}^{\gamma\gamma} &=& -i\frac{\langle 24\rangle^3\langle14\rangle}
{\langle13\rangle\langle14\rangle\langle23\rangle\langle24\rangle},
\end{eqnarray}
while for process~(\ref{eq:schannel1}) we choose,
\begin{equation}
\label{eq:s+-++}
{\cal S}_{s+-++}^{\P,[i]} = i\frac{\langle 23 \rangle[13]}{\langle 34 \rangle^2}.
\end{equation}
The factors for the other $s$-channel processes, (\ref{eq:schannel3},\ref{eq:schannel4}) are obtained by
Bose symmetry (exchanging labels 3 and 4).   Note that for the two-gluon process, this symmetry also exchanges the first two
colour structures, $[1]$ and $[2]$.  The spinor prefactors for the $u$- and $t-$ channels are also obtained
by crossing symmetry,
\begin{equation}
{\cal S}_{u\lambda_2\lambda_3\lambda_1\lambda_4}^{\P,[i]} = 
{\cal S}_{s\lambda_2\lambda_1\lambda_3\lambda_4}^{\P,[i]} ~~(p_1 \leftrightarrow p_3),\qquad\qquad
{\cal S}_{t\lambda_2\lambda_4\lambda_3\lambda_1}^{\P,[i]} = 
{\cal S}_{s\lambda_2\lambda_1\lambda_3\lambda_4}^{\P,[i]} ~~(p_1 \leftrightarrow p_4).
\end{equation}

\subsection{Tree level helicity amplitudes}

At tree-level, inserting the results for the tensor coefficients for the various
processes given in Section~\ref{sec:treelevel},
the only non-vanishing helicity amplitudes in the $s$-channel are,
\begin{eqnarray}
 |\widehat {\cal M}_s^{gg,(0),[1]}\rangle_{+-+-} =|\widehat {\cal M}_s^{gg,(0),[1]}\rangle_{+--+} &=& -2,\nonumber\\
 |\widehat {\cal M}_s^{gg,(0),[2]}\rangle_{+-+-} = |\widehat {\cal M}_s^{gg,(0),[2]}\rangle_{+--+} &=& -2,\nonumber\\
 |\widehat {\cal M}_s^{g\gamma,(0)}\rangle_{+-+-}=|\widehat {\cal M}^{g\gamma,(0)}\rangle_{+--+} &=&
 2,\nonumber \\
 |\widehat {\cal M}_s^{\gamma\gamma,(0)}\rangle_{+-+-}=|\widehat {\cal M}^{\gamma\gamma,(0)}\rangle_{+--+} &=& 2.
\end{eqnarray}
The appropriate spinor factors are obtained from Eq.~(\ref{eq:s+-+-}) while the colour factors are given in 
Eqs.~(\ref{eq:tfactor}) and (\ref{eq:tfactor2}).  Tree helicity amplitudes for the other channels are
obtained by the appropriate crossing symmetry.

\subsection{One-loop helicity amplitudes}
\label{subsec:oneloopamps}
The one-loop helicity amplitudes are straightforwardly obtained by inserting the all-orders results for 
the tensor coefficients given in Section~\ref{subsec:oneloop} into Eq.~(\ref{eq:helamp}),
expanding the one-loop bubble and box
integrals around $\epsilon = 0$, and renormalising according to Eq.~(\ref{eq:helren}).
Explicit expansions in $\epsilon$ for the $d=4-2\epsilon$ one-loop bubble and $d=6-2\epsilon$ one-loop box
graphs can be found, for example, in the Appendix of Ref.~\cite{qqgg}.
   
The finite remainders of the one-loop amplitudes defined through Eq.~(\ref{eq:remainder1})
for the two-gluon process can be 
decomposed according to the number of colours and massless quark flavours,
\begin{eqnarray}
|\widehat {\cal M}_c^{gg,(1),[i],finite}\rangle
&=& N A_c^{gg,(1),[i]} + \frac{1}{N} B_c^{gg,(1),[i]} + \NF
C_c^{gg,(1),[i]}-\beta_0\Ls\,|\widehat {\cal M}_c^{gg,(0),[i]}\rangle,\nonumber \\
\end{eqnarray}
for $i=1,2$ and
\begin{equation}
|\widehat {\cal M}_c^{gg,(1),[3],finite}\rangle
= D_c^{gg,(1),[3]} + \frac{\NF}{N} E_c^{gg,(1),[3]}
\end{equation}
for $i=3$ and where
\begin{equation}\Ls = \log\left(\frac{s}{\mu^2}\right)\end{equation}
and $\beta_0$ is defined in Eq.~(\ref{betas}).
For the other two processes we have,
\begin{eqnarray}
|\widehat {\cal M}_c^{g\gamma,(1),[1],finite}\rangle
&=& N A_c^{g\gamma,(1),[1]} + \frac{1}{N} B_c^{g\gamma,(1),[1]} + \NF
C_c^{g\gamma,(1),[1]}-\frac{1}{2}\beta_0\Ls\,|\widehat {\cal M}_c^{g\gamma,(0),[1]}\rangle, \nonumber \\
&&\\
|\widehat {\cal M}_c^{\gamma\gamma,(1),[1],finite}\rangle
&=& N A_c^{\gamma\gamma,(1),[1]} 
+ \frac{1}{N} B_c^{\gamma\gamma,(1),[1]} + \NF
C_c^{\gamma\gamma,(1),[1]}.\nonumber\\
\end{eqnarray}
For clarity, the dependence on the helicities has been suppressed.  
Explicit expressions for $A,\ldots,E$
for $q\bar q \to gg$ in the physical region, $s = s_{12} > 0$, 
$t = s_{23} < 0$ and $u = s_{13} < 0$, are given in Appendix~\ref{app:oneloop}.
Explicit results for $q\bar q \to gg$ scattering in the physical region, $s = s_{12} > 0$, 
$t = s_{23} < 0$ and $u = s_{13} < 0$ are given in Appendix~\ref{app:oneloop}.
Formulae for the other channels and processes are available from the authors as FORM outputs.
We have checked that the finite one-loop helicity amplitudes presented here are in agreement with the
helicity amplitudes for $q\bar q \to gg$ (and crossed processes) given for the HV scheme in
Ref.~\cite{BFDqqgg}.

\subsection{Two-loop helicity amplitudes}

The main results of this paper are the two-loop amplitudes for processes
(\ref{eq:schannel1})--(\ref{eq:tchannel4}).
As in the tree and one-loop cases, the helicity amplitudes can be directly extracted from the appropriate
linear combination, Eq.~(\ref{eq:helamp}), of unrenormalised two-loop tensor coefficients obtained by
direct evaluation of the projectors given in Section~\ref{subsec:projectors} acting on two-loop graphs.
Renormalisation is achieved via  Eq.~(\ref{eq:helren}) and the two-loop master integrals expanded around
$\epsilon=0$. The finite remainder of the two-loop amplitudes are defined through 
Eq.~(\ref{eq:remainder2}) and, for each process can be 
decomposed according to the number of colours and massless quark flavours,
\begin{eqnarray}
|\widehat {\cal M}_c^{gg,(2),[i],finite}\rangle
&=& N^2 A_c^{gg,(2),[i]} +  B_c^{gg,(2),[i]} + 
\frac{1}{N^2} C_c^{gg,(2),[i]}
+N\NF D_c^{gg,(2),[i]}
 \nonumber \\
&& 
+  \frac{\NF}{N} E_c^{gg,(2),[i]}
+ \NF^2 F_c^{gg,(2),[i]} + \frac{\NF^2}{N^2} G_c^{gg,(2),[i]}
\nonumber\\
&&-2\beta_0\Ls\,|\widehat {\cal M}_c^{gg,(1),[i],finite}\rangle
-(\beta_1 \Ls + \beta_0^2\Ls^2)\,|\widehat {\cal M}_c^{gg,(0),[i]}\rangle,
\end{eqnarray}
for $i=1,2$ and
\begin{eqnarray}
|\widehat {\cal M}_c^{gg,(2),[3],finite}\rangle
&=&N H_c^{gg,(2),[3]} + \frac{1}{N} I_c^{gg,(2),[3]} 
+\NF J_c^{gg,(2),[3]} \nonumber \\
&&+  \frac{\NF}{N^2} K_c^{gg,(2),[3]}+ \frac{\NF^2}{N} L_c^{gg,(2),[3]} 
\nonumber \\
&&  -2\beta_0\Ls\,|\widehat {\cal M}_c^{gg,(1),[3],finite}\rangle
-(\beta_1 \Ls + \beta_0^2\Ls^2)\,|\widehat {\cal M}_c^{gg,(0),[3]}\rangle.\nonumber \\
\end{eqnarray}
For the other processes we have,
\begin{eqnarray}
|\widehat {\cal M}_c^{g\gamma,(2),[1],finite}\rangle
&=& N^2 A_c^{g\gamma,(2),[1]} +  B_c^{g\gamma,(2),[1]} + 
\frac{1}{N^2}\,C_c^{g\gamma,(2),[1]}
+N\NF D_c^{g\gamma,(2),[1]}\nonumber\\
&& +  \frac{\NF}{N}\,E_c^{g\gamma,(2),[1]} + 
\NF^2\,F_c^{g\gamma,(2),[1]}
+\biggl( \sum_q Q_q\biggr)\left(\frac{4}{N}-N\right)\,G_c^{g\gamma,(2),[1]}
\nonumber\\
&&-\frac{3}{2}\beta_0\Ls\,|\widehat {\cal M}_c^{g\gamma,(1),[1],finite}\rangle
-\biggl(\frac{1}{2}\beta_1 \Ls + \frac{3}{8}\beta_0^2\Ls^2\biggr)\,|\widehat {\cal M}_c^{g\gamma,(0),[1]}\rangle, \\
|\widehat {\cal M}_c^{\gamma\gamma,(2),[1],finite}\rangle
&=& N^2 A_c^{\gamma\gamma,(2),[1]} +  B_c^{\gamma\gamma,(2),[1]} + 
\frac{1}{N^2}\,C_c^{\gamma\gamma,(2),[1]}
+N\NF D_c^{\gamma\gamma,(2),[1]}\nonumber\\
&& +  \frac{\NF}{N}\,E_c^{\gamma\gamma,(2),[1]} + 
\biggl( \sum_q Q_q^2\biggr)\CF \,F_c^{\gamma\gamma,(2),[1]}
-\beta_0\Ls\,|\widehat {\cal M}_c^{\gamma\gamma,(1),[1],finite}\rangle,\nonumber \\
\end{eqnarray}
where $Q_q$ is the electric charge of a massless quark of flavour $q$.

As in the previous section, the dependence on the helicities has been suppressed.  
Explicit expressions for $A,\ldots,L$
for $q\bar q \to gg$ process in the physical region, $s = s_{12} > 0$, 
$t = s_{23} < 0$ and $u = s_{13} < 0$, are given in Appendix~\ref{app:twoloop}.
Formulae for the other channels and processes are available from the authors as FORM outputs.
We have checked that the finite two-loop helicity amplitudes presented here are in agreement with the
helicity amplitudes for $q\bar q \to gg$ (and crossed processes) given for the HV scheme in
Ref.~\cite{BFDqqgg}.

\section{Comparison with previous results}
\label{sec:matelem}

In recovering the square, or interference, of amplitudes in a particular channel $c$,
the contraction of the $n$-loop colour vector $|{\cal M}^{\P,(n)}\rangle$ 
with a conjugate $m$-loop
colour vector $\langle {\cal M}^{\P,(m)}| $ obeys the rule
\begin{eqnarray}
\label{eq:makesquare}
\lefteqn{\langle {\cal M}^{\P,(m)} | {\cal M}^{\P,(n)} \rangle }\nonumber \\
&=&
\sum_{\rm helicities} ~\sum_{\rm colours}
 |{\cal M}_c^{\P,(n)}\rangle_{\lambda_1\lambda_2\lambda_3\lambda_4}^*
  \, |{\cal M}_c^{\P,(n)}\rangle_{\lambda_1\lambda_2\lambda_3\lambda_4} 
  \,
\end{eqnarray}
where   $|{\cal M}_c^{\P,(n)}\rangle_{\lambda_1\lambda_2\lambda_3\lambda_4} $ contains all of the spinor
and colour information for a particular helicity.
Expanding the colour and spinor factors $\C$ and ${\cal S}$, and 
dropping the explicit dependence on the helicities, we see that,
\begin{eqnarray}
\label{eq:makesquare2}
\lefteqn{\langle {\cal M}^{\P,(m)} | {\cal M}^{\P,(n)} \rangle }\nonumber \\
&=&
\sum_{\rm helicities} ~\sum_{\rm colours}
 ~\sum_{i,j} \C_c^{\P,[i] *}\, \C_c^{\P,[j]} \; {\cal S}_c^{\P,[i] *} \;{\cal S}_c^{\P,[j]} \;
     |\widehat {\cal M}_c^{\P,(n),[i]}\rangle^*
  \, |\widehat {\cal M}_c^{\P,(n),[j]}\rangle 
  \, \nonumber \\
&=&  
\sum_{\rm helicities}  
 ~\sum_{i,j}   \, {\cal C\!C}^{\P}_{ij} \; {\cal S\!S}^{\P}_{ij} \; |\widehat {\cal M}_c^{\P,(n),[i]}\rangle^*
  \, |\widehat {\cal M}_c^{\P,(n),[j]}\rangle 
.
\end{eqnarray}
The symmetric colour sum
matrix ${\cal C\!C}^{\P} = \sum_{\rm colours}
\C_c^{\P,[i]*}\,\C_c^{\P,[j]}$ is given by
\begin{equation}
{\cal C\!C}^{gg} = \frac{N^2-1}{2}\left(
\begin{array}{ccc}
\frac{N^2-1}{2N} & -\frac{1}{2N} & 1 \\
-\frac{1}{2N} &\frac{N^2-1}{2N} & 1 \\
 1 & 1 & 2N\end{array}
\),
\end{equation}
and,
\begin{equation}
{\cal C\!C}^{g\gamma} = \frac{N^2-1}{2}, \qquad {\cal C\!C}^{\gamma\gamma} = N.
\end{equation}
In the $s$-channel, the process and helicity dependent spinor matrices ${\cal S\!S}^{\P}_c = {\cal S}_c^{\P,[i] *} \;{\cal S}_c^{\P,[j]}$ are
given by,
\begin{equation}
{\cal S\!S}^{gg}_{s+-+-} = \frac{u}{s^2t}\left(
\begin{array}{ccc}
u^2 & ut & us \\
tu & t^2  & ts \\
su & ts & s^2
\end{array}
\),
\qquad {\cal S\!S}^{g\gamma}_{s+-+-} = {\cal S\!S}^{\gamma\gamma}_{s+-+-} = \frac{u}{t},
\end{equation}
and
\begin{equation}
{\cal S\!S}^{gg}_{s+-++} = \frac{ut}{s^2}\left(
\begin{array}{ccc}
1 & 1 & 1 \\
1 & 1  & 1 \\
1 & 1 & 1
\end{array}
\),\qquad
{\cal S\!S}^{g\gamma}_{s+-++} = {\cal S\!S}^{\gamma\gamma}_{s+-++}
= \frac{ut}{s^2}.
\end{equation}
Similar matrices for the other helicities are obtained by Bose symmetry, while corresponding matrices for
the $u$ and $t$-channels are obtained by crossing symmetry,
\begin{equation}
{\cal S\!S}_u^{\P} = -{\cal S\!S}_s^{\P} (s \leftrightarrow t),
\qquad
{\cal S\!S}_t^{\P} = -{\cal S\!S}_s^{\P} (s \leftrightarrow u).
\end{equation}
The explicit minus sign accounts for the exchange of a single fermion between the initial and final states.
 
Eqs.~(\ref{eq:makesquare}) and (\ref{eq:makesquare2}) can be used to recreate the full interference of
tree and two-loop amplitudes. However, we notice that we can immediately separate the singularities from
the finite parts using Eq.~(\ref{eq:remainder2}). The finite remainder, denoted by ${\cal
F}_{inite}^{2\times 0}(s,t,u)$
in Refs.~\cite{qqgg,qqpp} is thus given by,
\begin{equation}
2 \Re \sum_{\rm helicities}  
 ~\sum_{i,j}   \, {\cal C\!C}^{\P}_{ij} \; {\cal S\!S}^{\P}_{ij} \; |\widehat {\cal M}_c^{\P,(0),[i]}\rangle^*
  \, |\widehat {\cal M}_c^{\P,(2),[j],finite}\rangle .
\end{equation}
Note that by using helicities, we are implicitly treating the external particle states in
4-dimensions - the HV scheme.    However, the singular contributions to the
renormalised amplitudes defined in Eqs.~(\ref{eq:remainder1}), (\ref{eq:remainder2}) are given
relative to the tree amplitude and therefore any additional terms that would be produced in
CDR are automatically removed.  Treating the external
states differently for both the $n$-loop and tree amplitudes does not alter the finite
contribution.  Direct application of Eq.~(\ref{eq:makesquare}) for the real part of the tree and
two-loop interference ($n=0$ and $m=2$) in the HV scheme therefore generates the same finite parts given in
section 3.2 of Ref.~\cite{qqgg} for  quark-gluon scattering and in Section 4 and Appendix A of
Ref.~\cite{qqpp} obtained using CDR.

Similarly, the square of one-loop graphs is obtained using $n=1$ and $m=1$ in
Eq.~(\ref{eq:makesquare}). In principle, the one-loop amplitude should be expanded through to ${\cal
O}(\epsilon^2)$.   However, as noted in \cite{qqgg,BFDgggg,metythesis}, the finite self interference only contains
logarithms (up to the fourth power), but does not contain triple and quartic polylogs.  These terms
naturally arise when the ${\cal O}(\epsilon)$ and ${\cal O}(\epsilon^2)$ terms in the expansion of the
one-loop box are multiplied by singularities from the conjugate amplitude.  Such terms appear in the
Catani pole structure, $\langle {\cal M}^{(1)} | \bom I^{(1)} | {\cal M}^{(0)}\rangle$ where $| {\cal
M}^{(1)}\rangle$ does need to be expanded through to ${\cal O}(\epsilon^2)$. 
The finite parts of the renormalised one-loop self
interference remaining after Catani's prediction for the singularities has been removed should therefore
be
identical to those obtained by  summing the finite parts using Eq.~(\ref{eq:makesquare}).   
 In this instance, the finite remainder denoted by ${\cal
F}_{inite}^{1\times 1}(s,t,u)$ in \cite{qqpp} is obtained from
\begin{equation}
\sum_{\rm helicities}  
 ~\sum_{i,j}   \, {\cal C\!C}^{\P}_{ij} \; {\cal S\!S}^{\P}_{ij} \; |\widehat {\cal
 M}_c^{\P,(1),[i],finite}\rangle^*
  \, |\widehat {\cal M}_c^{\P,(1),[j],finite}\rangle .
\end{equation}
We have checked that this procedure correctly reproduces the coefficients given in Section 5 and 
Appendix B of Ref.~\cite{qqpp} for the processes involving one or more photons.
Unfortunately, the results of Ref.~\cite{qqgg} are not presented in quite the same way and a direct
comparison is more difficult.\footnote{In fact there is a misprint in the renormalisation term, ${\cal R_A}$, of Eq.~4.6 of
Ref.~\cite{qqgg}.}
However, we have checked the finite remainders after the Catani poles have been subtracted, both directly from the square of the
one-loop amplitude (the CDR scheme) and the one-loop square reconstructed from the helicity amplitudes (the HV scheme) with
the HV results of Ref.~\cite{BFDqqgg} and again find agreement.

\section{Conclusions}
\label{sec:conclusions}

In this paper we have presented the analytic expressions for the one- and two-loop virtual  helicity
amplitudes for quark-gauge boson scattering, $q\bar q \to gg$, $q\bar q \to g\gamma$ and $q\bar q
\to \gamma\gamma$  and those processes related by crossing symmetry. These amplitudes were obtained
by the construction of $d$-dimensional projection operators that extract the coefficients of the
most general tensor representation for the amplitude, order by order in perturbation theory. Once
the renormalised tensor coefficients are known, the renormalised helicity amplitudes can be
constructed straightforwardly as a linear combination of the coefficients.    

At the level of the diagrammatic calculation, we applied the projection operators directly to the
Feynman diagrams that contribute at each order in $\as$. The projectors saturate the tensor
structure of the Feynman diagram and yield a set of scalar integrals that can be related to a basis
set of master integrals with the application of widely used reduction algorithms.  In fact, we are 
using exactly the same tools as we did in~\cite{qqgg, qqpp}, except in that case, the projector was
merely the conjugate tree amplitude.   Because the projectors exist in $d$-dimensions, conventional
dimensional regularisation is preserved and there is no ambiguity in dealing with $\gamma_5$.   Once
the general tensor coefficients are determined, the external particles can be treated in
4-dimensions (the 't Hooft-Veltman scheme) and standard helicity methods used to construct the
helicity amplitudes.   

By summing over helicities and colours, the full interference of tree and two-loop graphs can be
reconstructed.   In previous work~\cite{qqgg, qqpp}, the finite part has been separated from infrared singular parts
using the Catani formalism~\cite{catani}.  In this procedure, the singular operators multiply tree
and one-loop amplitudes.  Changing the scheme for the external particles therefore changes the
overall pole contribution.   However, the finite remainder left after subtracting the poles is
invariant under changing from the conventional dimensional regularisation scheme to the 't
Hooft-Veltman scheme.  Therefore, the finite parts of the helicity amplitudes presented here precisely reproduce the 
finite parts of the interference of tree and two-loop graphs  given in~\cite{qqgg, qqpp}.  
Similarly, the finite parts of the self interference of one-loop amplitudes can be reconstructed
purely from the finite parts of the one-loop helicity amplitudes.
In addition, the finite remainders of the two-loop amplitudes for quark-gluon scattering in the HV scheme
agree with the independent calculation of Ref.~\cite{BFDqqgg}. 

For the case of quark-gluon scattering, the presence of four coloured external
particles gives rise to additional colour correlations proportional to
$1/\epsilon$. While the existence of such additional non-trivial colour
correlations was pointed out in Ref.~\cite{catani} and are expected on general
grounds~\cite{stermanTY}, the precise form of the colour structure of  ${\bom
H}^{\P,(2)}$ was not predicted.  So far, the results for gluon-gluon 
scattering~\cite{BFDgggg} as well as those for quark-gluon scattering presented
in this paper  and  in  Ref.~\cite{BFDqqgg}, support the form given in
Eqs.~(\ref{eq:h2coloura}) and (\ref{eq:h2colourb}). These colour correlations
vanish when the interference of tree and two-loop amplitudes is constructed.

The more general goal is to evaluate the NNLO QCD corrections to  the associated scattering
processes which should yield a significant improvement in theoretical stability, see for
example~\cite{nigelRCLL}.  However, the reality  is that the phenomenology is still some way off
and  there is still a significant amount of work remaining to be done. The two-loop
corrections to the $2 \to 2$ partonic process presented here, need to be combined with infrared singular
contributions from both
the one-loop $2 \to 3$ partonic process, when one of the partons becomes unresolved~\cite{sone1,sone2,sone3} and
the tree-level $2 \to 4$ partonic process 
when two
partons become unresolved~\cite{tc1,tc2,tc3,ds1,ds2}. These different contributions need to be put together in such a way to
make the explicit singularities analytically cancel, so that the four-dimensional limit can be safely taken, 
and then by encoding the finite remainders in a stable numerical implementation.  There has been recent progress in
identifying appropriate counterterms~\cite{Kosower,subsNNLO} and 
we are optimistic that the technical problems will soon be overcome enabling predictions of
both dijet, jet and photon  and diphoton production at the Tevatron and the LHC at an accuracy of 
${\cal O}(5\%
)$  which is expected to be competitive with the experimental measurements.

\section*{Acknowledgements}

We thank Babis Anastasiou for helpful discussions and Lance Dixon, Zvi Bern and Abilio De Freitas for
communicating their results to us at an early stage and help in comparing our results.
M.E. Tejeda-Yeomans thanks the IPPP for hospitality.

\newpage
\appendix
\section{Finite two-loop contributions}
\label{app:twoloop}
In this appendix we give explicit formulae for the coefficients $A,\ldots,L$ for the $q\bar q \to gg$ process.
Expressions for quark-gluon scattering as well as the processes involving photons are available from the authors
as a FORM output file. 

As usual, the polylogarithms ${\rm Li}_n(w)$ are defined by
\begin{eqnarray}
 {\rm Li}_n(w) &=& \int_0^w \frac{dt}{t} {\rm Li}_{n-1}(t) \qquad {\rm ~for~}
 n=2,3,4\nonumber \\
 {\rm Li}_2(w) &=& -\int_0^w \frac{dt}{t} \log(1-t).
\label{eq:lidef}
\end{eqnarray} 
Using the standard polylogarithm identities~\cite{kolbig},
we retain the polylogarithms with arguments $x$, $1-x$ and
$(x-1)/x$, where
\begin{equation}
\label{eq:xydef}
x = -\frac{t}{s}, \qquad y = -\frac{u}{s} = 1-x, \qquad z=-\frac{u}{t} = \frac{x-1}{x}.
\end{equation}
For convenience, we also introduce the following logarithms
\begin{equation}
\label{eq:xydef1}
\lnx = \log\left(\frac{-t}{s}\right),
\qquad \lny = \log\left(\frac{-u}{s}\right),
\qquad \Ls = \log\left(\frac{s}{\mu^2}\right),
\end{equation}
where $\mu$ is the renormalisation scale.

\subsection{$q(p_2,+) + \bar q(p_1,-)\to g(p_3,+)+g(p_4,+)$}
\begin{eqnarray}
{A^{gg,(2),[1]}_{s+-++} }&=&{ }\Biggl ({}-{25\over 24}\,{\Lx^2}-{35\over 18}\,{\Lx}-{175\over 54}\Biggr ){}\,{\one }  
+{}\Biggl ({}-{17\over 8}\,{\Lx^2}-{5\over 2}\,{\Lx}\Biggr ){}\,{\tou }  
-{5\over 4}\,{\Lx^2}\,{\ttouu } \nonumber \\ &&
+{}\Biggl ({}-{1\over 8}\,{\Lx^2}+{11\over 8}\,{\Lx}-{475\over 144}\Biggr ){}\,{\sot } \nonumber \\ &&
+{}\Biggl ({}\Biggl ({}-{25\over 12}\,{\Lx}-{19\over 6}\Biggr ){}\,{\one }  
+{}\Biggl ({}-{17\over 4}\,{\Lx}-{5\over 2}\Biggr ){}\,{\tou }  
-{5\over 2}\,{\Lx}\,{\ttouu }  
+{}\Biggl ({}-{1\over 4}\,{\Lx}-{11\over 24}\Biggr ){}\,{\sot }  
\Biggr ){}\,{\IPI }\nonumber \\ &&
{}\\
{B^{gg,(2),[1]}_{s+-++} }&=&{ }\Biggl ({1\over 4}\,{\Lx^2}-{7\over 4}\,{\Lx}\,{\Ly}+{1\over 2}\,{\Lx}+{7\over 8}\,{\Ly^2}-{3\over 2}\,{\Ly}+{7\over 8}\,{\pi^2}-{2\over 3}\Biggr ){}\,{\one } 
+{}\Biggl ({}-{9\over 8}\,{\Lx^2}-{\Lx}\Biggr ){}\,{\tou } \nonumber \\ &&
-{1\over 2}\,{\Lx^2}\,{\ttouu } 
+{}\Biggl ({3\over 4}\,{\Lx^2}-{3\over 2}\,{\Lx}\,{\Ly}+{3\over 4}\,{\Ly^2}+{3\over 4}\,{\pi^2}\Biggr ){}\,{\tos } 
+{}\Biggl ({7\over 4}\,{\Lx}+{1\over 4}\,{\Ly^2}-{3\over 4}\,{\Ly}-{139\over 36}\Biggr ){}\,{\sot } \nonumber \\ &&
+{1\over 8}\,{\Ly^2}\,{\ssott } \nonumber \\ &&
+{}\Biggl ({}\Biggl ({}-{5\over 4}\,{\Lx}-{1}\Biggr ){}\,{\one }  
+{}\Biggl ({}-{9\over 4}\,{\Lx}-{1}\Biggr ){}\,{\tou }  
-{\Lx}\,{\ttouu }  
+{}\Biggl ({1\over 2}\,{\Ly}-{5\over 6}\Biggr ){}\,{\sot }  
+{1\over 4}\,{\Ly}\,{\ssott }  
\Biggr ){}\,{\IPI }\nonumber \\ &&
{}\\
{C^{gg,(2),[1]}_{s+-++} }&=&{ }\Biggl ({7\over 8}\,{\Lx^2}-{7\over 4}\,{\Lx}\,{\Ly}+{2}\,{\Lx}+{7\over 8}\,{\Ly^2}-{3\over 2}\,{\Ly}+{7\over 8}\,{\pi^2}\Biggr ){}\,{\one }  
+{}\Biggl ({1\over 4}\,{\Lx^2}+{1\over 2}\,{\Lx}\Biggr ){}\,{\tou } \nonumber \\ &&
+{1\over 4}\,{\Lx^2}\,{\ttouu } 
+{}\Biggl ({3\over 4}\,{\Lx^2}-{3\over 2}\,{\Lx}\,{\Ly}+{3\over 4}\,{\Ly^2}+{3\over 4}\,{\pi^2}\Biggr ){}\,{\tos } \nonumber \\ &&
+{}\Biggl ({1\over 8}\,{\Lx^2}+{3\over 8}\,{\Lx}+{1\over 4}\,{\Ly^2}-{3\over 4}\,{\Ly}-{17\over 16}\Biggr ){}\,{\sot }  
+{1\over 8}\,{\Ly^2}\,{\ssott } \nonumber \\ &&
+{}\Biggl ({1\over 2}\,{\one }  
+{}\Biggl ({1\over 2}\,{\Lx}+{1\over 2}\Biggr ){}\,{\tou }  
+{1\over 2}\,{\Lx}\,{\ttouu }  
+{}\Biggl ({1\over 4}\,{\Lx}+{1\over 2}\,{\Ly}-{3\over 8}\Biggr ){}\,{\sot }  
+{1\over 4}\,{\Ly}\,{\ssott }  
\Biggr ){}\,{\IPI }\nonumber \\ &&
{}\\
{D^{gg,(2),[1]}_{s+-++} }&=&{ }\Biggl ({5\over 12}\,{\Lx^2}+{7\over 18}\,{\Lx}+{397\over 108}\Biggr ){}\,{\one }  
+{}\Biggl ({3\over 4}\,{\Lx^2}+{\Lx}\Biggr ){}\,{\tou }  
+{1\over 2}\,{\Lx^2}\,{\ttouu }  
+{}\Biggl ({}-{1\over 4}\,{\Lx}+{37\over 36}\Biggr ){}\,{\sot }  \nonumber \\ &&
+{}\Biggl ({}\Biggl ({5\over 6}\,{\Lx}+{11\over 6}\Biggr ){}\,{\one } 
+{}\Biggl ({3\over 2}\,{\Lx}+{1}\Biggr ){}\,{\tou } 
+{\Lx}\,{\ttouu }  
+{1\over 12}\,{\sot }  
\Biggr ){}\,{\IPI }\nonumber \\ &&
{}\\
{E^{gg,(2),[1]}_{s+-++} }&=&{ }\Biggl ({}-{1\over 2}\,{\Lx}\,{\Ly}+{3\over 2}\,{\Lx}+{1\over 4}\,{\Ly^2}-{3}\,{\Ly}+{1\over 4}\,{\pi^2}+{11\over 12}\Biggr ){}\,{\one } 
+{}\Biggl ({}-{\Lx^2}-{3\over 2}\,{\Lx}\Biggr ){}\,{\tou } \nonumber \\ &&
-{3\over 4}\,{\Lx^2}\,{\ttouu }  
+{}\Biggl ({1\over 2}\,{\Lx^2}-{\Lx}\,{\Ly}+{1\over 2}\,{\Ly^2}+{1\over 2}\,{\pi^2}\Biggr ){}\,{\tos }  
+{}\Biggl ({}-{1\over 4}\,{\Lx}+{1\over 2}\,{\Ly^2}-{3\over 2}\,{\Ly}+{19\over 36}\Biggr ){}\,{\sot } \nonumber \\ &&
+{3\over 4}\,{\Ly^2}\,{\ssott } \nonumber \\ &&
+{}\Biggl ({}\Biggl ({}-{1\over 2}\,{\Lx}-{3\over 2}\Biggr ){}\,{\one }  
+{}\Biggl ({}-{2}\,{\Lx}-{3\over 2}\Biggr ){}\,{\tou }  
-{3\over 2}\,{\Lx}\,{\ttouu }  
+{}\Biggl ({\Ly}-{17\over 12}\Biggr ){}\,{\sot }  
+{3\over 2}\,{\Ly}\,{\ssott }  
\Biggr ){}\,{\IPI }\nonumber \\ &&
{}\\
{F^{gg,(2),[1]}_{s+-++} }&=&{ }-{1\over 6}\,{\IPI } 
{}\,{\one }  
+{}\Biggl ({1\over 18}\,{\Lx}-{5\over 27}\Biggr ){}\,{\one } \nonumber \\ &&
\\
{G^{gg,(2),[1]}_{s+-++} }&=&{ 0}
\end{eqnarray}
\begin{eqnarray}
{A^{gg,(2),[2]}_{s+-++} }&=&{ }\Biggl ({1\over 24}\,{\Ly^2}+{59\over 72}\,{\Ly}-{25\over 432}\Biggr ){}\,{\one }  
+{}\Biggl ({}-{1\over 8}\,{\Ly^2}+{11\over 8}\,{\Ly}-{475\over 144}\Biggr ){}\,{\tou }  
+{}\Biggl ({3\over 8}\,{\Ly^2}-{5\over 2}\,{\Ly}\Biggr ){}\,{\sot } \nonumber \\ &&
+{5\over 4}\,{\Ly^2}\,{\ssott } \nonumber \\ &&
+{}\Biggl ({}\Biggl ({1\over 12}\,{\Ly}+{5\over 24}\Biggr ){}\,{\one } 
+{}\Biggl ({}-{1\over 4}\,{\Ly}-{11\over 24}\Biggr ){}\,{\tou }  
+{}\Biggl ({3\over 4}\,{\Ly}-{5\over 2}\Biggr ){}\,{\sot }  
+{5\over 2}\,{\Ly}\,{\ssott } 
\Biggr ){}\,{\IPI }\nonumber \\ &&
{}\\
{B^{gg,(2),[2]}_{s+-++} }&=&{ }\Biggl ({1\over 4}\,{\Lx}\,{\Ly}+{3\over 4}\,{\Lx}-{1\over 8}\,{\Ly^2}+{1\over 4}\,{\Ly}-{1\over 8}\,{\pi^2}-{115\over 36}\Biggr ){}\,{\one } 
+{}\Biggl ({}-{3\over 4}\,{\Lx}+{7\over 4}\,{\Ly}-{139\over 36}\Biggr ){}\,{\tou }  \nonumber \\ &&
-{1\over 8}\,{\Lx^2}\,{\ttouu }  
+{}\Biggl ({3\over 4}\,{\Lx^2}-{3\over 2}\,{\Lx}\,{\Ly}+{3\over 4}\,{\Ly^2}+{3\over 4}\,{\pi^2}\Biggr ){}\,{\tos }  
+{}\Biggl ({}-{1\over 8}\,{\Ly^2}-{\Ly}\Biggr ){}\,{\sot }  
+{1\over 2}\,{\Ly^2}\,{\ssott } \nonumber \\ &&
+{}\Biggl ({}\Biggl ({1\over 4}\,{\Lx}-{5\over 6}\Biggr ){}\,{\one }  
-{5\over 6}\,{\tou }  
-{1\over 4}\,{\Lx}\,{\ttouu }  
+{}\Biggl ({}-{1\over 4}\,{\Ly}-{1}\Biggr ){}\,{\sot }  
+{\Ly}\,{\ssott }  
\Biggr ){}\,{\IPI }\nonumber \\ &&
{}\\
{C^{gg,(2),[2]}_{s+-++} }&=&{ }\Biggl ({1\over 4}\,{\Lx}\,{\Ly}+{3\over 4}\,{\Lx}-{9\over 8}\,{\Ly}-{1\over 8}\,{\pi^2}-{17\over 16}\Biggr ){}\,{\one }  
+{}\Biggl ({}-{3\over 4}\,{\Lx}+{1\over 8}\,{\Ly^2}+{3\over 8}\,{\Ly}-{17\over 16}\Biggr ){}\,{\tou } \nonumber \\ &&
-{1\over 8}\,{\Lx^2}\,{\ttouu }  
+{}\Biggl ({3\over 4}\,{\Lx^2}-{3\over 2}\,{\Lx}\,{\Ly}+{3\over 4}\,{\Ly^2}+{3\over 4}\,{\pi^2}\Biggr ){}\,{\tos } \nonumber \\ &&
+{}\Biggl ({}-{1\over 4}\,{\Ly^2}+{1\over 2}\,{\Ly}\Biggr ){}\,{\sot }  
-{1\over 4}\,{\Ly^2}\,{\ssott } \nonumber \\ &&
+{}\Biggl ({}\Biggl ({1\over 4}\,{\Lx}+{1\over 4}\,{\Ly}-{3\over 8}\Biggr ){}\,{\one }  
+{}\Biggl ({1\over 4}\,{\Ly}-{3\over 8}\Biggr ){}\,{\tou }  
-{1\over 4}\,{\Lx}\,{\ttouu }  
+{}\Biggl ({}-{1\over 2}\,{\Ly}+{1\over 2}\Biggr ){}\,{\sot }  
-{1\over 2}\,{\Ly}\,{\ssott }  
\Biggr ){}\,{\IPI }\nonumber \\ &&
{}\\
{D^{gg,(2),[2]}_{s+-++} }&=&{ }\Biggl ({}-{1\over 6}\,{\Ly^2}+{13\over 36}\,{\Ly}-{143\over 54}\Biggr ){}\,{\one }  
+{}\Biggl ({}-{1\over 4}\,{\Ly}+{37\over 36}\Biggr ){}\,{\tou }  
+{}\Biggl ({}-{1\over 4}\,{\Ly^2}+{\Ly}\Biggr ){}\,{\sot }  
-{1\over 2}\,{\Ly^2}\,{\ssott } \nonumber \\ &&
+{}\Biggl ({}\Biggl ({}-{1\over 3}\,{\Ly}-{3\over 4}\Biggr ){}\,{\one }  
+{1\over 12}\,{\tou }  
+{}\Biggl ({}-{1\over 2}\,{\Ly}+{1}\Biggr ){}\,{\sot }  
-{\Ly}\,{\ssott }  
\Biggr ){}\,{\IPI }\nonumber \\ &&
{}\\
{E^{gg,(2),[2]}_{s+-++} }&=&{ }\Biggl ({}-{1\over 2}\,{\Lx}\,{\Ly}+{3\over 2}\,{\Lx}+{1\over 4}\,{\Ly^2}-{13\over 4}\,{\Ly}+{1\over 4}\,{\pi^2}-{7\over 18}\Biggr ){}\,{\one }  
+{}\Biggl ({}-{\Lx^2}-{3\over 2}\,{\Lx}-{1\over 4}\,{\Ly}+{19\over 36}\Biggr ){}\,{\tou } \nonumber \\ &&
-{3\over 4}\,{\Lx^2}\,{\ttouu }  
+{}\Biggl ({1\over 2}\,{\Lx^2}-{\Lx}\,{\Ly}+{1\over 2}\,{\Ly^2}+{1\over 2}\,{\pi^2}\Biggr ){}\,{\tos }  
+{}\Biggl ({1\over 2}\,{\Ly^2}-{3\over 2}\,{\Ly}\Biggr ){}\,{\sot } 
+{3\over 4}\,{\Ly^2}\,{\ssott } \nonumber \\ &&
+{}\Biggl ({}\Biggl ({}-{1\over 2}\,{\Lx}-{17\over 12}\Biggr ){}\,{\one } 
+{}\Biggl ({}-{2}\,{\Lx}-{17\over 12}\Biggr ){}\,{\tou }  
-{3\over 2}\,{\Lx}\,{\ttouu }  
+{}\Biggl ({\Ly}-{3\over 2}\Biggr ){}\,{\sot }  
+{3\over 2}\,{\Ly}\,{\ssott }  
\Biggr ){}\,{\IPI }\nonumber \\ &&
{}\\
{F^{gg,(2),[2]}_{s+-++} }&=&{ 1\over 6}\,{\IPI } 
{}\,{\one } 
+{}\Biggl ({}-{1\over 18}\,{\Ly}+{5\over 27}\Biggr ){}\,{\one } \nonumber \\ &&
\\
{G^{gg,(2),[2]}_{s+-++} }&=&{ 0}
\end{eqnarray}
\begin{eqnarray}
{H^{gg,(2),[3]}_{s+-++} }&=&{ }\Biggl ({}-{1\over 48}\,{\Lx^2}-{1\over 2}\,{\Lx}\,{\Ly}+{11\over 18}\,{\Lx}+{5\over 24}\,{\Ly^2}-{16\over 9}\,{\Ly}+{1\over 4}\,{\pi^2}\Biggr ){}\,{\one } \nonumber \\ &&
+{}\Biggl ({}-{11\over 16}\,{\Lx^2}-{1\over 4}\,{\Lx}\,{\Ly}-{7\over 6}\,{\Lx}+{1\over 8}\,{\pi^2}\Biggr ){}\,{\tou } 
-{5\over 8}\,{\Lx^2}\,{\ttouu } 
+{5\over 8}\,{\Ly^2}\,{\ssott } \nonumber \\ &&
+{}\Biggl ({1\over 4}\,{\Lx^2}-{1\over 2}\,{\Lx}\,{\Ly}+{1\over 4}\,{\Ly^2}+{1\over 4}\,{\pi^2}\Biggr ){}\,{\tos } 
+{}\Biggl ({}-{1\over 4}\,{\Lx}\,{\Ly}+{9\over 16}\,{\Ly^2}-{7\over 6}\,{\Ly}+{1\over 8}\,{\pi^2}\Biggr ){}\,{\sot } \nonumber \\ &&
+{}\Biggl ({}\Biggl ({}-{13\over 24}\,{\Lx}-{1\over 12}\,{\Ly}-{7\over 6}\Biggr ){}\,{\one }  
+{}\Biggl ({}-{13\over 8}\,{\Lx}-{1\over 4}\,{\Ly}-{7\over 6}\Biggr ){}\,{\tou } 
-{5\over 4}\,{\Lx}\,{\ttouu } \nonumber \\ &&
+{}\Biggl ({}-{1\over 4}\,{\Lx}+{7\over 8}\,{\Ly}-{7\over 6}\Biggr ){}\,{\sot }  
+{5\over 4}\,{\Ly}\,{\ssott }  
\Biggr ){}\,{\IPI }\nonumber \\ &&
{}\\
{I^{gg,(2),[3]}_{s+-++} }&=&{ }\Biggl ({}-{1\over 16}\,{\Lx^2}+{1\over 4}\,{\Lx}\,{\Ly}-{11\over 12}\,{\Lx}-{1\over 4}\,{\Ly^2}+{\Ly}-{1\over 8}\,{\pi^2}\Biggr ){}\,{\one } \nonumber \\ &&
+{}\Biggl ({3\over 16}\,{\Lx^2}-{1\over 4}\,{\Lx}\,{\Ly}+{1\over 12}\,{\Lx}+{1\over 8}\,{\pi^2}\Biggr ){}\,{\tou } \nonumber \\ &&
+{}\Biggl ({}-{1\over 2}\,{\Lx^2}+{\Lx}\,{\Ly}-{1\over 2}\,{\Ly^2}-{1\over 2}\,{\pi^2}\Biggr ){}\,{\tos } \nonumber \\ &&
+{}\Biggl ({}-{1\over 4}\,{\Lx}\,{\Ly}+{3\over 16}\,{\Ly^2}+{1\over 12}\,{\Ly}+{1\over 8}\,{\pi^2}\Biggr ){}\,{\sot } \nonumber \\ &&
+{}\Biggl ({}\Biggl ({1\over 8}\,{\Lx}-{1\over 4}\,{\Ly}+{1\over 12}\Biggr ){}\,{\one }  
+{}\Biggl ({1\over 8}\,{\Lx}-{1\over 4}\,{\Ly}+{1\over 12}\Biggr ){}\,{\tou }  
+{}\Biggl ({}-{1\over 4}\,{\Lx}+{1\over 8}\,{\Ly}+{1\over 12}\Biggr ){}\,{\sot }  
\Biggr ){}\,{\IPI }\nonumber \\ &&
{}\\
{J^{gg,(2),[3]}_{s+-++} }&=&{ }\Biggl ({}-{1\over 24}\,{\Lx^2}+{1\over 2}\,{\Lx}\,{\Ly}-{55\over 72}\,{\Lx}-{1\over 3}\,{\Ly^2}+{47\over 36}\,{\Ly}-{1\over 4}\,{\pi^2}\Biggr ){}\,{\one }  
+{}\Biggl ({3\over 8}\,{\Lx^2}+{13\over 24}\,{\Lx}\Biggr ){}\,{\tou } \nonumber \\ &&
+{1\over 4}\,{\Lx^2}\,{\ttouu } 
+{}\Biggl ({}-{1\over 2}\,{\Lx^2}+{\Lx}\,{\Ly}-{1\over 2}\,{\Ly^2}-{1\over 2}\,{\pi^2}\Biggr ){}\,{\tos }  
+{}\Biggl ({}-{1\over 8}\,{\Ly^2}+{13\over 24}\,{\Ly}\Biggr ){}\,{\sot } \nonumber \\ &&
-{1\over 4}\,{\Ly^2}\,{\ssott } \nonumber \\ &&
+{}\Biggl ({}\Biggl ({5\over 12}\,{\Lx}-{1\over 6}\,{\Ly}+{13\over 24}\Biggr ){}\,{\one }  
+{}\Biggl ({3\over 4}\,{\Lx}+{13\over 24}\Biggr ){}\,{\tou }  
+{1\over 2}\,{\Lx}\,{\ttouu } 
+{}\Biggl ({}-{1\over 4}\,{\Ly}+{13\over 24}\Biggr ){}\,{\sot } \nonumber \\ &&
-{1\over 2}\,{\Ly}\,{\ssott }  
\Biggr ){}\,{\IPI }\nonumber \\ &&
{}\\
{K^{gg,(2),[3]}_{s+-++} }&=&{ }\Biggl ({1\over 4}\,{\Lx}\,{\Ly}-{17\over 24}\,{\Lx}-{1\over 8}\,{\Ly^2}+{3\over 2}\,{\Ly}-{1\over 8}\,{\pi^2}\Biggr ){}\,{\one }  
+{}\Biggl ({1\over 2}\,{\Lx^2}+{19\over 24}\,{\Lx}\Biggr ){}\,{\tou } \nonumber \\ &&
+{3\over 8}\,{\Lx^2}\,{\ttouu }  
+{}\Biggl ({}-{1\over 4}\,{\Lx^2}+{1\over 2}\,{\Lx}\,{\Ly}-{1\over 4}\,{\Ly^2}-{1\over 4}\,{\pi^2}\Biggr ){}\,{\tos } \nonumber \\ &&
+{}\Biggl ({}-{1\over 4}\,{\Ly^2}+{19\over 24}\,{\Ly}\Biggr ){}\,{\sot }  
-{3\over 8}\,{\Ly^2}\,{\ssott } \nonumber \\ &&
+{}\Biggl ({}\Biggl ({1\over 4}\,{\Lx}+{19\over 24}\Biggr ){}\,{\one }  
+{}\Biggl ({\Lx}+{19\over 24}\Biggr ){}\,{\tou }  
+{3\over 4}\,{\Lx}\,{\ttouu }  
+{}\Biggl ({}-{1\over 2}\,{\Ly}+{19\over 24}\Biggr ){}\,{\sot }  
-{3\over 4}\,{\Ly}\,{\ssott }  
\Biggr ){}\,{\IPI }\nonumber \\ &&
{}\\
{L^{gg,(2),[3]}_{s+-++} }&=&{ }\Biggl ({1\over 36}\,{\Lx}-{1\over 36}\,{\Ly}\Biggr ){}\,{\one }  
\end{eqnarray}
\subsection{$q(p_2,+) + \bar q(p_1,-)\to g(p_3,+)+g(p_4,-)$}
\begin{eqnarray}
{A^{gg,(2),[1]}_{s+-+-} }&=&{ }\Biggl ({1\over 2}\,{\Lidx}+{35\over 12}\,{\Licx}-{1\over 4}\,{\Libx}\,{\Lx^2}-{35\over 12}\,{\Libx}\,{\Lx}+{1\over 3}\,{\Libx}\,{\pi^2}-{1\over 16}\,{\Lx^4}\nonumber \\ &&-{1\over 6}\,{\Lx^3}\,{\Ly}+{77\over 72}\,{\Lx^3}-{35\over 24}\,{\Lx^2}\,{\Ly}+{1\over 3}\,{\Lx^2}\,{\pi^2}-{505\over 144}\,{\Lx^2}+{1\over 3}\,{\Lx}\,{\Ly}\,{\pi^2}+{2}\,{\Lx}\,{\zeta_3}\nonumber \\ &&+{143\over 72}\,{\Lx}\,{\pi^2}-{2273\over 432}\,{\Lx}-{241\over 24}\,{\zeta_3}-{19\over 288}\,{\pi^4}+{5\over 4}\,{\pi^2}+{36077\over 1728}\Biggr ){}\,{\one } \nonumber \\ &&
+{}\Biggl ({}-{1\over 4}\,{\Licx}+{1\over 4}\,{\Libx}\,{\Lx}-{7\over 24}\,{\Lx^3}+{1\over 8}\,{\Lx^2}\,{\Ly}+{31\over 16}\,{\Lx^2}+{2\over 3}\,{\Lx}\,{\pi^2}+{203\over 144}\,{\Lx}\nonumber \\ &&-{1\over 2}\,{\zeta_3}+{17\over 36}\,{\pi^2}+{475\over 144}\Biggr ){}\,{\tou } \nonumber \\ &&
+{}\Biggl ({1\over 2}\,{\Licx}-{1\over 2}\,{\Libx}\,{\Lx}+{1\over 3}\,{\Lx^3}-{1\over 4}\,{\Lx^2}\,{\Ly}+{11\over 16}\,{\Lx^2}-{5\over 6}\,{\Lx}\,{\pi^2}+{475\over 72}\,{\Lx}\nonumber \\ &&+{\zeta_3}-{23\over 18}\,{\pi^2}\Biggr ){}\,{\ttouu } \nonumber \\ &&
+{}\Biggl ({3\over 2}\,{\Lidx}-{\Licx}\,{\Lx}+{1\over 4}\,{\Libx}\,{\Lx^2}+{7\over 48}\,{\Lx^4}-{43\over 36}\,{\Lx^3}-{2\over 3}\,{\Lx^2}\,{\pi^2}\nonumber \\ &&+{475\over 144}\,{\Lx^2}+{\Lx}\,{\zeta_3}-{23\over 18}\,{\Lx}\,{\pi^2}-{1\over 60}\,{\pi^4}\Biggr ){}\,{\tttouuu } \nonumber \\ &&
+{}\Biggl ({}\Biggl ({}-{1\over 2}\,{\Libx}\,{\Lx}-{35\over 12}\,{\Libx}-{1\over 4}\,{\Lx^3}-{1\over 2}\,{\Lx^2}\,{\Ly}+{11\over 8}\,{\Lx^2}-{35\over 12}\,{\Lx}\,{\Ly}\nonumber \\ &&+{1\over 6}\,{\Lx}\,{\pi^2}-{7\over 24}\,{\Lx}+{2}\,{\zeta_3}+{11\over 24}\,{\pi^2}+{823\over 432}\Biggr ){}\,{\one } \nonumber \\ &&
+{}\Biggl ({1\over 4}\,{\Libx}-{7\over 8}\,{\Lx^2}+{1\over 4}\,{\Lx}\,{\Ly}+{49\over 24}\,{\Lx}+{1\over 12}\,{\pi^2}+{467\over 144}\Biggr ){}\,{\tou } \nonumber \\ &&
+{}\Biggl ({}-{1\over 2}\,{\Libx}+{\Lx^2}-{1\over 2}\,{\Lx}\,{\Ly}+{121\over 24}\,{\Lx}-{1\over 6}\,{\pi^2}+{475\over 72}\Biggr ){}\,{\ttouu } \nonumber \\ &&
+{}\Biggl ({}-{\Licx}+{1\over 2}\,{\Libx}\,{\Lx}+{7\over 12}\,{\Lx^3}-{7\over 4}\,{\Lx^2}-{1\over 6}\,{\Lx}\,{\pi^2}+{475\over 72}\,{\Lx}+{\zeta_3}\Biggr ){}\,{\tttouuu } 
\Biggr ){}\,{\IPI }\nonumber \\ &&
{}\\
{B^{gg,(2),[1]}_{s+-+-} }&=&{ }\Biggl ({2}\,{\Lidx}-{7}\,{\Lidy}+{\Lidz}+{\Licx}\,{\Lx}-{\Licx}\,{\Ly}+{9\over 2}\,{\Licx}+{2}\,{\Licy}\,{\Lx}\nonumber \\ &&+{4}\,{\Licy}\,{\Ly}+{6}\,{\Licy}-{5\over 2}\,{\Libx}\,{\Lx^2}-{9\over 2}\,{\Libx}\,{\Lx}+{1\over 6}\,{\Libx}\,{\pi^2}-{4}\,{\Liby}\,{\Lx}\,{\Ly}\nonumber \\ &&-{6}\,{\Liby}\,{\Ly}+{1\over 6}\,{\Lx^4}-{11\over 6}\,{\Lx^3}\,{\Ly}-{25\over 36}\,{\Lx^3}-{3\over 2}\,{\Lx^2}\,{\Ly^2}+{13\over 12}\,{\Lx^2}\,{\Ly}+{3\over 4}\,{\Lx^2}\,{\pi^2}\nonumber \\ &&-{95\over 36}\,{\Lx^2}+{5\over 6}\,{\Lx}\,{\Ly^3}-{13\over 2}\,{\Lx}\,{\Ly^2}-{1\over 2}\,{\Lx}\,{\Ly}\,{\pi^2}+{25\over 9}\,{\Lx}\,{\Ly}-{2}\,{\Lx}\,{\zeta_3}-{40\over 9}\,{\Lx}\,{\pi^2}+{97\over 12}\,{\Lx}\nonumber \\ &&-{1\over 12}\,{\Ly^4}+{5\over 3}\,{\Ly^3}-{1\over 4}\,{\Ly^2}\,{\pi^2}-{191\over 72}\,{\Ly^2}+{\Ly}\,{\zeta_3}+{7\over 6}\,{\Ly}\,{\pi^2}+{3\over 2}\,{\Ly}+{11\over 72}\,{\zeta_3}\nonumber \\ &&+{73\over 1440}\,{\pi^4}-{329\over 144}\,{\pi^2}-{14135\over 1296}\Biggr ){}\,{\one } \nonumber \\ &&
+{}\Biggl ({9\over 2}\,{\Lidx}-{15}\,{\Lidy}-{11}\,{\Lidz}+{7}\,{\Licx}\,{\Lx}-{7}\,{\Licx}\,{\Ly}+{13\over 2}\,{\Licx}\nonumber \\ &&+{14}\,{\Licy}\,{\Lx}+{6}\,{\Licy}-{15\over 4}\,{\Libx}\,{\Lx^2}-{13\over 2}\,{\Libx}\,{\Lx}-{17\over 6}\,{\Libx}\,{\pi^2}\nonumber \\ &&-{10}\,{\Liby}\,{\Lx}\,{\Ly}+{2}\,{\Liby}\,{\Ly^2}-{6}\,{\Liby}\,{\Ly}-{11\over 48}\,{\Lx^4}-{31\over 12}\,{\Lx^3}-{5}\,{\Lx^2}\,{\Ly^2}+{1\over 4}\,{\Lx^2}\,{\Ly}\nonumber \\ &&-{5\over 6}\,{\Lx^2}\,{\pi^2}+{305\over 72}\,{\Lx^2}+{3\over 2}\,{\Lx}\,{\Ly^3}-{29\over 6}\,{\Lx}\,{\Ly^2}-{5\over 6}\,{\Lx}\,{\Ly}\,{\pi^2}-{37\over 36}\,{\Lx}\,{\Ly}-{13}\,{\Lx}\,{\zeta_3}-{25\over 3}\,{\Lx}\,{\pi^2}\nonumber \\ &&+{383\over 36}\,{\Lx}-{5\over 4}\,{\Ly^2}\,{\pi^2}+{7}\,{\Ly}\,{\zeta_3}+{3\over 2}\,{\Ly}\,{\pi^2}+{3\over 4}\,{\Ly}-{10}\,{\zeta_3}+{19\over 45}\,{\pi^4}-{407\over 72}\,{\pi^2}+{139\over 36}\Biggr ){}\,{\tou } \nonumber \\ &&
+{}\Biggl ({}-{9\over 2}\,{\Lidx}-{12}\,{\Lidy}-{12}\,{\Lidz}+{12}\,{\Licx}\,{\Lx}-{6}\,{\Licx}\,{\Ly}+{7\over 2}\,{\Licx}\nonumber \\ &&+{12}\,{\Licy}\,{\Lx}-{15\over 4}\,{\Libx}\,{\Lx^2}-{7\over 2}\,{\Libx}\,{\Lx}-{3}\,{\Libx}\,{\pi^2}-{6}\,{\Liby}\,{\Lx}\,{\Ly}-{5\over 16}\,{\Lx^4}\nonumber \\ &&+{\Lx^3}\,{\Ly}-{17\over 12}\,{\Lx^3}-{3}\,{\Lx^2}\,{\Ly^2}-{5\over 2}\,{\Lx^2}\,{\Ly}-{\Lx^2}\,{\pi^2}+{173\over 24}\,{\Lx^2}-{2}\,{\Lx}\,{\Ly}\,{\pi^2}+{3\over 2}\,{\Lx}\,{\Ly}\nonumber \\ &&-{12}\,{\Lx}\,{\zeta_3}-{77\over 12}\,{\Lx}\,{\pi^2}+{305\over 36}\,{\Lx}+{6}\,{\Ly}\,{\zeta_3}+{1\over 3}\,{\Ly}\,{\pi^2}-{11\over 2}\,{\zeta_3}+{11\over 20}\,{\pi^4}-{103\over 36}\,{\pi^2}\Biggr ){}\,{\ttouu } \nonumber \\ &&
+{}\Biggl ({}-{2}\,{\Lidx}+{\Licx}\,{\Lx}+{3\over 2}\,{\Licx}-{3\over 2}\,{\Libx}\,{\Lx}-{1\over 3}\,{\Libx}\,{\pi^2}+{1\over 8}\,{\Lx^4}\nonumber \\ &&-{43\over 36}\,{\Lx^3}-{1\over 4}\,{\Lx^2}\,{\pi^2}+{359\over 72}\,{\Lx^2}-{\Lx}\,{\zeta_3}-{91\over 36}\,{\Lx}\,{\pi^2}-{3\over 2}\,{\zeta_3}+{7\over 90}\,{\pi^4}\Biggr ){}\,{\tttouuu } \nonumber \\ &&
+{}\Biggl ({}-{3\over 8}\,{\Lx^2}+{3\over 4}\,{\Lx}\,{\Ly}-{3\over 8}\,{\Ly^2}-{3\over 8}\,{\pi^2}\Biggr ){}\,{\tos } \nonumber \\ &&
+{}\Biggl ({}\Biggl ({6}\,{\Licy}-{\Libx}\,{\Lx}+{3\over 2}\,{\Libx}-{4}\,{\Liby}\,{\Ly}+{1\over 3}\,{\Lx^3}-{7\over 2}\,{\Lx^2}\,{\Ly}+{37\over 12}\,{\Lx^2}\nonumber \\ &&+{2}\,{\Lx}\,{\Ly^2}-{29\over 6}\,{\Lx}\,{\Ly}+{2\over 3}\,{\Lx}\,{\pi^2}+{3}\,{\Lx}-{1\over 6}\,{\Ly^3}+{3\over 2}\,{\Ly^2}-{5\over 6}\,{\Ly}\,{\pi^2}-{91\over 36}\,{\Ly}-{\zeta_3}+{\pi^2}-{13\over 4}\Biggr ){}\,{\one } \nonumber \\ &&
+{}\Biggl ({14}\,{\Licy}+{5\over 2}\,{\Libx}\,{\Lx}-{1\over 2}\,{\Libx}-{6}\,{\Liby}\,{\Ly}+{5\over 12}\,{\Lx^3}-{2}\,{\Lx^2}\,{\Ly}+{5\over 4}\,{\Lx^2}\nonumber \\ &&+{3}\,{\Lx}\,{\Ly^2}-{19\over 6}\,{\Lx}\,{\Ly}-{2\over 3}\,{\Lx}\,{\pi^2}+{299\over 18}\,{\Lx}+{1\over 6}\,{\Ly^3}-{11\over 6}\,{\Ly^2}-{5\over 6}\,{\Ly}\,{\pi^2}-{37\over 36}\,{\Ly}-{6}\,{\zeta_3}\nonumber \\ &&+{4\over 3}\,{\pi^2}+{119\over 9}\Biggr ){}\,{\tou } \nonumber \\ &&
+{}\Biggl ({6}\,{\Licx}+{12}\,{\Licy}-{3\over 2}\,{\Libx}\,{\Lx}-{7\over 2}\,{\Libx}-{6}\,{\Liby}\,{\Ly}+{1\over 4}\,{\Lx^3}+{1\over 2}\,{\Lx^2}\nonumber \\ &&-{5}\,{\Lx}\,{\Ly}-{3\over 2}\,{\Lx}\,{\pi^2}+{235\over 12}\,{\Lx}+{3\over 2}\,{\Ly}-{6}\,{\zeta_3}+{7\over 12}\,{\pi^2}+{305\over 36}\Biggr ){}\,{\ttouu } \nonumber \\ &&
+{}\Biggl ({\Licx}-{3\over 2}\,{\Libx}+{1\over 3}\,{\Lx^3}-{\Lx^2}-{1\over 6}\,{\Lx}\,{\pi^2}+{359\over 36}\,{\Lx}-{\zeta_3}+{1\over 4}\,{\pi^2}\Biggr ){}\,{\tttouuu } 
\Biggr ){}\,{\IPI }\nonumber \\ &&
{}\\
{C^{gg,(2),[1]}_{s+-+-} }&=&{ }\Biggl ({}-{5\over 2}\,{\Lidx}+{\Lidy}+{\Lidz}+{\Licx}\,{\Lx}+{\Licx}\,{\Ly}-{3\over 4}\,{\Licx}+{3}\,{\Licy}\nonumber \\ &&-{1\over 4}\,{\Libx}\,{\Lx^2}+{3\over 4}\,{\Libx}\,{\Lx}-{1\over 6}\,{\Libx}\,{\pi^2}-{3}\,{\Liby}\,{\Ly}-{1\over 48}\,{\Lx^4}-{1\over 6}\,{\Lx^3}\,{\Ly}-{3\over 8}\,{\Lx^3}\nonumber \\ &&+{1\over 4}\,{\Lx^2}\,{\Ly^2}+{3\over 8}\,{\Lx^2}\,{\Ly}+{1\over 3}\,{\Lx^2}\,{\pi^2}+{5\over 16}\,{\Lx^2}-{3\over 2}\,{\Lx}\,{\Ly^2}-{1\over 6}\,{\Lx}\,{\Ly}\,{\pi^2}-{\Lx}\,{\zeta_3}+{\Lx}\,{\pi^2}+{39\over 16}\,{\Lx}\nonumber \\ &&-{11\over 8}\,{\Ly^2}-{\Ly}\,{\zeta_3}+{3\over 2}\,{\Ly}+{9\over 4}\,{\zeta_3}+{31\over 360}\,{\pi^4}+{5\over 16}\,{\pi^2}-{187\over 64}\Biggr ){}\,{\one } \nonumber \\ &&
+{}\Biggl ({}-{7\over 2}\,{\Lidx}+{7}\,{\Lidy}+{7}\,{\Lidz}-{\Licx}\,{\Lx}+{7}\,{\Licx}\,{\Ly}+{3\over 4}\,{\Licx}+{9}\,{\Licy}\nonumber \\ &&-{3\over 4}\,{\Libx}\,{\Lx^2}-{3\over 4}\,{\Libx}\,{\Lx}+{1\over 6}\,{\Libx}\,{\pi^2}-{9}\,{\Liby}\,{\Ly}-{1\over 16}\,{\Lx^4}-{7\over 6}\,{\Lx^3}\,{\Ly}-{31\over 24}\,{\Lx^3}\nonumber \\ &&+{7\over 4}\,{\Lx^2}\,{\Ly^2}+{5\over 8}\,{\Lx^2}\,{\Ly}+{2\over 3}\,{\Lx^2}\,{\pi^2}+{27\over 16}\,{\Lx^2}-{9\over 2}\,{\Lx}\,{\Ly^2}-{7\over 6}\,{\Lx}\,{\Ly}\,{\pi^2}-{5\over 4}\,{\Lx}\,{\Ly}+{\Lx}\,{\zeta_3}\nonumber \\ &&+{2}\,{\Lx}\,{\pi^2}+{97\over 16}\,{\Lx}-{3\over 2}\,{\Ly^2}-{7}\,{\Ly}\,{\zeta_3}-{2\over 3}\,{\Ly}\,{\pi^2}+{3\over 4}\,{\Ly}-{1\over 2}\,{\zeta_3}+{1\over 90}\,{\pi^4}+{65\over 24}\,{\pi^2}+{17\over 16}\Biggr ){}\,{\tou } \nonumber \\ &&
+{}\Biggl ({3\over 2}\,{\Lidx}+{12}\,{\Lidy}+{12}\,{\Lidz}-{6}\,{\Licx}\,{\Lx}+{12}\,{\Licx}\,{\Ly}+{6}\,{\Licy}\nonumber \\ &&-{3\over 4}\,{\Libx}\,{\Lx^2}+{\Libx}\,{\pi^2}-{6}\,{\Liby}\,{\Ly}-{1\over 16}\,{\Lx^4}-{2}\,{\Lx^3}\,{\Ly}-{11\over 12}\,{\Lx^3}+{3}\,{\Lx^2}\,{\Ly^2}+{1\over 4}\,{\Lx^2}\,{\Ly}\nonumber \\ &&+{1\over 4}\,{\Lx^2}\,{\pi^2}+{35\over 16}\,{\Lx^2}-{3}\,{\Lx}\,{\Ly^2}-{2}\,{\Lx}\,{\Ly}\,{\pi^2}-{3\over 2}\,{\Lx}\,{\Ly}+{6}\,{\Lx}\,{\zeta_3}+{9\over 4}\,{\Lx}\,{\pi^2}+{23\over 8}\,{\Lx}\nonumber \\ &&-{12}\,{\Ly}\,{\zeta_3}-{2\over 3}\,{\Ly}\,{\pi^2}+{5\over 2}\,{\zeta_3}-{11\over 60}\,{\pi^4}+{23\over 12}\,{\pi^2}\Biggr ){}\,{\ttouu } \nonumber \\ &&
+{}\Biggl ({5\over 2}\,{\Lidx}+{6}\,{\Lidy}+{6}\,{\Lidz}-{4}\,{\Licx}\,{\Lx}+{6}\,{\Licx}\,{\Ly}-{3\over 2}\,{\Licx}\nonumber \\ &&-{1\over 4}\,{\Libx}\,{\Lx^2}+{3\over 2}\,{\Libx}\,{\Lx}+{2\over 3}\,{\Libx}\,{\pi^2}-{1\over 48}\,{\Lx^4}-{\Lx^3}\,{\Ly}+{3\over 2}\,{\Lx^2}\,{\Ly^2}-{1\over 12}\,{\Lx^2}\,{\pi^2}\nonumber \\ &&+{11\over 16}\,{\Lx^2}-{\Lx}\,{\Ly}\,{\pi^2}+{4}\,{\Lx}\,{\zeta_3}+{5\over 4}\,{\Lx}\,{\pi^2}-{6}\,{\Ly}\,{\zeta_3}+{3\over 2}\,{\zeta_3}-{5\over 36}\,{\pi^4}\Biggr ){}\,{\tttouuu } \nonumber \\ &&
+{}\Biggl ({}-{3\over 8}\,{\Lx^2}+{3\over 4}\,{\Lx}\,{\Ly}-{3\over 8}\,{\Ly^2}-{3\over 8}\,{\pi^2}\Biggr ){}\,{\tos } \nonumber \\ &&
+{}\Biggl ({}\Biggl ({2}\,{\Licx}-{1\over 2}\,{\Libx}\,{\Lx}+{15\over 4}\,{\Libx}-{1\over 4}\,{\Lx^3}-{9\over 8}\,{\Lx^2}+{3\over 4}\,{\Lx}\,{\Ly}+{5\over 8}\,{\Lx}\nonumber \\ &&-{11\over 4}\,{\Ly}-{2}\,{\zeta_3}-{1\over 4}\,{\pi^2}+{63\over 16}\Biggr ){}\,{\one } \nonumber \\ &&
+{}\Biggl ({6}\,{\Licx}-{3\over 2}\,{\Libx}\,{\Lx}+{33\over 4}\,{\Libx}-{3\over 4}\,{\Lx^3}-{23\over 8}\,{\Lx^2}+{5\over 4}\,{\Lx}\,{\Ly}+{17\over 8}\,{\Lx}\nonumber \\ &&-{17\over 4}\,{\Ly}-{6}\,{\zeta_3}-{3\over 4}\,{\pi^2}+{109\over 16}\Biggr ){}\,{\tou } \nonumber \\ &&
+{}\Biggl ({6}\,{\Licx}-{3\over 2}\,{\Libx}\,{\Lx}+{6}\,{\Libx}-{3\over 4}\,{\Lx^3}-{5\over 2}\,{\Lx^2}+{1\over 2}\,{\Lx}\,{\Ly}+{23\over 8}\,{\Lx}\nonumber \\ &&-{3\over 2}\,{\Ly}-{6}\,{\zeta_3}-{3\over 4}\,{\pi^2}+{23\over 8}\Biggr ){}\,{\ttouu } \nonumber \\ &&
+{}\Biggl ({2}\,{\Licx}-{1\over 2}\,{\Libx}\,{\Lx}+{3\over 2}\,{\Libx}-{1\over 4}\,{\Lx^3}-{3\over 4}\,{\Lx^2}+{11\over 8}\,{\Lx}-{2}\,{\zeta_3}-{1\over 4}\,{\pi^2}\Biggr ){}\,{\tttouuu } \nonumber \\ &&
\Biggr ){}\,{\IPI }\nonumber \\ &&
{}\\
{D^{gg,(2),[1]}_{s+-+-} }&=&{ }\Biggl ({}-{2\over 3}\,{\Licx}+{2\over 3}\,{\Libx}\,{\Lx}-{7\over 36}\,{\Lx^3}+{1\over 3}\,{\Lx^2}\,{\Ly}+{43\over 36}\,{\Lx^2}-{13\over 36}\,{\Lx}\,{\pi^2}-{5\over 9}\,{\Lx}+{17\over 12}\,{\zeta_3}\nonumber \\ &&-{77\over 108}\,{\pi^2}-{1307\over 432}\Biggr ){}\,{\one } \nonumber \\ &&
+{}\Biggl ({}-{1\over 2}\,{\Lx^2}-{19\over 18}\,{\Lx}-{5\over 36}\,{\pi^2}-{37\over 36}\Biggr ){}\,{\tou } 
+{}\Biggl ({}-{3\over 4}\,{\Lx^2}-{37\over 18}\,{\Lx}+{5\over 18}\,{\pi^2}\Biggr ){}\,{\ttouu } \nonumber \\ &&
+{}\Biggl ({7\over 36}\,{\Lx^3}-{37\over 36}\,{\Lx^2}+{5\over 18}\,{\Lx}\,{\pi^2}\Biggr ){}\,{\tttouuu } \nonumber \\ &&
+{}\Biggl ({}\Biggl ({2\over 3}\,{\Libx}-{1\over 4}\,{\Lx^2}+{2\over 3}\,{\Lx}\,{\Ly}-{1\over 18}\,{\Lx}-{1\over 12}\,{\pi^2}-{13\over 18}\Biggr ){}\,{\one } \nonumber \\ &&
+{}\Biggl ({}-{2\over 3}\,{\Lx}-{25\over 18}\Biggr ){}\,{\tou } 
+{}\Biggl ({}-{13\over 6}\,{\Lx}-{37\over 18}\Biggr ){}\,{\ttouu } 
+{}\Biggl ({1\over 4}\,{\Lx^2}-{37\over 18}\,{\Lx}\Biggr ){}\,{\tttouuu } 
\Biggr ){}\,{\IPI }\nonumber \\ &&
{}\\
{E^{gg,(2),[1]}_{s+-+-} }&=&{ }\Biggl ({}-{2\over 3}\,{\Licy}+{2\over 3}\,{\Liby}\,{\Ly}+{7\over 36}\,{\Lx^3}-{1\over 3}\,{\Lx^2}\,{\Ly}+{2\over 9}\,{\Lx^2}+{1\over 2}\,{\Lx}\,{\Ly^2}-{5\over 9}\,{\Lx}\,{\Ly}+{4\over 9}\,{\Lx}\,{\pi^2}\nonumber \\ &&-{11\over 6}\,{\Lx}-{1\over 6}\,{\Ly^3}-{17\over 18}\,{\Ly^2}+{7\over 18}\,{\Ly}\,{\pi^2}+{3}\,{\Ly}+{23\over 36}\,{\zeta_3}+{37\over 72}\,{\pi^2}+{3401\over 1296}\Biggr ){}\,{\one } \nonumber \\ &&
+{}\Biggl ({}-{4}\,{\Lidx}+{4}\,{\Lidy}+{4}\,{\Lidz}+{4}\,{\Licx}\,{\Ly}+{8}\,{\Licy}-{2\over 3}\,{\Libx}\,{\pi^2}-{8}\,{\Liby}\,{\Ly}\nonumber \\ &&+{1\over 6}\,{\Lx^4}-{2\over 3}\,{\Lx^3}\,{\Ly}+{1\over 2}\,{\Lx^3}+{\Lx^2}\,{\Ly^2}-{1\over 4}\,{\Lx^2}\,{\Ly}+{1\over 3}\,{\Lx^2}\,{\pi^2}+{1\over 9}\,{\Lx^2}-{11\over 3}\,{\Lx}\,{\Ly^2}-{2\over 3}\,{\Lx}\,{\Ly}\,{\pi^2}\nonumber \\ &&+{8\over 9}\,{\Lx}\,{\Ly}+{\Lx}\,{\pi^2}+{7\over 9}\,{\Lx}-{2}\,{\Ly^2}-{4}\,{\Ly}\,{\zeta_3}+{7\over 6}\,{\Ly}\,{\pi^2}+{5\over 3}\,{\Ly}-{8}\,{\zeta_3}+{7\over 45}\,{\pi^4}-{7\over 12}\,{\pi^2}-{19\over 36}\Biggr ){}\,{\tou } \nonumber \\ &&
+{}\Biggl ({}-{12}\,{\Lidx}+{12}\,{\Lidy}+{12}\,{\Lidz}+{12}\,{\Licx}\,{\Ly}+{8}\,{\Licy}-{2}\,{\Libx}\,{\pi^2}\nonumber \\ &&-{8}\,{\Liby}\,{\Ly}+{1\over 2}\,{\Lx^4}-{2}\,{\Lx^3}\,{\Ly}+{1\over 3}\,{\Lx^3}+{3}\,{\Lx^2}\,{\Ly^2}+{1\over 4}\,{\Lx^2}\,{\Ly}+{\Lx^2}\,{\pi^2}+{11\over 12}\,{\Lx^2}-{4}\,{\Lx}\,{\Ly^2}\nonumber \\ &&-{2}\,{\Lx}\,{\Ly}\,{\pi^2}-{2\over 3}\,{\Lx}\,{\Ly}+{5\over 6}\,{\Lx}\,{\pi^2}+{11\over 18}\,{\Lx}-{12}\,{\Ly}\,{\zeta_3}+{4\over 3}\,{\Ly}\,{\pi^2}-{8}\,{\zeta_3}+{7\over 15}\,{\pi^4}-{1\over 2}\,{\pi^2}\Biggr ){}\,{\ttouu } \nonumber \\ &&
+{}\Biggl ({}-{8}\,{\Lidx}+{8}\,{\Lidy}+{8}\,{\Lidz}+{8}\,{\Licx}\,{\Ly}-{2\over 3}\,{\Licx}+{2\over 3}\,{\Libx}\,{\Lx}\nonumber \\ &&-{4\over 3}\,{\Libx}\,{\pi^2}+{1\over 3}\,{\Lx^4}-{4\over 3}\,{\Lx^3}\,{\Ly}+{7\over 36}\,{\Lx^3}+{2}\,{\Lx^2}\,{\Ly^2}+{2\over 3}\,{\Lx^2}\,{\pi^2}-{1\over 36}\,{\Lx^2}-{4\over 3}\,{\Lx}\,{\Ly}\,{\pi^2}\nonumber \\ &&+{5\over 6}\,{\Lx}\,{\pi^2}-{8}\,{\Ly}\,{\zeta_3}+{2\over 3}\,{\zeta_3}+{14\over 45}\,{\pi^4}\Biggr ){}\,{\tttouuu } 
-{3\over 2}\,{\Ly^2}\,{\sot } \nonumber \\ &&
+{}\Biggl ({}\Biggl ({}-{2\over 3}\,{\Libx}-{1\over 12}\,{\Lx^2}-{1\over 3}\,{\Lx}\,{\Ly}-{10\over 9}\,{\Lx}-{1\over 3}\,{\Ly^2}-{22\over 9}\,{\Ly}+{3}\Biggr ){}\,{\one } \nonumber \\ &&
+{}\Biggl ({4}\,{\Licx}+{8}\,{\Libx}+{1\over 4}\,{\Lx^2}+{1\over 6}\,{\Lx}\,{\Ly}-{5\over 9}\,{\Lx}+{1\over 3}\,{\Ly^2}-{28\over 9}\,{\Ly}-{4}\,{\zeta_3}+{19\over 9}\Biggr ){}\,{\tou } \nonumber \\ &&
+{}\Biggl ({12}\,{\Licx}+{8}\,{\Libx}+{1\over 4}\,{\Lx^2}+{1\over 2}\,{\Lx}\,{\Ly}+{1\over 2}\,{\Lx}-{2\over 3}\,{\Ly}-{12}\,{\zeta_3}+{11\over 18}\Biggr ){}\,{\ttouu } \nonumber \\ &&
+{}\Biggl ({8}\,{\Licx}+{2\over 3}\,{\Libx}-{1\over 12}\,{\Lx^2}-{1\over 18}\,{\Lx}-{8}\,{\zeta_3}-{1\over 9}\,{\pi^2}\Biggr ){}\,{\tttouuu }  
-{3}\,{\Ly}\,{\sot } 
\Biggr ){}\,{\IPI }\nonumber \\ &&
{}\\
{F^{gg,(2),[1]}_{s+-+-} }&=&{ }\Biggl ({1\over 18}\,{\Lx}+{5\over 27}\Biggr ){}\,{\IPI } 
{}\,{\one } 
+{}\Biggl ({}-{1\over 12}\,{\Lx^2}+{5\over 27}\,{\Lx}+{2\over 27}\,{\pi^2}\Biggr ){}\,{\one } \nonumber \\ &&
\\
{G^{gg,(2),[1]}_{s+-+-} }&=&{ }\Biggl ({1\over 36}\,{\Lx}\,{\Ly}-{1\over 36}\,{\Ly^2}\Biggr ){}\,{\one } 
+{}\Biggl ({1\over 36}\,{\Lx^2}-{1\over 36}\,{\Lx}\,{\Ly}\Biggr ){}\,{\tou } \nonumber \\ &&
+{}\Biggl ({}\Biggl ({1\over 36}\,{\Lx}-{1\over 36}\,{\Ly}\Biggr ){}\,{\one } 
+{}\Biggl ({1\over 36}\,{\Lx}-{1\over 36}\,{\Ly}\Biggr ){}\,{\tou } 
\Biggr ){}\,{\IPI }\nonumber \\ &&
\end{eqnarray}
\begin{eqnarray}
{A^{gg,(2),[2]}_{s+-+-} }&=&{ }\Biggl ({}-{2}\,{\Lidx}+{\Lidy}+{2}\,{\Lidz}+{2}\,{\Licx}\,{\Ly}+{\Licy}\,{\Ly}+{41\over 3}\,{\Licy}-{2\over 3}\,{\Libx}\,{\pi^2}\nonumber \\ &&-{1\over 2}\,{\Liby}\,{\Ly^2}-{41\over 3}\,{\Liby}\,{\Ly}+{1\over 12}\,{\Lx^4}-{1\over 3}\,{\Lx^3}\,{\Ly}+{1\over 2}\,{\Lx^2}\,{\Ly^2}+{1\over 6}\,{\Lx^2}\,{\pi^2}-{1\over 6}\,{\Lx}\,{\Ly^3}\nonumber \\ &&-{41\over 6}\,{\Lx}\,{\Ly^2}-{1\over 3}\,{\Lx}\,{\Ly}\,{\pi^2}-{5\over 24}\,{\Ly^4}+{26\over 9}\,{\Ly^3}+{\Ly^2}\,{\pi^2}-{145\over 18}\,{\Ly^2}-{\Ly}\,{\zeta_3}+{247\over 72}\,{\Ly}\,{\pi^2}\nonumber \\ &&-{2\over 27}\,{\Ly}-{445\over 24}\,{\zeta_3}+{121\over 1440}\,{\pi^4}-{13\over 6}\,{\pi^2}+{30593\over 1728}\Biggr ){}\,{\one } \nonumber \\ &&
+{}\Biggl ({}-{12}\,{\Lidx}+{12}\,{\Lidy}+{12}\,{\Lidz}+{12}\,{\Licx}\,{\Ly}+{10}\,{\Licy}-{2}\,{\Libx}\,{\pi^2}\nonumber \\ &&-{10}\,{\Liby}\,{\Ly}+{1\over 2}\,{\Lx^4}-{2}\,{\Lx^3}\,{\Ly}+{3}\,{\Lx^2}\,{\Ly^2}+{\Lx^2}\,{\pi^2}-{5}\,{\Lx}\,{\Ly^2}-{2}\,{\Lx}\,{\Ly}\,{\pi^2}-{12}\,{\Ly}\,{\zeta_3}\nonumber \\ &&+{5\over 3}\,{\Ly}\,{\pi^2}-{10}\,{\zeta_3}+{7\over 15}\,{\pi^4}-{5\over 3}\,{\pi^2}+{1\over 8}\Biggr ){}\,{\tou } \nonumber \\ &&
+{}\Biggl ({}-{10}\,{\Lidx}+{10}\,{\Lidy}+{10}\,{\Lidz}+{10}\,{\Licx}\,{\Ly}-{5\over 3}\,{\Libx}\,{\pi^2}+{5\over 12}\,{\Lx^4}\nonumber \\ &&-{5\over 3}\,{\Lx^3}\,{\Ly}+{5\over 2}\,{\Lx^2}\,{\Ly^2}+{5\over 6}\,{\Lx^2}\,{\pi^2}-{5\over 3}\,{\Lx}\,{\Ly}\,{\pi^2}-{10}\,{\Ly}\,{\zeta_3}+{7\over 18}\,{\pi^4}\Biggr ){}\,{\ttouu } \nonumber \\ &&
+{}\Biggl ({}-{3\over 2}\,{\Lidy}+{\Licy}\,{\Ly}+{3}\,{\Licy}-{1\over 4}\,{\Liby}\,{\Ly^2}-{3}\,{\Liby}\,{\Ly}-{3\over 2}\,{\Lx}\,{\Ly^2}\nonumber \\ &&-{7\over 48}\,{\Ly^4}+{13\over 9}\,{\Ly^3}+{2\over 3}\,{\Ly^2}\,{\pi^2}+{713\over 144}\,{\Ly^2}-{\Ly}\,{\zeta_3}+{23\over 18}\,{\Ly}\,{\pi^2}-{5}\,{\Ly}-{3}\,{\zeta_3}+{1\over 60}\,{\pi^4}\Biggr ){}\,{\sot } \nonumber \\ &&
+{5\over 2}\,{\Ly^2}\,{\ssott } \nonumber \\ &&
+{}\Biggl ({}\Biggl ({2}\,{\Licx}+{\Licy}+{41\over 3}\,{\Libx}-{\Liby}\,{\Ly}-{1\over 2}\,{\Lx}\,{\Ly^2}-{5\over 6}\,{\Ly^3}+{5}\,{\Ly^2}+{1\over 3}\,{\Ly}\,{\pi^2}\nonumber \\ &&-{35\over 9}\,{\Ly}-{\zeta_3}-{29\over 72}\,{\pi^2}+{142\over 27}\Biggr ){}\,{\one } \nonumber \\ &&
+{}\Biggl ({12}\,{\Licx}+{10}\,{\Libx}-{12}\,{\zeta_3}\Biggr ){}\,{\tou } 
+{5}\,{\Ly}\,{\ssott }  
+{}\Biggl ({10}\,{\Licx}-{10}\,{\zeta_3}\Biggr ){}\,{\ttouu } \nonumber \\ &&
+{}\Biggl ({\Licy}+{3}\,{\Libx}-{1\over 2}\,{\Liby}\,{\Ly}-{7\over 12}\,{\Ly^3}+{5\over 2}\,{\Ly^2}+{1\over 6}\,{\Ly}\,{\pi^2}+{713\over 72}\,{\Ly}-{\zeta_3}-{5}\Biggr ){}\,{\sot } 
\Biggr ){}\,{\IPI }\nonumber \\ &&
{}\\
{B^{gg,(2),[2]}_{s+-+-} }&=&{ }\Biggl ({}-{6}\,{\Lidx}+{15}\,{\Lidy}+{20}\,{\Lidz}-{8}\,{\Licx}\,{\Lx}+{14}\,{\Licx}\,{\Ly}+{\Licx}\nonumber \\ &&-{12}\,{\Licy}\,{\Lx}+{5}\,{\Licy}\,{\Ly}+{7}\,{\Licy}+{\Libx}\,{\Lx^2}-{\Libx}\,{\Lx}+{7\over 3}\,{\Libx}\,{\pi^2}\nonumber \\ &&+{6}\,{\Liby}\,{\Lx}\,{\Ly}-{5\over 2}\,{\Liby}\,{\Ly^2}-{7}\,{\Liby}\,{\Ly}+{3\over 4}\,{\Lx^4}-{3}\,{\Lx^3}\,{\Ly}+{1\over 6}\,{\Lx^3}+{11\over 2}\,{\Lx^2}\,{\Ly^2}\nonumber \\ &&+{1\over 3}\,{\Lx^2}\,{\Ly}+{5\over 12}\,{\Lx^2}\,{\pi^2}-{41\over 18}\,{\Lx^2}-{5\over 6}\,{\Lx}\,{\Ly^3}-{31\over 6}\,{\Lx}\,{\Ly^2}-{4\over 3}\,{\Lx}\,{\Ly}\,{\pi^2}+{19\over 18}\,{\Lx}\,{\Ly}+{12}\,{\Lx}\,{\zeta_3}\nonumber \\ &&-{5\over 6}\,{\Lx}\,{\pi^2}+{3\over 2}\,{\Lx}-{1\over 24}\,{\Ly^4}+{5\over 3}\,{\Ly^3}+{5\over 6}\,{\Ly^2}\,{\pi^2}-{145\over 36}\,{\Ly^2}-{14}\,{\Ly}\,{\zeta_3}-{5\over 6}\,{\Ly}\,{\pi^2}+{71\over 12}\,{\Ly}\nonumber \\ &&+{731\over 72}\,{\zeta_3}-{383\over 1440}\,{\pi^4}+{343\over 144}\,{\pi^2}-{18815\over 1296}\Biggr ){}\,{\one } \nonumber \\ &&
+{}\Biggl ({3}\,{\Lidx}+{27}\,{\Lidy}+{27}\,{\Lidz}-{18}\,{\Licx}\,{\Lx}+{21}\,{\Licx}\,{\Ly}-{5\over 2}\,{\Licx}\nonumber \\ &&-{12}\,{\Licy}\,{\Lx}+{7}\,{\Licy}+{3}\,{\Libx}\,{\Lx^2}+{5\over 2}\,{\Libx}\,{\Lx}+{7\over 2}\,{\Libx}\,{\pi^2}+{6}\,{\Liby}\,{\Lx}\,{\Ly}\nonumber \\ &&-{7}\,{\Liby}\,{\Ly}+{7\over 8}\,{\Lx^4}-{7\over 2}\,{\Lx^3}\,{\Ly}-{1\over 2}\,{\Lx^3}+{27\over 4}\,{\Lx^2}\,{\Ly^2}+{9\over 4}\,{\Lx^2}\,{\Ly}-{\Lx^2}\,{\pi^2}-{13\over 8}\,{\Lx^2}\nonumber \\ &&-{7\over 2}\,{\Lx}\,{\Ly^2}-{1\over 2}\,{\Lx}\,{\Ly}\,{\pi^2}-{3}\,{\Lx}\,{\Ly}+{18}\,{\Lx}\,{\zeta_3}-{5\over 12}\,{\Lx}\,{\pi^2}+{3\over 4}\,{\Lx}-{21}\,{\Ly}\,{\zeta_3}-{5\over 6}\,{\Ly}\,{\pi^2}\nonumber \\ &&+{15\over 2}\,{\zeta_3}-{37\over 60}\,{\pi^4}+{5\over 4}\,{\pi^2}+{1\over 4}\Biggr ){}\,{\tou } \nonumber \\ &&
+{}\Biggl ({5}\,{\Lidx}+{7}\,{\Lidy}+{7}\,{\Lidz}-{6}\,{\Licx}\,{\Lx}+{7}\,{\Licx}\,{\Ly}-{7\over 2}\,{\Licx}+{7\over 2}\,{\Libx}\,{\Lx}\nonumber \\ &&+{5\over 6}\,{\Libx}\,{\pi^2}+{1\over 24}\,{\Lx^4}-{7\over 6}\,{\Lx^3}\,{\Ly}+{\Lx^3}+{7\over 4}\,{\Lx^2}\,{\Ly^2}+{1\over 4}\,{\Lx^2}\,{\Ly}-{5\over 12}\,{\Lx^2}\,{\pi^2}-{11\over 8}\,{\Lx^2}\nonumber \\ &&-{7\over 6}\,{\Lx}\,{\Ly}\,{\pi^2}+{6}\,{\Lx}\,{\zeta_3}+{5\over 12}\,{\Lx}\,{\pi^2}-{7}\,{\Ly}\,{\zeta_3}+{7\over 2}\,{\zeta_3}-{7\over 36}\,{\pi^4}\Biggr ){}\,{\ttouu } \nonumber \\ &&
+{}\Biggl ({3\over 8}\,{\Lx^2}-{3\over 4}\,{\Lx}\,{\Ly}+{3\over 8}\,{\Ly^2}+{3\over 8}\,{\pi^2}\Biggr ){}\,{\tos } \nonumber \\ &&
+{}\Biggl ({3}\,{\Lidx}-{7}\,{\Lidy}+{\Lidz}-{2}\,{\Licx}\,{\Ly}+{6}\,{\Licx}+{\Licy}\,{\Lx}+{4}\,{\Licy}\,{\Ly}\nonumber \\ &&+{3}\,{\Licy}-{2}\,{\Libx}\,{\Lx^2}-{6}\,{\Libx}\,{\Lx}-{1\over 6}\,{\Libx}\,{\pi^2}-{4}\,{\Liby}\,{\Lx}\,{\Ly}-{3}\,{\Liby}\,{\Ly}\nonumber \\ &&+{1\over 24}\,{\Lx^4}-{5\over 3}\,{\Lx^3}\,{\Ly}-{7\over 4}\,{\Lx^2}\,{\Ly^2}+{1\over 3}\,{\Lx^2}\,{\Ly}+{4\over 3}\,{\Lx^2}\,{\pi^2}+{5\over 6}\,{\Lx}\,{\Ly^3}-{5}\,{\Lx}\,{\Ly^2}-{5\over 6}\,{\Lx}\,{\Ly}\,{\pi^2}\nonumber \\ &&+{25\over 9}\,{\Lx}\,{\Ly}-{\Lx}\,{\zeta_3}-{5\over 3}\,{\Lx}\,{\pi^2}-{1\over 6}\,{\Ly^4}+{28\over 9}\,{\Ly^3}+{1\over 6}\,{\Ly^2}\,{\pi^2}-{199\over 72}\,{\Ly^2}\nonumber \\ &&+{2}\,{\Ly}\,{\zeta_3}+{35\over 18}\,{\Ly}\,{\pi^2}-{2}\,{\Ly}-{3}\,{\zeta_3}+{7\over 72}\,{\pi^4}\Biggr ){}\,{\sot } 
+{\Ly^2}\,{\ssott } \nonumber \\ &&
+{}\Biggl ({}\Biggl ({6}\,{\Licx}-{7}\,{\Licy}-{4}\,{\Libx}\,{\Lx}+{6}\,{\Libx}+{\Liby}\,{\Ly}+{2\over 3}\,{\Lx^3}-{\Lx^2}\,{\Ly}+{4\over 3}\,{\Lx^2}\nonumber \\ &&-{3\over 2}\,{\Lx}\,{\Ly^2}-{8\over 3}\,{\Lx}\,{\Ly}+{1\over 2}\,{\Lx}\,{\pi^2}-{7\over 2}\,{\Lx}-{1\over 6}\,{\Ly^3}+{10\over 3}\,{\Ly^2}-{7}\,{\Ly}-{2}\,{\zeta_3}-{5\over 6}\,{\pi^2}-{29\over 4}\Biggr ){}\,{\one } \nonumber \\ &&
+{}\Biggl ({3}\,{\Licx}-{12}\,{\Licy}+{19\over 2}\,{\Libx}+{6}\,{\Liby}\,{\Ly}+{\Lx^3}-{1\over 2}\,{\Lx^2}+{9\over 2}\,{\Lx}\,{\Ly}+{1\over 2}\,{\Lx}\,{\pi^2}\nonumber \\ &&-{25\over 4}\,{\Lx}-{3}\,{\Ly}-{3}\,{\zeta_3}-{17\over 12}\,{\pi^2}+{3\over 4}\Biggr ){}\,{\tou } \nonumber \\ &&
+{}\Biggl ({\Licx}+{7\over 2}\,{\Libx}+{3\over 2}\,{\Lx^2}+{1\over 2}\,{\Lx}\,{\Ly}-{11\over 4}\,{\Lx}-{\zeta_3}-{7\over 12}\,{\pi^2}\Biggr ){}\,{\ttouu } \nonumber \\ &&
+{}\Biggl ({}-{2}\,{\Licx}+{5}\,{\Licy}-{3}\,{\Libx}-{4}\,{\Liby}\,{\Ly}-{1\over 6}\,{\Lx^3}-{3}\,{\Lx^2}\,{\Ly}+{10\over 3}\,{\Lx^2}+{2}\,{\Lx}\,{\Ly^2}\nonumber \\ &&-{19\over 3}\,{\Lx}\,{\Ly}+{5\over 6}\,{\Lx}\,{\pi^2}+{25\over 9}\,{\Lx}-{1\over 2}\,{\Ly^3}+{4}\,{\Ly^2}-{2\over 3}\,{\Ly}\,{\pi^2}-{11\over 4}\,{\Ly}+{\zeta_3}+{2}\,{\pi^2}-{2}\Biggr ){}\,{\sot } \nonumber \\ &&
+{2}\,{\Ly}\,{\ssott } 
\Biggr ){}\,{\IPI }\nonumber \\ &&
{}\\
{C^{gg,(2),[2]}_{s+-+-} }&=&{ }\Biggl ({}-{2}\,{\Lidx}+{2}\,{\Lidy}+{2}\,{\Lidz}+{2}\,{\Licx}\,{\Ly}+{\Licx}+{\Licy}-{\Libx}\,{\Lx}\nonumber \\ &&-{1\over 3}\,{\Libx}\,{\pi^2}-{\Liby}\,{\Ly}+{1\over 12}\,{\Lx^4}-{1\over 3}\,{\Lx^3}\,{\Ly}+{1\over 2}\,{\Lx^2}\,{\Ly^2}-{1\over 2}\,{\Lx^2}\,{\Ly}+{1\over 6}\,{\Lx^2}\,{\pi^2}+{1\over 2}\,{\Lx^2}\nonumber \\ &&-{1\over 2}\,{\Lx}\,{\Ly^2}-{1\over 3}\,{\Lx}\,{\Ly}\,{\pi^2}-{3\over 2}\,{\Lx}\,{\Ly}+{1\over 6}\,{\Lx}\,{\pi^2}+{3\over 2}\,{\Lx}+{3\over 4}\,{\Ly^2}-{2}\,{\Ly}\,{\zeta_3}+{1\over 6}\,{\Ly}\,{\pi^2}-{3\over 4}\,{\Ly}\nonumber \\ &&+{7\over 4}\,{\zeta_3}+{13\over 120}\,{\pi^4}-{5\over 48}\,{\pi^2}-{247\over 64}\Biggr ){}\,{\one } \nonumber \\ &&
+{}\Biggl ({}-{3}\,{\Lidx}+{3}\,{\Lidy}+{3}\,{\Lidz}+{3}\,{\Licx}\,{\Ly}+{1\over 2}\,{\Licx}+{\Licy}-{1\over 2}\,{\Libx}\,{\Lx}\nonumber \\ &&-{1\over 2}\,{\Libx}\,{\pi^2}-{\Liby}\,{\Ly}+{1\over 8}\,{\Lx^4}-{1\over 2}\,{\Lx^3}\,{\Ly}+{3\over 4}\,{\Lx^2}\,{\Ly^2}-{1\over 4}\,{\Lx^2}\,{\Ly}+{1\over 4}\,{\Lx^2}\,{\pi^2}-{1\over 8}\,{\Lx^2}\nonumber \\ &&-{1\over 2}\,{\Lx}\,{\Ly^2}-{1\over 2}\,{\Lx}\,{\Ly}\,{\pi^2}+{1\over 12}\,{\Lx}\,{\pi^2}+{3\over 4}\,{\Lx}-{3}\,{\Ly}\,{\zeta_3}+{1\over 6}\,{\Ly}\,{\pi^2}-{3\over 2}\,{\zeta_3}+{7\over 60}\,{\pi^4}-{1\over 4}\,{\pi^2}+{1\over 8}\Biggr ){}\,{\tou } \nonumber \\ &&
+{}\Biggl ({}-{\Lidx}+{\Lidy}+{\Lidz}+{\Licx}\,{\Ly}-{1\over 2}\,{\Licx}+{1\over 2}\,{\Libx}\,{\Lx}-{1\over 6}\,{\Libx}\,{\pi^2}\nonumber \\ &&+{1\over 24}\,{\Lx^4}-{1\over 6}\,{\Lx^3}\,{\Ly}+{1\over 4}\,{\Lx^2}\,{\Ly^2}+{1\over 4}\,{\Lx^2}\,{\Ly}+{1\over 12}\,{\Lx^2}\,{\pi^2}+{1\over 8}\,{\Lx^2}-{1\over 6}\,{\Lx}\,{\Ly}\,{\pi^2}-{1\over 12}\,{\Lx}\,{\pi^2}\nonumber \\ &&-{\Ly}\,{\zeta_3}+{1\over 2}\,{\zeta_3}+{7\over 180}\,{\pi^4}\Biggr ){}\,{\ttouu } \nonumber \\ &&
+{}\Biggl ({3\over 8}\,{\Lx^2}-{3\over 4}\,{\Lx}\,{\Ly}+{3\over 8}\,{\Ly^2}+{3\over 8}\,{\pi^2}\Biggr ){}\,{\tos } \nonumber \\ &&
+{}\Biggl ({}-{\Lidx}+{5\over 2}\,{\Lidy}+{\Lidz}-{\Licy}\,{\Lx}-{\Licy}\,{\Ly}-{3}\,{\Licy}-{1\over 6}\,{\Libx}\,{\pi^2}\nonumber \\ &&+{1\over 4}\,{\Liby}\,{\Ly^2}+{3}\,{\Liby}\,{\Ly}+{1\over 24}\,{\Lx^4}-{1\over 6}\,{\Lx^3}\,{\Ly}+{1\over 12}\,{\Lx^2}\,{\pi^2}+{3\over 2}\,{\Lx}\,{\Ly^2}-{1\over 6}\,{\Lx}\,{\Ly}\,{\pi^2}\nonumber \\ &&+{\Lx}\,{\zeta_3}+{1\over 16}\,{\Ly^4}-{1\over 4}\,{\Ly^2}\,{\pi^2}-{15\over 16}\,{\Ly^2}+{\Ly}\,{\zeta_3}-{1\over 2}\,{\Ly}\,{\pi^2}+{\Ly}+{3}\,{\zeta_3}-{1\over 120}\,{\pi^4}\Biggr ){}\,{\sot } \nonumber \\ &&
-{1\over 2}\,{\Ly^2}\,{\ssott } 
+{}\Biggl ({}\Biggl ({2}\,{\Licx}-{\Lx}\,{\Ly}-{1\over 2}\,{\Lx}-{2}\,{\zeta_3}+{1\over 6}\,{\pi^2}+{3\over 4}\Biggr ){}\,{\one } \nonumber \\ &&
+{}\Biggl ({3}\,{\Licx}+{1\over 2}\,{\Libx}-{1\over 2}\,{\Lx}\,{\Ly}-{1\over 4}\,{\Lx}-{3}\,{\zeta_3}+{1\over 12}\,{\pi^2}+{3\over 4}\Biggr ){}\,{\tou } \nonumber \\ &&
+{}\Biggl ({\Licx}+{1\over 2}\,{\Libx}+{1\over 2}\,{\Lx}\,{\Ly}+{1\over 4}\,{\Lx}-{\zeta_3}-{1\over 12}\,{\pi^2}\Biggr ){}\,{\ttouu } \nonumber \\ &&
+{}\Biggl ({}-{2}\,{\Licy}-{3}\,{\Libx}+{1\over 2}\,{\Liby}\,{\Ly}+{1\over 4}\,{\Ly^3}-{15\over 8}\,{\Ly}+{2}\,{\zeta_3}+{1}\Biggr ){}\,{\sot } \nonumber \\ &&
-{\Ly}\,{\ssott } 
\Biggr ){}\,{\IPI }\nonumber \\ &&
{}\\
{D^{gg,(2),[2]}_{s+-+-} }&=&{ }\Biggl ({2}\,{\Lidx}-{2}\,{\Lidy}-{2}\,{\Lidz}-{2}\,{\Licx}\,{\Ly}-{14\over 3}\,{\Licy}+{1\over 3}\,{\Libx}\,{\pi^2}+{14\over 3}\,{\Liby}\,{\Ly}\nonumber \\ &&-{1\over 12}\,{\Lx^4}+{1\over 3}\,{\Lx^3}\,{\Ly}-{1\over 2}\,{\Lx^2}\,{\Ly^2}-{1\over 6}\,{\Lx^2}\,{\pi^2}+{7\over 3}\,{\Lx}\,{\Ly^2}+{1\over 3}\,{\Lx}\,{\Ly}\,{\pi^2}-{7\over 18}\,{\Ly^3}+{71\over 36}\,{\Ly^2}\nonumber \\ &&+{2}\,{\Ly}\,{\zeta_3}-{47\over 36}\,{\Ly}\,{\pi^2}-{14\over 9}\,{\Ly}+{65\over 12}\,{\zeta_3}-{7\over 90}\,{\pi^4}+{10\over 27}\,{\pi^2}-{863\over 432}\Biggr ){}\,{\one } \nonumber \\ &&
+{}\Biggl ({6}\,{\Lidx}-{6}\,{\Lidy}-{6}\,{\Lidz}-{6}\,{\Licx}\,{\Ly}-{4}\,{\Licy}+{\Libx}\,{\pi^2}+{4}\,{\Liby}\,{\Ly}\nonumber \\ &&-{1\over 4}\,{\Lx^4}+{\Lx^3}\,{\Ly}-{3\over 2}\,{\Lx^2}\,{\Ly^2}-{1\over 2}\,{\Lx^2}\,{\pi^2}+{2}\,{\Lx}\,{\Ly^2}+{\Lx}\,{\Ly}\,{\pi^2}+{6}\,{\Ly}\,{\zeta_3}-{2\over 3}\,{\Ly}\,{\pi^2}+{4}\,{\zeta_3}\nonumber \\ &&-{7\over 30}\,{\pi^4}+{2\over 3}\,{\pi^2}\Biggr ){}\,{\tou } \nonumber \\ &&
+{}\Biggl ({4}\,{\Lidx}-{4}\,{\Lidy}-{4}\,{\Lidz}-{4}\,{\Licx}\,{\Ly}+{2\over 3}\,{\Libx}\,{\pi^2}-{1\over 6}\,{\Lx^4}+{2\over 3}\,{\Lx^3}\,{\Ly}\nonumber \\ &&-{\Lx^2}\,{\Ly^2}-{1\over 3}\,{\Lx^2}\,{\pi^2}+{2\over 3}\,{\Lx}\,{\Ly}\,{\pi^2}+{4}\,{\Ly}\,{\zeta_3}-{7\over 45}\,{\pi^4}\Biggr ){}\,{\ttouu } \nonumber \\ &&
+{}\Biggl ({}-{7\over 36}\,{\Ly^3}-{71\over 36}\,{\Ly^2}-{5\over 18}\,{\Ly}\,{\pi^2}+{2}\,{\Ly}\Biggr ){}\,{\sot } \nonumber \\ &&
-{\Ly^2}\,{\ssott } \nonumber \\ &&
+{}\Biggl ({}\Biggl ({}-{2}\,{\Licx}-{14\over 3}\,{\Libx}-{1\over 2}\,{\Ly^2}+{1\over 2}\,{\Ly}+{2}\,{\zeta_3}+{1\over 36}\,{\pi^2}-{25\over 18}\Biggr ){}\,{\one } \nonumber \\ &&
+{}\Biggl ({}-{6}\,{\Licx}-{4}\,{\Libx}+{6}\,{\zeta_3}\Biggr ){}\,{\tou } 
+{}\Biggl ({}-{4}\,{\Licx}+{4}\,{\zeta_3}\Biggr ){}\,{\ttouu } \nonumber \\ &&
+{}\Biggl ({}-{1\over 4}\,{\Ly^2}-{71\over 18}\,{\Ly}+{2}\Biggr ){}\,{\sot } 
-{2}\,{\Ly}\,{\ssott } 
\Biggr ){}\,{\IPI }\nonumber \\ &&
{}\\
{E^{gg,(2),[2]}_{s+-+-} }&=&{ }\Biggl ({}-{4}\,{\Lidx}+{4}\,{\Lidy}+{4}\,{\Lidz}+{4}\,{\Licx}\,{\Ly}+{26\over 3}\,{\Licy}-{2\over 3}\,{\Libx}\,{\pi^2}-{26\over 3}\,{\Liby}\,{\Ly}\nonumber \\ &&+{1\over 6}\,{\Lx^4}-{2\over 3}\,{\Lx^3}\,{\Ly}+{1\over 12}\,{\Lx^3}+{\Lx^2}\,{\Ly^2}-{1\over 12}\,{\Lx^2}\,{\Ly}+{1\over 3}\,{\Lx^2}\,{\pi^2}+{14\over 9}\,{\Lx^2}-{23\over 6}\,{\Lx}\,{\Ly^2}\nonumber \\ &&-{2\over 3}\,{\Lx}\,{\Ly}\,{\pi^2}-{7\over 9}\,{\Lx}\,{\Ly}+{3}\,{\Lx}-{1\over 6}\,{\Ly^3}+{1\over 18}\,{\Ly^2}-{4}\,{\Ly}\,{\zeta_3}+{7\over 9}\,{\Ly}\,{\pi^2}-{2}\,{\Ly}-{313\over 36}\,{\zeta_3}\nonumber \\ &&+{7\over 45}\,{\pi^4}-{85\over 72}\,{\pi^2}+{4085\over 1296}\Biggr ){}\,{\one } \nonumber \\ &&
+{}\Biggl ({}-{12}\,{\Lidx}+{12}\,{\Lidy}+{12}\,{\Lidz}+{12}\,{\Licx}\,{\Ly}+{8}\,{\Licy}-{2}\,{\Libx}\,{\pi^2}\nonumber \\ &&-{8}\,{\Liby}\,{\Ly}+{1\over 2}\,{\Lx^4}-{2}\,{\Lx^3}\,{\Ly}+{1\over 4}\,{\Lx^3}+{3}\,{\Lx^2}\,{\Ly^2}-{1\over 4}\,{\Lx^2}\,{\Ly}+{\Lx^2}\,{\pi^2}+{2}\,{\Lx^2}-{4}\,{\Lx}\,{\Ly^2}\nonumber \\ &&-{2}\,{\Lx}\,{\Ly}\,{\pi^2}-{2\over 3}\,{\Lx}\,{\Ly}+{5\over 3}\,{\Lx}-{12}\,{\Ly}\,{\zeta_3}+{4\over 3}\,{\Ly}\,{\pi^2}-{8}\,{\zeta_3}+{7\over 15}\,{\pi^4}-{7\over 9}\,{\pi^2}\Biggr ){}\,{\tou } \nonumber \\ &&
+{}\Biggl ({}-{8}\,{\Lidx}+{8}\,{\Lidy}+{8}\,{\Lidz}+{8}\,{\Licx}\,{\Ly}-{2\over 3}\,{\Licx}+{2\over 3}\,{\Libx}\,{\Lx}\nonumber \\ &&-{4\over 3}\,{\Libx}\,{\pi^2}+{1\over 3}\,{\Lx^4}-{4\over 3}\,{\Lx^3}\,{\Ly}+{2}\,{\Lx^2}\,{\Ly^2}+{2\over 3}\,{\Lx^2}\,{\pi^2}+{1\over 2}\,{\Lx^2}-{4\over 3}\,{\Lx}\,{\Ly}\,{\pi^2}+{5\over 9}\,{\Lx}\,{\pi^2}\nonumber \\ &&-{8}\,{\Ly}\,{\zeta_3}+{2\over 3}\,{\zeta_3}+{14\over 45}\,{\pi^4}\Biggr ){}\,{\ttouu } \nonumber \\ &&
+{}\Biggl ({2\over 3}\,{\Licy}-{2\over 3}\,{\Liby}\,{\Ly}-{1\over 3}\,{\Lx^2}\,{\Ly}+{1\over 2}\,{\Lx}\,{\Ly^2}-{5\over 9}\,{\Lx}\,{\Ly}+{1\over 6}\,{\Lx}\,{\pi^2}-{13\over 36}\,{\Ly^3}\nonumber \\ &&+{49\over 12}\,{\Ly^2}-{\Ly}\,{\pi^2}-{3}\,{\Ly}-{2\over 3}\,{\zeta_3}\Biggr ){}\,{\sot } \nonumber \\ &&
+{3\over 2}\,{\Ly^2}\,{\ssott } \nonumber \\ &&
+{}\Biggl ({}\Biggl ({4}\,{\Licx}+{26\over 3}\,{\Libx}+{1\over 6}\,{\Lx^2}+{5\over 6}\,{\Lx}\,{\Ly}+{7\over 3}\,{\Lx}-{2\over 3}\,{\Ly}-{4}\,{\zeta_3}+{19\over 6}\Biggr ){}\,{\one } \nonumber \\ &&
+{}\Biggl ({12}\,{\Licx}+{8}\,{\Libx}+{1\over 2}\,{\Lx^2}-{1\over 2}\,{\Lx}\,{\Ly}+{10\over 3}\,{\Lx}-{2\over 3}\,{\Ly}-{12}\,{\zeta_3}+{5\over 3}\Biggr ){}\,{\tou } \nonumber \\ &&
+{}\Biggl ({8}\,{\Licx}+{2\over 3}\,{\Libx}-{1\over 3}\,{\Lx^2}+{\Lx}-{8}\,{\zeta_3}-{1\over 9}\,{\pi^2}\Biggr ){}\,{\ttouu } \nonumber \\ &&
+{}\Biggl ({2\over 3}\,{\Libx}-{1\over 3}\,{\Lx^2}+{\Lx}\,{\Ly}-{5\over 9}\,{\Lx}+{1\over 12}\,{\Ly^2}+{137\over 18}\,{\Ly}-{3}\Biggr ){}\,{\sot } 
+{3}\,{\Ly}\,{\ssott } 
\Biggr ){}\,{\IPI }\nonumber \\ &&
{}\\
{F^{gg,(2),[2]}_{s+-+-} }&=&{ }\Biggl ({1\over 18}\,{\Ly}+{5\over 27}\Biggr ){}\,{\IPI }
+{}\Biggl ({}-{1\over 12}\,{\Ly^2}+{5\over 27}\,{\Ly}+{2\over 27}\,{\pi^2}\Biggr ){}\,{\one } \nonumber \\ &&
\\
{G^{gg,(2),[2]}_{s+-+-} }&=&{ }\Biggl ({}-{1\over 36}\,{\Lx^2}+{1\over 18}\,{\Lx}\,{\Ly}-{1\over 36}\,{\Ly^2}\Biggr ){}\,{\one }  
+{}\Biggl ({1\over 36}\,{\Lx}-{1\over 36}\,{\Ly}\Biggr ){}\,{\IPI }
{}\,{\sot }  
+{}\Biggl ({1\over 36}\,{\Lx}\,{\Ly}-{1\over 36}\,{\Ly^2}\Biggr ){}\,{\sot } \nonumber \\ &&
\end{eqnarray}
\begin{eqnarray}
{H^{gg,(2),[3]}_{s+-+-} }&=&{ }
\Biggl ({}-{4}\,{\Lidx}-{13\over 2}\,{\Lidy}+{4}\,{\Lidz}+{\Licx}\,{\Lx}+{7\over 2}\,{\Licx}\,{\Ly}-{13\over 6}\,{\Licx}-{5\over 2}\,{\Licy}\,{\Lx}\nonumber \\ &&+{15\over 2}\,{\Licy}\,{\Ly}+{9\over 4}\,{\Licy}-{\Libx}\,{\Lx^2}+{13\over 6}\,{\Libx}\,{\Lx}+{11\over 6}\,{\Libx}\,{\pi^2}+{2}\,{\Liby}\,{\Lx}\,{\Ly}\nonumber \\ &&-{9\over 4}\,{\Liby}\,{\Ly^2}-{9\over 4}\,{\Liby}\,{\Ly}+{1\over 6}\,{\Lx^4}-{\Lx^3}\,{\Ly}+{7\over 4}\,{\Lx^2}\,{\Ly^2}+{4\over 3}\,{\Lx^2}\,{\Ly}-{1\over 6}\,{\Lx^2}\,{\pi^2}-{1\over 4}\,{\Lx}\,{\Ly^3}\nonumber \\ &&-{23\over 12}\,{\Lx}\,{\Ly^2}+{11\over 6}\,{\Lx}\,{\Ly}+{5\over 2}\,{\Lx}\,{\zeta_3}-{91\over 24}\,{\Lx}\,{\pi^2}-{1\over 16}\,{\Ly^4}+{11\over 18}\,{\Ly^3}-{1\over 4}\,{\Ly^2}\,{\pi^2}+{161\over 72}\,{\Ly^2}-{3}\,{\Ly}\,{\zeta_3}\nonumber \\ &&+{43\over 48}\,{\Ly}\,{\pi^2}-{211\over 108}\,{\Ly}-{9\over 4}\,{\zeta_3}+{43\over 120}\,{\pi^4}-{391\over 72}\,{\pi^2}\Biggr ){}\,{\one } \nonumber \\ &&
+{}\Biggl ({25\over 4}\,{\Lidx}-{7\over 2}\,{\Licx}\,{\Lx}-{\Licx}\,{\Ly}-{13\over 3}\,{\Licx}-{2}\,{\Licy}\,{\Lx}-{29\over 2}\,{\Licy}\nonumber \\ &&+{3\over 8}\,{\Libx}\,{\Lx^2}+{13\over 3}\,{\Libx}\,{\Lx}+{5\over 3}\,{\Libx}\,{\pi^2}+{\Liby}\,{\Lx}\,{\Ly}+{29\over 2}\,{\Liby}\,{\Ly}+{3\over 32}\,{\Lx^4}\nonumber \\ &&+{1\over 6}\,{\Lx^3}\,{\Ly}-{11\over 12}\,{\Lx^3}+{7\over 6}\,{\Lx^2}\,{\Ly}-{13\over 8}\,{\Lx^2}\,{\pi^2}+{25\over 24}\,{\Lx^2}+{29\over 4}\,{\Lx}\,{\Ly^2}+{2\over 3}\,{\Lx}\,{\Ly}\,{\pi^2}+{17\over 4}\,{\Lx}\,{\Ly}\nonumber \\ &&+{7\over 2}\,{\Lx}\,{\zeta_3}-{23\over 9}\,{\Lx}\,{\pi^2}-{5\over 6}\,{\Lx}+{\Ly}\,{\zeta_3}-{25\over 12}\,{\Ly}\,{\pi^2}+{46\over 3}\,{\zeta_3}-{25\over 72}\,{\pi^4}-{77\over 72}\,{\pi^2}\Biggr ){}\,{\tou } \nonumber \\ &&
+{}\Biggl ({11}\,{\Lidx}-{33\over 2}\,{\Lidy}-{33\over 2}\,{\Lidz}+{7\over 2}\,{\Licx}\,{\Lx}-{31\over 2}\,{\Licx}\,{\Ly}+{31\over 12}\,{\Licx}\nonumber \\ &&+{2}\,{\Licy}\,{\Lx}-{3\over 4}\,{\Libx}\,{\Lx^2}-{31\over 12}\,{\Libx}\,{\Lx}+{19\over 12}\,{\Libx}\,{\pi^2}-{\Liby}\,{\Lx}\,{\Ly}-{11\over 16}\,{\Lx^4}\nonumber \\ &&+{31\over 12}\,{\Lx^3}\,{\Ly}-{1\over 12}\,{\Lx^3}-{33\over 8}\,{\Lx^2}\,{\Ly^2}+{1\over 3}\,{\Lx^2}\,{\Ly}-{7\over 8}\,{\Lx^2}\,{\pi^2}+{15\over 8}\,{\Lx^2}+{25\over 12}\,{\Lx}\,{\Ly}\,{\pi^2}\nonumber \\ &&-{7\over 2}\,{\Lx}\,{\zeta_3}-{191\over 72}\,{\Lx}\,{\pi^2}+{31\over 2}\,{\Ly}\,{\zeta_3}-{31\over 12}\,{\zeta_3}-{139\over 360}\,{\pi^4}\Biggr ){}\,{\ttouu } \nonumber \\ &&
+{}\Biggl ({5\over 2}\,{\Lidx}-{5\over 2}\,{\Lidy}+{8}\,{\Lidz}-{13\over 2}\,{\Licx}\,{\Lx}+{6}\,{\Licx}\,{\Ly}-{31\over 6}\,{\Licx}\nonumber \\ &&-{6}\,{\Licy}\,{\Lx}+{13\over 2}\,{\Licy}\,{\Ly}+{31\over 6}\,{\Licy}+{5\over 4}\,{\Libx}\,{\Lx^2}+{31\over 6}\,{\Libx}\,{\Lx}+{11\over 3}\,{\Libx}\,{\pi^2}\nonumber \\ &&+{4}\,{\Liby}\,{\Lx}\,{\Ly}-{5\over 4}\,{\Liby}\,{\Ly^2}-{31\over 6}\,{\Liby}\,{\Ly}+{19\over 48}\,{\Lx^4}-{17\over 12}\,{\Lx^3}\,{\Ly}-{11\over 18}\,{\Lx^3}+{3}\,{\Lx^2}\,{\Ly^2}\nonumber \\ &&+{7\over 4}\,{\Lx^2}\,{\Ly}+{5\over 12}\,{\Lx^2}\,{\pi^2}+{43\over 18}\,{\Lx^2}+{1\over 12}\,{\Lx}\,{\Ly^3}-{7\over 4}\,{\Lx}\,{\Ly^2}+{5\over 6}\,{\Lx}\,{\Ly}\,{\pi^2}+{1\over 2}\,{\Lx}\,{\Ly}+{11\over 2}\,{\Lx}\,{\zeta_3}\nonumber \\ &&-{23\over 16}\,{\Lx}\,{\pi^2}+{19\over 27}\,{\Lx}-{1\over 16}\,{\Ly^4}+{11\over 18}\,{\Ly^3}+{1\over 4}\,{\Ly^2}\,{\pi^2}-{26\over 9}\,{\Ly^2}-{11\over 2}\,{\Ly}\,{\zeta_3}+{23\over 16}\,{\Ly}\,{\pi^2}\nonumber \\ &&-{19\over 27}\,{\Ly}-{41\over 180}\,{\pi^4}-{1\over 4}\,{\pi^2}\Biggr ){}\,{\tos } 
+{3\over 2}\,{\Ly^2}\,{\sot } \nonumber \\ &&
+{}\Biggl ({}\Biggl ({9\over 2}\,{\Licx}+{5}\,{\Licy}-{4}\,{\Libx}\,{\Lx}+{53\over 12}\,{\Libx}-{5\over 2}\,{\Liby}\,{\Ly}+{1\over 6}\,{\Lx^3}-{5\over 4}\,{\Lx^2}\,{\Ly}\nonumber \\ &&+{1\over 4}\,{\Lx^2}+{19\over 4}\,{\Lx}\,{\Ly}-{1\over 2}\,{\Lx}\,{\pi^2}+{11\over 6}\,{\Lx}-{1\over 4}\,{\Ly^3}-{19\over 24}\,{\Ly^2}-{3\over 2}\,{\Ly}\,{\pi^2}+{149\over 12}\,{\Ly}-{1\over 2}\,{\zeta_3}\nonumber \\ &&-{187\over 144}\,{\pi^2}-{211\over 108}\Biggr ){}\,{\one } \nonumber \\ &&
+{}\Biggl ({}-{9\over 2}\,{\Licx}-{2}\,{\Licy}-{1\over 4}\,{\Libx}\,{\Lx}-{61\over 6}\,{\Libx}+{\Liby}\,{\Ly}+{7\over 8}\,{\Lx^3}-{\Lx^2}+{7\over 3}\,{\Lx}\,{\Ly}\nonumber \\ &&-{1\over 3}\,{\Lx}\,{\pi^2}+{19\over 3}\,{\Lx}+{17\over 4}\,{\Ly}+{9\over 2}\,{\zeta_3}-{5\over 9}\,{\pi^2}-{5\over 6}\Biggr ){}\,{\tou } \nonumber \\ &&
+{}\Biggl ({}-{12}\,{\Licx}+{2}\,{\Licy}-{1\over 2}\,{\Libx}\,{\Lx}-{31\over 12}\,{\Libx}-{\Liby}\,{\Ly}-{1\over 4}\,{\Lx^3}+{11\over 8}\,{\Lx^2}\nonumber \\ &&+{2\over 3}\,{\Lx}\,{\Ly}-{1\over 6}\,{\Lx}\,{\pi^2}+{15\over 4}\,{\Lx}+{12}\,{\zeta_3}+{31\over 72}\,{\pi^2}\Biggr ){}\,{\ttouu } \nonumber \\ &&
+{}\Biggl ({}-{1\over 2}\,{\Licx}+{1\over 2}\,{\Licy}-{3\over 2}\,{\Libx}\,{\Lx}+{31\over 3}\,{\Libx}+{3\over 2}\,{\Liby}\,{\Ly}+{5\over 12}\,{\Lx^3}\nonumber \\ &&-{5\over 4}\,{\Lx^2}\,{\Ly}-{5\over 6}\,{\Lx^2}+{5\over 4}\,{\Lx}\,{\Ly^2}+{31\over 6}\,{\Lx}\,{\Ly}+{\Lx}\,{\pi^2}-{5\over 6}\,{\Lx}-{5\over 12}\,{\Ly^3}+{5\over 6}\,{\Ly^2}-{\Ly}\,{\pi^2}\nonumber \\ &&+{5\over 6}\,{\Ly}-{31\over 36}\,{\pi^2}\Biggr ){}\,{\tos } 
+{3}\,{\Ly}\,{\sot } 
\Biggr ){}\,{\IPI }\nonumber \\ &&
{}\\
{I^{gg,(2),[3]}_{s+-+-} }&=&
{ }\Biggl ({}-{2}\,{\Lidx}-{5\over 2}\,{\Lidy}+{\Licx}\,{\Lx}+{1\over 2}\,{\Licx}\,{\Ly}-{3\over 2}\,{\Licx}+{1\over 2}\,{\Licy}\,{\Lx}+{3\over 2}\,{\Licy}\,{\Ly}\nonumber \\ &&+{9\over 4}\,{\Licy}+{3\over 2}\,{\Libx}\,{\Lx}-{1\over 6}\,{\Libx}\,{\pi^2}-{1\over 4}\,{\Liby}\,{\Ly^2}-{9\over 4}\,{\Liby}\,{\Ly}-{1\over 12}\,{\Lx^3}\,{\Ly}\nonumber \\ &&+{3\over 8}\,{\Lx^2}\,{\Ly^2}-{5\over 12}\,{\Lx^2}\,{\Ly}+{5\over 8}\,{\Lx^2}\,{\pi^2}-{1\over 6}\,{\Lx}\,{\Ly^3}-{7\over 6}\,{\Lx}\,{\Ly^2}-{1\over 3}\,{\Lx}\,{\Ly}\,{\pi^2}+{1\over 2}\,{\Lx}\,{\Ly}\nonumber \\ &&-{1\over 2}\,{\Lx}\,{\zeta_3}+{47\over 24}\,{\Lx}\,{\pi^2}+{1\over 48}\,{\Ly^4}+{13\over 24}\,{\Ly^2}\,{\pi^2}-{9\over 8}\,{\Ly^2}-{1\over 2}\,{\Ly}\,{\zeta_3}+{17\over 24}\,{\Ly}\,{\pi^2}+{47\over 12}\,{\Ly}-{9\over 4}\,{\zeta_3}\nonumber \\ &&+{1\over 36}\,{\pi^4}+{3\over 8}\,{\pi^2}\Biggr ){}\,{\one } \nonumber \\ &&
+{}\Biggl ({}-{23\over 4}\,{\Lidx}-{3}\,{\Lidy}-{3}\,{\Lidz}+{11\over 2}\,{\Licx}\,{\Lx}-{\Licx}\,{\Ly}-{4}\,{\Licx}+{4}\,{\Licy}\,{\Lx}\nonumber \\ &&+{1\over 2}\,{\Licy}-{9\over 8}\,{\Libx}\,{\Lx^2}+{4}\,{\Libx}\,{\Lx}-{11\over 6}\,{\Libx}\,{\pi^2}-{2}\,{\Liby}\,{\Lx}\,{\Ly}-{1\over 2}\,{\Liby}\,{\Ly}\nonumber \\ &&-{5\over 32}\,{\Lx^4}+{1\over 6}\,{\Lx^3}\,{\Ly}-{3\over 4}\,{\Lx^2}\,{\Ly^2}-{1\over 12}\,{\Lx^2}\,{\Ly}+{\Lx^2}\,{\pi^2}+{13\over 8}\,{\Lx^2}-{1\over 4}\,{\Lx}\,{\Ly^2}-{5\over 6}\,{\Lx}\,{\Ly}\,{\pi^2}\nonumber \\ &&-{7\over 12}\,{\Lx}\,{\Ly}-{11\over 2}\,{\Lx}\,{\zeta_3}+{10\over 3}\,{\Lx}\,{\pi^2}+{1\over 2}\,{\Lx}+{\Ly}\,{\zeta_3}+{5\over 12}\,{\Ly}\,{\pi^2}+{133\over 360}\,{\pi^4}+{11\over 24}\,{\pi^2}\Biggr ){}\,{\tou } \nonumber \\ &&
+{}\Biggl ({}-{4}\,{\Lidx}-{3\over 2}\,{\Lidy}-{3\over 2}\,{\Lidz}+{7\over 2}\,{\Licx}\,{\Lx}-{1\over 2}\,{\Licx}\,{\Ly}-{9\over 4}\,{\Licx}+{2}\,{\Licy}\,{\Lx}\nonumber \\ &&-{3\over 4}\,{\Libx}\,{\Lx^2}+{9\over 4}\,{\Libx}\,{\Lx}-{11\over 12}\,{\Libx}\,{\pi^2}-{\Liby}\,{\Lx}\,{\Ly}-{1\over 16}\,{\Lx^4}+{1\over 12}\,{\Lx^3}\,{\Ly}-{1\over 12}\,{\Lx^3}\nonumber \\ &&-{3\over 8}\,{\Lx^2}\,{\Ly^2}+{1\over 3}\,{\Lx^2}\,{\Ly}+{3\over 8}\,{\Lx^2}\,{\pi^2}+{1\over 8}\,{\Lx^2}-{5\over 12}\,{\Lx}\,{\Ly}\,{\pi^2}-{7\over 2}\,{\Lx}\,{\zeta_3}+{11\over 8}\,{\Lx}\,{\pi^2}\nonumber \\ &&+{1\over 2}\,{\Ly}\,{\zeta_3}+{9\over 4}\,{\zeta_3}+{71\over 360}\,{\pi^4}\Biggr ){}\,{\ttouu } \nonumber \\ &&
+{}\Biggl ({1\over 2}\,{\Lidx}-{1\over 2}\,{\Lidy}-{1\over 2}\,{\Licx}\,{\Lx}+{1\over 2}\,{\Licy}\,{\Ly}+{1\over 4}\,{\Libx}\,{\Lx^2}-{1\over 3}\,{\Libx}\,{\pi^2}\nonumber \\ &&-{1\over 4}\,{\Liby}\,{\Ly^2}-{1\over 48}\,{\Lx^4}+{1\over 12}\,{\Lx^3}\,{\Ly}+{1\over 12}\,{\Lx^2}\,{\pi^2}+{1\over 2}\,{\Lx^2}-{1\over 12}\,{\Lx}\,{\Ly^3}-{1\over 6}\,{\Lx}\,{\Ly}\,{\pi^2}\nonumber \\ &&+{\Lx}\,{\Ly}-{23\over 6}\,{\Lx}+{1\over 48}\,{\Ly^4}-{1\over 12}\,{\Ly^2}\,{\pi^2}-{3\over 2}\,{\Ly^2}+{23\over 6}\,{\Ly}+{1\over 36}\,{\pi^4}-{1\over 2}\,{\pi^2}\Biggr ){}\,{\tos } \nonumber \\ &&
+{1\over 4}\,{\Ly^2}\,{\sot } 
+{}\Biggl ({}-{1\over 4}\,{\Lx^2}+{1\over 2}\,{\Lx}\,{\Ly}-{1\over 4}\,{\Ly^2}-{1\over 4}\,{\pi^2}\Biggr ){}\,{\ttoss } \nonumber \\ &&
+{}\Biggl ({}\Biggl ({3\over 2}\,{\Licx}+{2}\,{\Licy}+{15\over 4}\,{\Libx}-{1\over 2}\,{\Liby}\,{\Ly}-{1\over 4}\,{\Lx^3}+{1\over 4}\,{\Lx^2}\,{\Ly}-{7\over 6}\,{\Lx^2}\nonumber \\ &&-{11\over 12}\,{\Lx}\,{\Ly}+{1\over 4}\,{\Lx}\,{\pi^2}+{1\over 2}\,{\Lx}-{1\over 6}\,{\Ly^3}-{1\over 24}\,{\Ly^2}+{1\over 4}\,{\Ly}\,{\pi^2}-{7\over 4}\,{\Ly}-{\zeta_3}-{1\over 8}\,{\pi^2}+{47\over 12}\Biggr ){}\,{\one } \nonumber \\ &&
+{}\Biggl ({9\over 2}\,{\Licx}+{4}\,{\Licy}-{1\over 4}\,{\Libx}\,{\Lx}+{9\over 2}\,{\Libx}-{2}\,{\Liby}\,{\Ly}-{5\over 8}\,{\Lx^3}-{25\over 12}\,{\Lx^2}-{1\over 6}\,{\Lx}\,{\Ly}\nonumber \\ &&-{1\over 12}\,{\Lx}\,{\pi^2}+{8\over 3}\,{\Lx}-{7\over 12}\,{\Ly}-{9\over 2}\,{\zeta_3}-{1\over 2}\,{\pi^2}+{1\over 2}\Biggr ){}\,{\tou } \nonumber \\ &&
+{}\Biggl ({3}\,{\Licx}+{2}\,{\Licy}-{1\over 2}\,{\Libx}\,{\Lx}+{9\over 4}\,{\Libx}-{\Liby}\,{\Ly}-{1\over 4}\,{\Lx^3}-{25\over 24}\,{\Lx^2}+{2\over 3}\,{\Lx}\,{\Ly}\nonumber \\ &&-{1\over 6}\,{\Lx}\,{\pi^2}+{1\over 4}\,{\Lx}-{3}\,{\zeta_3}-{3\over 8}\,{\pi^2}\Biggr ){}\,{\ttouu } \nonumber \\ &&
+{}\Biggl ({}-{1\over 2}\,{\Licx}+{1\over 2}\,{\Licy}+{1\over 2}\,{\Libx}\,{\Lx}-{1\over 2}\,{\Liby}\,{\Ly}-{1\over 12}\,{\Lx^3}+{1\over 4}\,{\Lx^2}\,{\Ly}\nonumber \\ &&-{1\over 4}\,{\Lx}\,{\Ly^2}+{2}\,{\Lx}+{1\over 12}\,{\Ly^3}-{2}\,{\Ly}\Biggr ){}\,{\tos } 
+{1\over 2}\,{\Ly}\,{\sot } 
\Biggr ){}\,{\IPI }\nonumber \\ &&
{}\\
{J^{gg,(2),[3]}_{s+-+-} }&=&{ }\Biggl ({2\over 3}\,{\Licx}+{1\over 3}\,{\Licy}-{2\over 3}\,{\Libx}\,{\Lx}-{1\over 3}\,{\Liby}\,{\Ly}-{5\over 24}\,{\Lx^2}\,{\Ly}+{1\over 24}\,{\Lx}\,{\Ly^2}\nonumber \\ &&-{29\over 72}\,{\Lx}\,{\Ly}+{5\over 12}\,{\Lx}\,{\pi^2}-{1\over 9}\,{\Ly^3}-{13\over 24}\,{\Ly^2}-{17\over 72}\,{\Ly}\,{\pi^2}+{19\over 72}\,{\Ly}-{1\over 3}\,{\zeta_3}+{17\over 12}\,{\pi^2}\Biggr ){}\,{\one } \nonumber \\ &&
+{}\Biggl ({}-{\Lidx}+{\Lidy}+{\Lidz}+{\Licx}\,{\Ly}+{2}\,{\Licy}-{1\over 6}\,{\Libx}\,{\pi^2}-{2}\,{\Liby}\,{\Ly}\nonumber \\ &&+{1\over 24}\,{\Lx^4}-{1\over 6}\,{\Lx^3}\,{\Ly}+{1\over 6}\,{\Lx^3}+{1\over 4}\,{\Lx^2}\,{\Ly^2}-{1\over 24}\,{\Lx^2}\,{\Ly}+{1\over 12}\,{\Lx^2}\,{\pi^2}-{5\over 12}\,{\Lx^2}-{\Lx}\,{\Ly^2}\nonumber \\ &&-{1\over 6}\,{\Lx}\,{\Ly}\,{\pi^2}+{1\over 12}\,{\Lx}\,{\Ly}+{3\over 4}\,{\Lx}\,{\pi^2}+{1\over 2}\,{\Lx}-{\Ly}\,{\zeta_3}+{1\over 3}\,{\Ly}\,{\pi^2}-{2}\,{\zeta_3}+{7\over 180}\,{\pi^4}-{5\over 12}\,{\pi^2}\Biggr ){}\,{\tou } \nonumber \\ &&
+{}\Biggl ({}-{2}\,{\Lidx}+{2}\,{\Lidy}+{2}\,{\Lidz}+{2}\,{\Licx}\,{\Ly}-{1\over 3}\,{\Libx}\,{\pi^2}+{1\over 12}\,{\Lx^4}-{1\over 3}\,{\Lx^3}\,{\Ly}\nonumber \\ &&+{1\over 2}\,{\Lx^2}\,{\Ly^2}+{1\over 24}\,{\Lx^2}\,{\Ly}+{1\over 6}\,{\Lx^2}\,{\pi^2}+{1\over 4}\,{\Lx^2}-{1\over 3}\,{\Lx}\,{\Ly}\,{\pi^2}-{1\over 12}\,{\Lx}\,{\pi^2}-{2}\,{\Ly}\,{\zeta_3}+{7\over 90}\,{\pi^4}\Biggr ){}\,{\ttouu } \nonumber \\ &&
+{}\Biggl ({1\over 3}\,{\Licx}-{1\over 3}\,{\Licy}-{1\over 3}\,{\Libx}\,{\Lx}+{1\over 3}\,{\Liby}\,{\Ly}+{1\over 9}\,{\Lx^3}-{1\over 4}\,{\Lx^2}\,{\Ly}-{29\over 24}\,{\Lx^2}\nonumber \\ &&+{1\over 4}\,{\Lx}\,{\Ly^2}+{\Lx}\,{\Ly}+{47\over 72}\,{\Lx}\,{\pi^2}+{5\over 18}\,{\Lx}-{1\over 9}\,{\Ly^3}+{5\over 24}\,{\Ly^2}-{47\over 72}\,{\Ly}\,{\pi^2}-{5\over 18}\,{\Ly}-{1\over 2}\,{\pi^2}\Biggr ){}\,{\tos } \nonumber \\ &&
-{1\over 2}\,{\Ly^2}\,{\sot } \nonumber \\ &&
+{}\Biggl ({}-{1\over 4}\,{\Lx^2}+{1\over 2}\,{\Lx}\,{\Ly}-{1\over 4}\,{\Ly^2}-{1\over 4}\,{\pi^2}\Biggr ){}\,{\ttoss } \nonumber \\ &&
+{}\Biggl ({}\Biggl ({}-{1\over 3}\,{\Libx}+{1\over 8}\,{\Lx^2}-{2\over 3}\,{\Lx}\,{\Ly}-{29\over 72}\,{\Lx}+{5\over 24}\,{\Ly^2}-{77\over 24}\,{\Ly}+{17\over 72}\,{\pi^2}+{19\over 72}\Biggr ){}\,{\one } \nonumber \\ &&
+{}\Biggl ({\Licx}+{2}\,{\Libx}-{1\over 24}\,{\Lx^2}-{1\over 12}\,{\Lx}\,{\Ly}-{3\over 4}\,{\Lx}+{1\over 12}\,{\Ly}-{\zeta_3}+{1\over 2}\Biggr ){}\,{\tou } \nonumber \\ &&
+{}\Biggl ({2}\,{\Licx}+{1\over 24}\,{\Lx^2}+{1\over 12}\,{\Lx}\,{\Ly}+{1\over 2}\,{\Lx}-{2}\,{\zeta_3}\Biggr ){}\,{\ttouu } \nonumber \\ &&
+{}\Biggl ({}-{2\over 3}\,{\Libx}-{1\over 12}\,{\Lx^2}-{1\over 3}\,{\Lx}\,{\Ly}+{11\over 36}\,{\Lx}+{1\over 12}\,{\Ly^2}-{11\over 36}\,{\Ly}+{1\over 18}\,{\pi^2}\Biggr ){}\,{\tos } \nonumber \\ &&
-{\Ly}\,{\sot } 
\Biggr ){}\,{\IPI }\nonumber \\ &&
{}\\
{K^{gg,(2),[3]}_{s+-+-} }&=&{ }\Biggl ({}-{1\over 3}\,{\Licy}+{1\over 3}\,{\Liby}\,{\Ly}+{1\over 24}\,{\Lx^2}\,{\Ly}+{1\over 24}\,{\Lx}\,{\Ly^2}+{1\over 8}\,{\Lx}\,{\Ly}-{1\over 12}\,{\Lx}\,{\pi^2}-{3\over 2}\,{\Ly^2}+{7\over 36}\,{\Ly}\,{\pi^2}\nonumber \\ &&+{29\over 24}\,{\Ly}+{1\over 3}\,{\zeta_3}-{1\over 8}\,{\pi^2}\Biggr ){}\,{\one } \nonumber \\ &&
+{}\Biggl ({}-{2}\,{\Lidx}+{2}\,{\Lidy}+{2}\,{\Lidz}+{2}\,{\Licx}\,{\Ly}+{1\over 3}\,{\Licx}+{4}\,{\Licy}-{1\over 3}\,{\Libx}\,{\Lx}\nonumber \\ &&-{1\over 3}\,{\Libx}\,{\pi^2}-{4}\,{\Liby}\,{\Ly}+{1\over 12}\,{\Lx^4}-{1\over 3}\,{\Lx^3}\,{\Ly}+{1\over 2}\,{\Lx^2}\,{\Ly^2}+{1\over 12}\,{\Lx^2}\,{\Ly}+{1\over 6}\,{\Lx^2}\,{\pi^2}\nonumber \\ &&+{3\over 4}\,{\Lx^2}-{2}\,{\Lx}\,{\Ly^2}-{1\over 3}\,{\Lx}\,{\Ly}\,{\pi^2}-{1\over 4}\,{\Lx}\,{\Ly}-{4\over 9}\,{\Lx}\,{\pi^2}+{5\over 6}\,{\Lx}-{2}\,{\Ly}\,{\zeta_3}+{2\over 3}\,{\Ly}\,{\pi^2}\nonumber \\ &&-{13\over 3}\,{\zeta_3}+{7\over 90}\,{\pi^4}-{17\over 36}\,{\pi^2}\Biggr ){}\,{\tou } \nonumber \\ &&
+{}\Biggl ({}-{4}\,{\Lidx}+{4}\,{\Lidy}+{4}\,{\Lidz}+{4}\,{\Licx}\,{\Ly}-{1\over 3}\,{\Licx}+{1\over 3}\,{\Libx}\,{\Lx}-{2\over 3}\,{\Libx}\,{\pi^2}\nonumber \\ &&+{1\over 6}\,{\Lx^4}-{2\over 3}\,{\Lx^3}\,{\Ly}+{\Lx^2}\,{\Ly^2}+{1\over 24}\,{\Lx^2}\,{\Ly}+{1\over 3}\,{\Lx^2}\,{\pi^2}+{1\over 4}\,{\Lx^2}-{2\over 3}\,{\Lx}\,{\Ly}\,{\pi^2}+{7\over 36}\,{\Lx}\,{\pi^2}\nonumber \\ &&-{4}\,{\Ly}\,{\zeta_3}+{1\over 3}\,{\zeta_3}+{7\over 45}\,{\pi^4}\Biggr ){}\,{\ttouu } \nonumber \\ &&
+{}\Biggl ({1\over 3}\,{\Licx}-{1\over 3}\,{\Licy}-{1\over 3}\,{\Libx}\,{\Lx}+{1\over 3}\,{\Liby}\,{\Ly}+{1\over 4}\,{\Lx^2}-{1\over 2}\,{\Lx}\,{\Ly}-{5\over 18}\,{\Lx}\,{\pi^2}-{1\over 3}\,{\Lx}\nonumber \\ &&+{1\over 4}\,{\Ly^2}+{5\over 18}\,{\Ly}\,{\pi^2}+{1\over 3}\,{\Ly}+{1\over 4}\,{\pi^2}\Biggr ){}\,{\tos } \nonumber \\ &&
-{3\over 4}\,{\Ly^2}\,{\sot } \nonumber \\ &&
+{}\Biggl ({}\Biggl ({}-{1\over 3}\,{\Libx}+{1\over 24}\,{\Lx^2}-{1\over 6}\,{\Lx}\,{\Ly}+{1\over 8}\,{\Lx}-{1\over 8}\,{\Ly^2}-{23\over 8}\,{\Ly}+{29\over 24}\Biggr ){}\,{\one } \nonumber \\ &&
+{}\Biggl ({2}\,{\Licx}+{11\over 3}\,{\Libx}+{1\over 4}\,{\Lx^2}+{1\over 6}\,{\Lx}\,{\Ly}+{5\over 4}\,{\Lx}-{1\over 4}\,{\Ly}-{2}\,{\zeta_3}+{1\over 18}\,{\pi^2}+{5\over 6}\Biggr ){}\,{\tou } \nonumber \\ &&
+{}\Biggl ({4}\,{\Licx}+{1\over 3}\,{\Libx}-{1\over 8}\,{\Lx^2}+{1\over 12}\,{\Lx}\,{\Ly}+{1\over 2}\,{\Lx}-{4}\,{\zeta_3}-{1\over 18}\,{\pi^2}\Biggr ){}\,{\ttouu } \nonumber \\ &&
+{}\Biggl ({}-{2\over 3}\,{\Libx}+{1\over 6}\,{\Lx^2}-{1\over 3}\,{\Lx}\,{\Ly}-{1\over 6}\,{\Ly^2}+{1\over 18}\,{\pi^2}\Biggr ){}\,{\tos } \nonumber \\ &&
-{3\over 2}\,{\Ly}\,{\sot } 
\Biggr ){}\,{\IPI }\nonumber \\ &&
{}\\
{L^{gg,(2),[3]}_{s+-+-} }&=&{ }\Biggl ({}-{1\over 72}\,{\Lx}\,{\Ly}-{1\over 36}\,{\Ly^2}+{5\over 54}\,{\Ly}-{5\over 72}\,{\pi^2}\Biggr ){}\,{\one } 
+{}\Biggl ({1\over 36}\,{\Lx^2}-{5\over 54}\,{\Lx}-{1\over 36}\,{\Ly^2}+{5\over 54}\,{\Ly}\Biggr ){}\,{\tos } \nonumber \\ &&
+{}\Biggl ({}\Biggl ({}-{1\over 72}\,{\Lx}+{1\over 24}\,{\Ly}+{5\over 54}\Biggr ){}\,{\one } 
+{}\Biggl ({}-{1\over 18}\,{\Lx}+{1\over 18}\,{\Ly}\Biggr ){}\,{\tos } 
\Biggr ){}\,{\IPI }\nonumber \\ &&
\end{eqnarray}

\subsection{$q(p_2,+) + \bar q(p_1,-)\to g(p_3,-)+g(p_4,+)$}

The finite remainders are obtained by Bose symmetry from the 
$q(p_2,+) + \bar q(p_1,-)\to g(p_3,+)+g(p_4,-)$ process.
\begin{eqnarray}
{A^{gg,(2),[1]}_{s+--+} } &=& {A^{gg,(2),[2]}_{s+-+-} } (t \leftrightarrow u),\nonumber \\
{A^{gg,(2),[2]}_{s+--+} } &=& {A^{gg,(2),[1]}_{s+-+-} } (t \leftrightarrow u),\nonumber \\
{A^{gg,(2),[3]}_{s+--+} } &=& {A^{gg,(2),[3]}_{s+-+-} } (t \leftrightarrow u),
\end{eqnarray}
etc.

\subsection{$q(p_2,+) + \bar q(p_1,-)\to g(p_3,-)+g(p_4,-)$}

The finite remainders are obtained by Bose symmetry from the 
$q(p_2,+) + \bar q(p_1,-)\to g(p_3,+)+g(p_4,-)$ process.
\begin{eqnarray}
{A^{gg,(2),[1]}_{s+---} } &=& {A^{gg,(2),[2]}_{s+-++} } (t \leftrightarrow u),\nonumber \\
{A^{gg,(2),[2]}_{s+---} } &=& {A^{gg,(2),[1]}_{s+-++} } (t \leftrightarrow u),\nonumber \\
{A^{gg,(2),[3]}_{s+---} } &=& {A^{gg,(2),[3]}_{s+-++} } (t \leftrightarrow u),
\end{eqnarray}
etc.

\newpage
\section{Finite one-loop contributions}
\label{app:oneloop}

In this appendix we list the coefficients $A,\ldots,E$ for the finite one-loop amplitudes defined in Section~\ref{subsec:oneloopamps} for $q\bar q \to gg$.
Expressions for quark-gluon scattering as well as the processes involving photons are available from the authors as a FORM output file.

\subsection{$q(p_2,+) + \bar q(p_1,-)\to g(p_3,+)+g(p_4,+)$}

\begin{eqnarray}
{A^{gg,(1),[1]}_{s+-++} }&=&{ }-{1\over 2}\,{\sot } -{1\over 3}\,{\one } \nonumber \\   
{B^{gg,(1),[1]}_{s+-++} }&=&{ }-{1\over 2}\,{\sot } \nonumber \\  
{C^{gg,(1),[1]}_{s+-++} }&=&{ 1\over 3}\,{\one } \nonumber \\   
{A^{gg,(1),[2]}_{s+-++} }&=&{ }-{1\over 2}\,{\tou } -{1\over 6}\,{\one } \nonumber \\   
{B^{gg,(1),[2]}_{s+-++} }&=&{ }-{1\over 2}\,{\tou } -{1\over 2}\,{\one } \nonumber \\  
{C^{gg,(1),[2]}_{s+-++} }&=&{ }-{1\over 3}\,{\one } \nonumber \\  
{D^{gg,(1),[3]}_{s+-++} }=
{E^{gg,(1),[3]}_{s+-++} }&=&{ 0} 
\end{eqnarray}

\subsection{$q(p_2,+) + \bar q(p_1,-)\to g(p_3,+)+g(p_4,-)$}
\begin{eqnarray}
{A^{gg,(1),[1]}_{s+-+-} }&=&{ }\Biggl ({}-{1\over 2}\,{\Lx^2}+{11\over 6}\,{\Lx}+{7\over 2}\Biggr ){}\,{\one } 
+{}\Biggl ({}-{1\over 2}\,{\Lx}+{1\over 2}\Biggr ){}\,{\tou } 
+{\Lx}\,{\ttouu }  
+{1\over 2}\,{\Lx^2}\,{\tttouuu } \nonumber \\ &&
+{}\Biggl ({}\Biggl ({}-{\Lx}-{11\over 6}\Biggr ){}\,{\one } 
-{1\over 2}\,{\tou } 
+{\ttouu } 
+{\Lx}\,{\tttouuu } 
\Biggr ){}\,{\IPI }\\
{B^{gg,(1),[1]}_{s+-+-} }&=&{ }\Biggl ({1\over 2}\,{\Lx^2}+{3\over 2}\,{\Lx}-{7\over 2}\Biggr ){}\,{\one } 
+{}\Biggl ({3\over 2}\,{\Lx^2}+{5\over 2}\,{\Lx}+{1\over 2}\Biggr ){}\,{\tou } 
+{}\Biggl ({3\over 2}\,{\Lx^2}+{\Lx}\Biggr ){}\,{\ttouu } 
+{1\over 2}\,{\Lx^2}\,{\tttouuu } \nonumber \\ &&
+{}\Biggl ({}\Biggl ({\Lx}+{3\over 2}\Biggr ){}\,{\one }  
+{}\Biggl ({3}\,{\Lx}+{5\over 2}\Biggr ){}\,{\tou } 
+{}\Biggl ({3}\,{\Lx}+{1}\Biggr ){}\,{\ttouu } 
+{\Lx}\,{\tttouuu }  
\Biggr ){}\,{\IPI } \\ 
{C^{gg,(1),[1]}_{s+-+-} }&=&{ }-{1\over 3}\,{\Lx}\,{\one } 
+{1\over 3}\,{\IPI }\\ 
{A^{gg,(1),[2]}_{s+-+-} }&=&{ }\Biggl ({}-{\Ly^2}+{10\over 3}\,{\Ly}+{3}\Biggr ){}\,{\one } 
-{1\over 2}\,{\Ly^2}\,{\sot } 
+{}\Biggl ({}\Biggl ({}-{2}\,{\Ly}-{1\over 3}\Biggr ){}\,{\one } 
-{\Ly}\,{\sot } 
\Biggr ){}\,{\IPI }\\ 
{B^{gg,(1),[2]}_{s+-+-} }&=&{ }-{4}\,{\one }  
-{1\over 2}\,{\Ly^2}\,{\sot }  
-{\Ly}\,{\IPI } 
{}\,{\sot }  \\ 
{C^{gg,(1),[2]}_{s+-+-} }&=&{ }-{1\over 3}\,{\Ly}\,{\one } 
+{1\over 3}\,{\IPI }\nonumber \\ &&
{}\,{\one }  
\\
{D^{gg,(1),[3]}_{s+-+-} }&=&{ }\Biggl ({\Lx}\,{\Ly}-{1\over 2}\,{\Ly^2}+{5\over 3}\,{\Ly}-{1\over 2}\,{\pi^2}\Biggr ){}\,{\one } 
+{3\over 4}\,{\Lx^2}\,{\tou } \nonumber \\ &&
+{}\Biggl ({1\over 2}\,{\Lx^2}-{5\over 3}\,{\Lx}-{1\over 2}\,{\Ly^2}+{5\over 3}\,{\Ly}\Biggr ){}\,{\tos } \nonumber \\ &&
+{}\Biggl ({}\Biggl ({\Lx}+{5\over 3}\Biggr ){}\,{\one }  
+{3\over 2}\,{\Lx}\,{\tou } 
+{}\Biggl ({\Lx}-{\Ly}\Biggr ){}\,{\tos }  
\Biggr ){}\,{\IPI }
{}\\
{E^{gg,(1),[3]}_{s+-+-} }&=&{ }-{1\over 6}\,{\Ly}\,{\one } 
-{1\over 6}\,{\IPI } 
{}\,{\one }  
+{}\Biggl ({1\over 6}\,{\Lx}-{1\over 6}\,{\Ly}\Biggr ){}\,{\tos }
\end{eqnarray}

\subsection{$q(p_2,+) + \bar q(p_1,-)\to g(p_3,-)+g(p_4,+)$}

The finite remainders are obtained by Bose symmetry from the 
$q(p_2,+) + \bar q(p_1,-)\to g(p_3,+)+g(p_4,-)$ process.
\begin{eqnarray}
{A^{gg,(1),[1]}_{s+--+} } &=& {A^{gg,(1),[2]}_{s+-+-} } (t \leftrightarrow u),\nonumber \\
{A^{gg,(1),[2]}_{s+--+} } &=& {A^{gg,(1),[1]}_{s+-+-} } (t \leftrightarrow u),\nonumber \\
{A^{gg,(1),[3]}_{s+--+} } &=& {A^{gg,(1),[3]}_{s+-+-} } (t \leftrightarrow u),
\end{eqnarray}
etc.

\subsection{$q(p_2,+) + \bar q(p_1,-)\to g(p_3,-)+g(p_4,-)$}

The finite remainders are obtained by Bose symmetry from the 
$q(p_2,+) + \bar q(p_1,-)\to g(p_3,+)+g(p_4,-)$ process.
\begin{eqnarray}
{A^{gg,(1),[1]}_{s+---} } &=& {A^{gg,(1),[2]}_{s+-++} } (t \leftrightarrow u),\nonumber \\
{A^{gg,(1),[2]}_{s+---} } &=& {A^{gg,(1),[1]}_{s+-++} } (t \leftrightarrow u),\nonumber \\
{A^{gg,(1),[3]}_{s+---} } &=& {A^{gg,(1),[3]}_{s+-++} } (t \leftrightarrow u),
\end{eqnarray}
etc.
\newpage
\bibliographystyle{JHEP}
\bibliography{qqgghel}

\providecommand{\href}[2]{#2}\begingroup\raggedright\begin{thebibliography}{10}

\bibitem{qqgg}
C.~Anastasiou, E.~W.~N. Glover, C.~Oleari, and M.~E. Tejeda-Yeomans, {\it
  Two-loop QCD corrections to massless quark gluon scattering},  {\em Nucl.
  Phys.} {\bf B605} (2001) 486--516,
  [\href{http://xxx.lanl.gov/abs/hep-ph/0101304}{{\tt hep-ph/0101304}}].

\bibitem{qqpp}
C.~Anastasiou, E.~W.~N. Glover, and M.~E. Tejeda-Yeomans, {\it Two-loop QED and
  QCD corrections to massless fermion boson scattering},  {\em Nucl. Phys.}
  {\bf B629} (2002) 255--289,
  [\href{http://xxx.lanl.gov/abs/hep-ph/0201274}{{\tt hep-ph/0201274}}].

\bibitem{BFDqqgg}
Z.~Bern, A.~De~Freitas, and L.~Dixon, {\it Two-loop helicity amplitudes for
  quark-gluon scattering in QCD and gluino-gluon scattering in supersymmetric
  Yang-Mills theory},  \href{http://xxx.lanl.gov/abs/hep-ph/0304168}{{\tt
  hep-ph/0304168}}.

\bibitem{nigelRCLL}
E.~W.~N. Glover, {\it Progress in NNLO calculations for scattering processes},
  \href{http://xxx.lanl.gov/abs/hep-ph/0211412}{{\tt hep-ph/0211412}}.

\bibitem{6g1}
J.~F. Gunion and J.~Kalinowski, {\it A direct calculation of the six gluon
  subprocess},  {\em Phys. Rev.} {\bf D34} (1986) 2119.

\bibitem{6g2}
S.~J. Parke and T.~R. Taylor, {\it Gluonic two goes to four},  {\em Nucl.
  Phys.} {\bf B269} (1986) 410.

\bibitem{6g3}
F.~A. Berends and W.~Giele, {\it The six gluon process as an example of
  Weyl-van der Waerden spinor calculus},  {\em Nucl. Phys.} {\bf B294} (1987)
  700.

\bibitem{6g4}
M.~L. Mangano, S.~Parke, and Z.~Xu, {\it Duality and multi - gluon scattering},
   {\em Nucl. Phys.} {\bf B298} (1988) 653.

\bibitem{4g2q1}
Z.~Kunszt, {\it Combined use of the CALKUL method and N=1 supersymmetry to
  calculate QCD six parton processes},  {\em Nucl. Phys.} {\bf B271} (1986)
  333.

\bibitem{4g2q2}
S.~J. Parke and T.~R. Taylor, {\it The cross-section for hard processes
  involving two quarks and four gluons},  {\em Phys. Rev.} {\bf D35} (1987)
  313.

\bibitem{2g4q}
J.~F. Gunion and Z.~Kunszt, {\it Four jet processes: Gluon-gluon scattering to
  nonidentical quark - anti-quark pairs},  {\em Phys. Lett.} {\bf B159} (1985)
  167.

\bibitem{6q}
J.~F. Gunion and Z.~Kunszt, {\it Six quark subprocesses in QCD},  {\em Phys.
  Lett.} {\bf B176} (1986) 163.

\bibitem{5g}
Z.~Bern, L.~J. Dixon, and D.~A. Kosower, {\it One loop corrections to five
  gluon amplitudes},  {\em Phys. Rev. Lett.} {\bf 70} (1993) 2677--2680,
  [\href{http://xxx.lanl.gov/abs/hep-ph/9302280}{{\tt hep-ph/9302280}}].

\bibitem{3g2q}
Z.~Bern, L.~J. Dixon, and D.~A. Kosower, {\it One loop corrections to two quark
  three gluon amplitudes},  {\em Nucl. Phys.} {\bf B437} (1995) 259--304,
  [\href{http://xxx.lanl.gov/abs/hep-ph/9409393}{{\tt hep-ph/9409393}}].

\bibitem{1g4q}
Z.~Kunszt, A.~Signer, and Z.~Trocsanyi, {\it One loop radiative corrections to
  the helicity amplitudes of QCD processes involving four quarks and one
  gluon},  {\em Phys. Lett.} {\bf B336} (1994) 529--536,
  [\href{http://xxx.lanl.gov/abs/hep-ph/9405386}{{\tt hep-ph/9405386}}].

\bibitem{moms1}
S.~A. Larin, P.~Nogueira, T.~van Ritbergen, and J.~A.~M. Vermaseren, {\it The
  3-loop QCD calculation of the moments of deep inelastic structure functions},
   {\em Nucl. Phys.} {\bf B492} (1997) 338--378,
  [\href{http://xxx.lanl.gov/abs/hep-ph/9605317}{{\tt hep-ph/9605317}}].

\bibitem{moms2}
S.~A. Larin, T.~van Ritbergen, and J.~A.~M. Vermaseren, {\it The next-to-next-to-leading QCD approximation for nonsinglet moments of deep inelastic
  structure functions},  {\em Nucl. Phys.} {\bf B427} (1994) 41--52.

\bibitem{moms3}
A.~Retey and J.~A.~M. Vermaseren, {\it Some higher moments of deep inelastic
  structure functions at next-to-next-to-leading order of perturbative QCD},
  {\em Nucl. Phys.} {\bf B604} (2001) 281--311,
  [\href{http://xxx.lanl.gov/abs/hep-ph/0007294}{{\tt hep-ph/0007294}}].

\bibitem{Gra1}
J.~A. Gracey, {\it Anomalous dimension of nonsinglet wilson operators at O(1/NF) in deep inelastic scattering},  {\em Phys. Lett.} {\bf B322} (1994)
  141--146, [\href{http://xxx.lanl.gov/abs/hep-ph/9401214}{{\tt
  hep-ph/9401214}}].

\bibitem{NV1}
W.~L. van Neerven and A.~Vogt, {\it NNLO evolution of deep-inelastic structure
  functions: The non-singlet case},  {\em Nucl. Phys.} {\bf B568} (2000)
  263--286, [\href{http://xxx.lanl.gov/abs/hep-ph/9907472}{{\tt
  hep-ph/9907472}}].

\bibitem{NV2}
W.~L. van Neerven and A.~Vogt, {\it NNLO evolution of deep-inelastic structure
  functions: The singlet case},  {\em Nucl. Phys.} {\bf B588} (2000) 345--373,
  [\href{http://xxx.lanl.gov/abs/hep-ph/0006154}{{\tt hep-ph/0006154}}].

\bibitem{NVplb}
W.~L. van Neerven and A.~Vogt, {\it Improved approximations for the three-loop
  splitting functions in QCD},  {\em Phys. Lett.} {\bf B490} (2000) 111--118,
  [\href{http://xxx.lanl.gov/abs/hep-ph/0007362}{{\tt hep-ph/0007362}}].

\bibitem{VMVdis}
J.~A.~M. Vermaseren, S.~Moch, and A.~Vogt, {\it First results for three-loop
  deep-inelastic structure functions in QCD},
  \href{http://xxx.lanl.gov/abs/hep-ph/0211296}{{\tt hep-ph/0211296}}.

\bibitem{carola}
C.~F. Berger, {\it Higher orders in O$(\alpha_s)/(1-x)_+$ of
  non-singlet partonic splitting functions},  {\em Phys. Rev.} {\bf D66} (2002)
  116002, [\href{http://xxx.lanl.gov/abs/hep-ph/0209107}{{\tt
  hep-ph/0209107}}].

\bibitem{KST}
Z.~Kunszt, A.~Signer, and Z.~Trocsanyi, {\it One loop helicity amplitudes for
  all 2 $\to$ 2 processes in QCD and N=1 supersymmetric Yang-Mills theory},
  {\em Nucl. Phys.} {\bf B411} (1994) 397--442,
  [\href{http://xxx.lanl.gov/abs/hep-ph/9305239}{{\tt hep-ph/9305239}}].

\bibitem{BFDggpp}
Z.~Bern, A.~De~Freitas, and L.~J. Dixon, {\it Two-loop amplitudes for gluon
  fusion into two photons},  {\em JHEP} {\bf 09} (2001) 037,
  [\href{http://xxx.lanl.gov/abs/hep-ph/0109078}{{\tt hep-ph/0109078}}].

\bibitem{BFDGpppp}
Z.~Bern, A.~De~Freitas, L.~J. Dixon, A.~Ghinculov, and H.~L. Wong, {\it QCD and
  QED corrections to light-by-light scattering},  {\em JHEP} {\bf 11} (2001)
  031, [\href{http://xxx.lanl.gov/abs/hep-ph/0109079}{{\tt hep-ph/0109079}}].

\bibitem{BGMBpppp}
T.~Binoth, E.~W.~N. Glover, P.~Marquard, and J.~J. van~der Bij, {\it Two-loop
  corrections to light-by-light scattering in supersymmetric QED},  {\em JHEP}
  {\bf 05} (2002) 060, [\href{http://xxx.lanl.gov/abs/hep-ph/0202266}{{\tt
  hep-ph/0202266}}].

\bibitem{BDKgggg}
Z.~Bern, L.~J. Dixon, and D.~A. Kosower, {\it A two-loop four-gluon helicity
  amplitude in QCD},  {\em JHEP} {\bf 01} (2000) 027,
  [\href{http://xxx.lanl.gov/abs/hep-ph/0001001}{{\tt hep-ph/0001001}}].

\bibitem{BFDgggg}
Z.~Bern, A.~De~Freitas, and L.~Dixon, {\it Two-loop helicity amplitudes for
  gluon gluon scattering in QCD and supersymmetric Yang-Mills theory},  {\em
  JHEP} {\bf 03} (2002) 018,
  [\href{http://xxx.lanl.gov/abs/hep-ph/0201161}{{\tt hep-ph/0201161}}].

\bibitem{helamp3j}
L.~W. Garland, T.~Gehrmann, E.~W.~N. Glover, A.~Koukoutsakis, and E.~Remiddi,
  {\it Two-loop QCD helicity amplitudes for $e^+ e^- \to 3$~jets},  {\em Nucl.
  Phys.} {\bf B642} (2002) 227--262,
  [\href{http://xxx.lanl.gov/abs/hep-ph/0206067}{{\tt hep-ph/0206067}}].

\bibitem{houchesQCD}
W.~Giele {\em et.~al.}, {\it The QCD/SM working group: Summary report},
  \href{http://xxx.lanl.gov/abs/hep-ph/0204316}{{\tt hep-ph/0204316}}.

\bibitem{thomasRCLL}
T.~Gehrmann, {\it Jet physics at two-loop accuracy},
  \href{http://xxx.lanl.gov/abs/hep-ph/0210157}{{\tt hep-ph/0210157}}.

\bibitem{thomasHAD}
T.~Gehrmann, {\it Theoretical developments on hard QCD processes at colliders},
   \href{http://xxx.lanl.gov/abs/hep-ph/0211169}{{\tt hep-ph/0211169}}.

\bibitem{laporta}
S.~Laporta, {\it High-precision calculation of multi-loop Feynman integrals by
  difference equations},  {\em Int. J. Mod. Phys.} {\bf A15} (2000) 5087--5159,
  [\href{http://xxx.lanl.gov/abs/hep-ph/0102033}{{\tt hep-ph/0102033}}].

\bibitem{IBP1}
K.~G. Chetyrkin, A.~L. Kataev, and F.~V. Tkachov, {\it New approach to
  evaluation of multiloop Feynman integrals: The gegenbauer polynomial x space
  technique},  {\em Nucl. Phys.} {\bf B174} (1980) 345--377.

\bibitem{IBP2}
K.~G. Chetyrkin and F.~V. Tkachov, {\it Integration by parts: The algorithm to
  calculate beta functions in 4 loops},  {\em Nucl. Phys.} {\bf B192} (1981)
  159--204.

\bibitem{diffeqLI}
T.~Gehrmann and E.~Remiddi, {\it Differential equations for two-loop four-point
  functions},  {\em Nucl. Phys.} {\bf B580} (2000) 485--518,
  [\href{http://xxx.lanl.gov/abs/hep-ph/9912329}{{\tt hep-ph/9912329}}].

\bibitem{xtri1}
R.~J. Gonsalves, {\it Dimensionally regularized two loop on-shell quark form-
  factor},  {\em Phys. Rev.} {\bf D28} (1983) 1542.

\bibitem{xtri2}
G.~Kramer and B.~Lampe, {\it Integrals for two loop calculations in massless
  QCD},  {\em J. Math. Phys.} {\bf 28} (1987) 945.

\bibitem{AGO1}
C.~Anastasiou, E.~W.~N. Glover, and C.~Oleari, {\it Scalar one-loop integrals
  using the negative-dimension approach},  {\em Nucl. Phys.} {\bf B572} (2000)
  307--360, [\href{http://xxx.lanl.gov/abs/hep-ph/9907494}{{\tt
  hep-ph/9907494}}].

\bibitem{AGO2}
C.~Anastasiou, E.~W.~N. Glover, and C.~Oleari, {\it Application of the
  negative-dimension approach to massless scalar box integrals},  {\em Nucl.
  Phys.} {\bf B565} (2000) 445--467,
  [\href{http://xxx.lanl.gov/abs/hep-ph/9907523}{{\tt hep-ph/9907523}}].

\bibitem{AGO3}
C.~Anastasiou, E.~W.~N. Glover, and C.~Oleari, {\it The two-loop scalar and
  tensor pentabox graph with light-like legs},  {\em Nucl. Phys.} {\bf B575}
  (2000) 416--436, [\href{http://xxx.lanl.gov/abs/hep-ph/9912251}{{\tt
  hep-ph/9912251}}]. Erratum-ibid.B585:763-770,2000.

\bibitem{planarA}
V.~A. Smirnov, {\it Analytical result for dimensionally regularized massless
  on-shell double box},  {\em Phys. Lett.} {\bf B460} (1999) 397--404,
  [\href{http://xxx.lanl.gov/abs/hep-ph/9905323}{{\tt hep-ph/9905323}}].

\bibitem{nonplanarA}
J.~B. Tausk, {\it Non-planar massless two-loop Feynman diagrams with four on-
  shell legs},  {\em Phys. Lett.} {\bf B469} (1999) 225--234,
  [\href{http://xxx.lanl.gov/abs/hep-ph/9909506}{{\tt hep-ph/9909506}}].

\bibitem{planarB}
V.~A. Smirnov and O.~L. Veretin, {\it Analytical results for dimensionally
  regularized massless on-shell double boxes with arbitrary indices and
  numerators},  {\em Nucl. Phys.} {\bf B566} (2000) 469--485,
  [\href{http://xxx.lanl.gov/abs/hep-ph/9907385}{{\tt hep-ph/9907385}}].

\bibitem{nonplanarB}
C.~Anastasiou, T.~Gehrmann, C.~Oleari, E.~Remiddi, and J.~B. Tausk, {\it The
  tensor reduction and master integrals of the two-loop massless crossed box
  with light-like legs},  {\em Nucl. Phys.} {\bf B580} (2000) 577--601,
  [\href{http://xxx.lanl.gov/abs/hep-ph/0003261}{{\tt hep-ph/0003261}}].

\bibitem{planarIR}
C.~Anastasiou, J.~B. Tausk, and M.~E. Tejeda-Yeomans, {\it The on-shell
  massless planar double box diagram with an irreducible numerator},  {\em
  Nucl. Phys. Proc. Suppl.} {\bf 89} (2000) 262--267,
  [\href{http://xxx.lanl.gov/abs/hep-ph/0005328}{{\tt hep-ph/0005328}}].

\bibitem{catani}
S.~Catani, {\it The singular behaviour of {QCD} amplitudes at two-loop order},
  {\em Phys. Lett.} {\bf B427} (1998) 161--171,
  [\href{http://xxx.lanl.gov/abs/hep-ph/9802439}{{\tt hep-ph/9802439}}].

\bibitem{hv1}
G.~'t~Hooft and M.~J.~G. Veltman, {\it Regularization and renormalization of
  gauge fields},  {\em Nucl. Phys.} {\bf B44} (1972) 189--213.

\bibitem{hv2}
G.~'t~Hooft, {\it Dimensional regularization and the renormalization group},
  {\em Nucl. Phys.} {\bf B61} (1973) 455--468.

\bibitem{dreg1}
C.~G. Bollini and J.~J. Giambiagi, {\it Dimensional renormalization: The number
  of dimensions as a regularizing parameter},  {\em Nuovo Cim.} {\bf B12}
  (1972) 20--25.

\bibitem{dreg2}
G.~M. Cicuta and E.~Montaldi, {\it Analytic renormalization via continuous
  space dimension},  {\em Nuovo Cim. Lett.} {\bf 4} (1972) 329--332.

\bibitem{QGRAF}
P.~Nogueira, {\it Automatic Feynman graph generation},  {\em J. Comput. Phys.}
  {\bf 105} (1993) 279--289.

\bibitem{MAPLE}
{\it {\rm MAPLE V} release 7}, . by Waterloo Maple Inc. Copyright 2001.

\bibitem{FORM3}
J.~A.~M. Vermaseren, {\it New features of FORM},
  \href{http://xxx.lanl.gov/abs/math-ph/0010025}{{\tt math-ph/0010025}}.

\bibitem{stermanTY}
G.~Sterman and M.~E. Tejeda-Yeomans, {\it Multi-loop amplitudes and
  resummation},  {\em Phys. Lett.} {\bf B552} (2003) 48--56,
  [\href{http://xxx.lanl.gov/abs/hep-ph/0210130}{{\tt hep-ph/0210130}}].

\bibitem{BDGbha}
Z.~Bern, L.~J. Dixon, and A.~Ghinculov, {\it Two-loop correction to Bhabha
  scattering},  {\em Phys. Rev.} {\bf D63} (2001) 053007,
  [\href{http://xxx.lanl.gov/abs/hep-ph/0010075}{{\tt hep-ph/0010075}}].

\bibitem{qqunlike}
C.~Anastasiou, E.~W.~N. Glover, C.~Oleari, and M.~E. Tejeda-Yeomans, {\it
  Two-loop QCD corrections to $q \bar q \to q' \bar q'$},  {\em Nucl. Phys.} {\bf
  B601} (2001) 318--340, [\href{http://xxx.lanl.gov/abs/hep-ph/0010212}{{\tt
  hep-ph/0010212}}].

\bibitem{qqlike}
C.~Anastasiou, E.~W.~N. Glover, C.~Oleari, and M.~E. Tejeda-Yeomans, {\it
  Two-loop QCD corrections to $q \bar q \to q \bar q$},  {\em Nucl. Phys.} {\bf
  B601} (2001) 341--360, [\href{http://xxx.lanl.gov/abs/hep-ph/0011094}{{\tt
  hep-ph/0011094}}].

\bibitem{gggg}
E.~W.~N. Glover, C.~Oleari, and M.~E. Tejeda-Yeomans, {\it Two-loop QCD
  corrections to gluon gluon scattering},  {\em Nucl. Phys.} {\bf B605} (2001)
  467--485, [\href{http://xxx.lanl.gov/abs/hep-ph/0102201}{{\tt
  hep-ph/0102201}}].

\bibitem{KOS}
N.~Kidonakis, G.~Oderda, and G.~Sterman, {\it Evolution of color exchange in
  {QCD} hard scattering},  {\em Nucl. Phys.} {\bf B531} (1998) 365--402,
  [\href{http://xxx.lanl.gov/abs/hep-ph/9803241}{{\tt hep-ph/9803241}}].

\bibitem{sone3}
S.~Catani and M.~Grazzini, {\it The soft-gluon current at one-loop order},
  {\em Nucl. Phys.} {\bf B591} (2000) 435--454,
  [\href{http://xxx.lanl.gov/abs/hep-ph/0007142}{{\tt hep-ph/0007142}}].

\bibitem{metythesis}
M.~E. Tejeda-Yeomans, {\it Parton-parton scattering at two-loops},  {\em
  University of Durham, Ph.D. Thesis} (2001).

\bibitem{sone1}
Z.~Bern, V.~Del~Duca, and C.~R. Schmidt, {\it The infrared behavior of one-loop
  gluon amplitudes at next-to-next-to-leading order},  {\em Phys. Lett.} {\bf
  B445} (1998) 168--177, [\href{http://xxx.lanl.gov/abs/hep-ph/9810409}{{\tt
  hep-ph/9810409}}].

\bibitem{sone2}
Z.~Bern, V.~Del~Duca, W.~B. Kilgore, and C.~R. Schmidt, {\it The infrared
  behavior of one-loop {QCD} amplitudes at next-to-next-to-leading order},
  {\em Phys. Rev.} {\bf D60} (1999) 116001,
  [\href{http://xxx.lanl.gov/abs/hep-ph/9903516}{{\tt hep-ph/9903516}}].

\bibitem{tc1}
J.~M. Campbell and E.~W.~N. Glover, {\it Double unresolved approximations to
  multiparton scattering amplitudes},  {\em Nucl. Phys.} {\bf B527} (1998)
  264--288, [\href{http://xxx.lanl.gov/abs/hep-ph/9710255}{{\tt
  hep-ph/9710255}}].

\bibitem{tc2}
S.~Catani and M.~Grazzini, {\it Collinear factorization and splitting functions
  for next- to-next-to-leading order {QCD} calculations},  {\em Phys. Lett.}
  {\bf B446} (1999) 143--152,
  [\href{http://xxx.lanl.gov/abs/hep-ph/9810389}{{\tt hep-ph/9810389}}].

\bibitem{tc3}
S.~Catani and M.~Grazzini, {\it Infrared factorization of tree level QCD
  amplitudes at the next-to-next-to-leading order and beyond},  {\em Nucl.
  Phys.} {\bf B570} (2000) 287--325,
  [\href{http://xxx.lanl.gov/abs/hep-ph/9908523}{{\tt hep-ph/9908523}}].

\bibitem{ds1}
F.~A. Berends and W.~T. Giele, {\it Multiple soft gluon radiation in parton
  processes},  {\em Nucl. Phys.} {\bf B313} (1989) 595.

\bibitem{ds2}
S.~Catani, {\it Proceedings of the workshop on new techniques for calculating
  higher order QCD corrections},  {\em ETH-TH} {\bf 93-01} (1992). Zurich.

\bibitem{Kosower}
D.~A. Kosower, {\it All-orders singular emission in gauge theories},
  \href{http://xxx.lanl.gov/abs/hep-ph/0301069}{{\tt hep-ph/0301069}}.

\bibitem{subsNNLO}
S.~Weinzierl, {\it Subtraction terms at NNLO},
  \href{http://xxx.lanl.gov/abs/hep-ph/0302180}{{\tt hep-ph/0302180}}.

\bibitem{kolbig}
K.~S. Kolbig, J.~A. Mignaco, and E.~Remiddi, {\it On Nielsen's generalised
  polylogarithms and their numerical calculation},  {\em B.I.T.} {\bf 10}
  (1970) 38.

\end{thebibliography}\endgroup

\end{document}